\documentclass{aa}
\usepackage{graphicx}
\usepackage{txfonts}
\usepackage{color}
\usepackage{hyperref}

\hypersetup{
      pdftex=true, 
   	  colorlinks=true,                                
      breaklinks=true,
      linkcolor=blue,                                 
      citecolor=blue,                                 
      filecolor=blue,                                 
      urlcolor=blue,
      unicode=false,                                  
      pdftoolbar=true,                                
      pdfmenubar=true,                                
      pdffitwindow=false,                             
      pdfstartview={Fit},                             
      pdftitle={VISION},                              
      pdfauthor={Stefan Meingast},                    
      pdfsubject={VISTA survey in Orion},                                     
      pdfcreator={Stefan Meingast},                   
      pdfkeywords={Techniques: image processing -- Methods: data analysis -- Stars: formation -- Stars: pre-main sequence}, 
      pdfnewwindow=true,                              
      pdfdisplaydoctitle=true                         
}

\begin{document} 

\title{VISION - Vienna survey in Orion}
\subtitle{I. VISTA Orion~A Survey\thanks{Based on observations made with ESO Telescopes at the La Silla Paranal Observatory under program ID 090.C-0797(A)}}

\author{Stefan Meingast\inst{1} 
                \and Jo\~ao Alves\inst{1}
        \and Diego Mardones\inst{2}
        \and Paula Teixeira\inst{1}
        \and Marco Lombardi\inst{3}
        \and Josefa Gro\ss schedl\inst{1}
        \and Joana Ascenso\inst{4,5}
        \and Herve Bouy\inst{6}
        \and Jan Forbrich\inst{1}
        \and Alyssa Goodman\inst{7}
        \and Alvaro Hacar\inst{1}
        \and Birgit Hasenberger\inst{1}
        \and Jouni Kainulainen\inst{8}
        \and Karolina Kubiak\inst{1}
        \and Charles Lada\inst{7}
        \and Elizabeth Lada\inst{9}
        \and André Moitinho\inst{10}
        \and Monika Petr-Gotzens\inst{11}
        \and Lara Rodrigues\inst{2}
        \and Carlos G. Rom\'an-Z\'u\~niga\inst{12}
        }
        
\institute{
        Department of Astrophysics, University of Vienna, T\"urkenschanzstrasse 17, 1180 Wien, Austria
    \and Departamento de Astronom\'ia, Universidad de Chile, Casilla 36-D, Santiago, Chile
    \and University of Milan, Department of Physics, via Celoria 16, 20133 Milan, Italy
    \and CENTRA, Instituto Superior Tecnico, Universidade de Lisboa, Av. Rovisco Pais 1, 1049-001 Lisbon, Portugal 
    \and Universidade do Porto, Departamento de Engenharia Física da Faculdade de Engenharia, Rua Dr. Roberto Frias, s/n, P-4200-465 Porto, Portugal
    \and Centro de Astrobiolog\'ia, INTA-CSIC, Depto Astrof\'isica, PO Box 78, 28691 Villanueva de la Ca\~nada, Madrid, Spain
    \and Harvard-Smithsonian Center for Astrophysics, 60 Garden Street, Cambridge, MA 02138, USA
    \and Max-Planck-Institute for Astronomy, Königstuhl 17, 69117 Heidelberg, Germany
    \and Astronomy Department, University of Florida, Gainesville, FL 32611, USA
    \and SIM/CENTRA, Faculdade de Ciencias de Universidade de Lisboa, Ed. C8, Campo Grande, P-1749-016 Lisboa, Portugal
    \and European Southern Observatory, Karl-Schwarzschild-Str. 2, 85748 Garching, Germany
    \and Instituto de Astronom\'ia, UNAM, Ensenada, C.P. 22860, Baja California, Mexico
    }
    
\date{Received September 15, 1996; accepted March 16, 1997}

\abstract
        {Orion~A hosts the nearest massive star factory, thus offering a unique opportunity to resolve the processes connected with the formation of both low- and high-mass stars. Here we present the most detailed and sensitive near-infrared (NIR) observations of the entire molecular cloud to date.}
    {With the unique combination of high image quality, survey coverage, and sensitivity, our NIR survey of Orion~A aims at establishing a solid empirical foundation for further studies of this important cloud. In this first paper we present the observations, data reduction, and source catalog generation. To demonstrate the data quality, we present a first application of our catalog to estimate the number of stars currently forming inside Orion~A and to verify the existence of a more evolved young foreground population.}
    {We used the European Southern Observatory's (ESO) Visible and Infrared Survey Telescope for Astronomy (VISTA) to survey the entire Orion~A molecular cloud in the NIR $J, H$, and $K_S$ bands, covering a total of $\sim$18.3 deg$^2$. We implemented all data reduction recipes independently of the ESO pipeline. Estimates of the young populations toward Orion~A are derived via the $K_S$-band luminosity function.}
    {Our catalog (799\,995 sources) increases the source counts compared to the Two Micron All Sky Survey by about an order of magnitude. The 90\% completeness limits are 20.4, 19.9, and 19.0 mag in $J, H$, and $K_S$, respectively. The reduced images have 20\% better resolution on average compared to pipeline products. We find between 2300 and 3000 embedded objects in Orion~A and confirm that there is an extended foreground population above the Galactic field, in agreement with previous work.}
    {The Orion~A VISTA catalog represents the most detailed NIR view of the nearest massive star-forming region and provides a fundamental basis for future studies of star formation processes toward Orion.}


\maketitle

\begin{figure*}[tp]
        \centering
    \resizebox{\hsize}{!}{\includegraphics[]{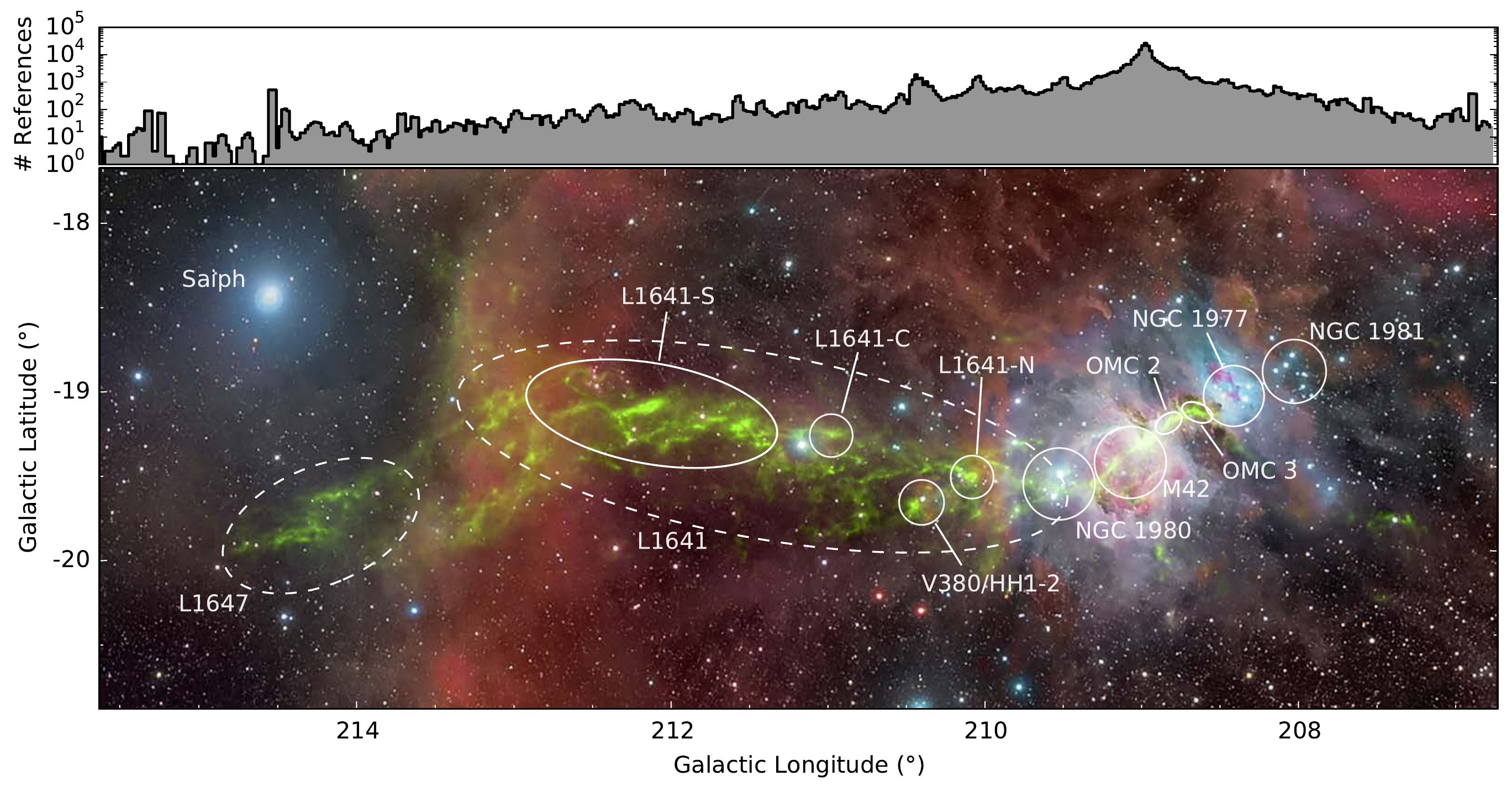}}
        \caption{Composite of optical data (image courtesy of Roberto Bernal Andreo; deepskycolors.com) overlayed with \textit{Planck-Herschel} column density measurements of Orion~A in green. Approximate positions of noteworthy objects and regions are marked and labeled. On top, a histogram (note the logarithmic scaling) shows the number of references for all objects in the SIMBAD database at a given galactic longitude with a 3 arcmin bin size. We see an extreme gradient in attention paid to the various portions of the cloud, with the peak coinciding with M42 and the ONC. Prominent objects (e.g., the V380/HH 1-2 region) produce a local spike in the reference histogram whereas the bulk of the molecular cloud has been studied in comparatively few articles. The coordinates of L1647 in the SIMBAD database ($l=212.13$, $b=-19.2$) do not match the original publication ($l=214.09$, $b=-20.04$).}
        \label{img:herschel_optical}
\end{figure*}

\begin{figure*}[p]
        \centering
    \resizebox{\hsize}{!}{\includegraphics[]{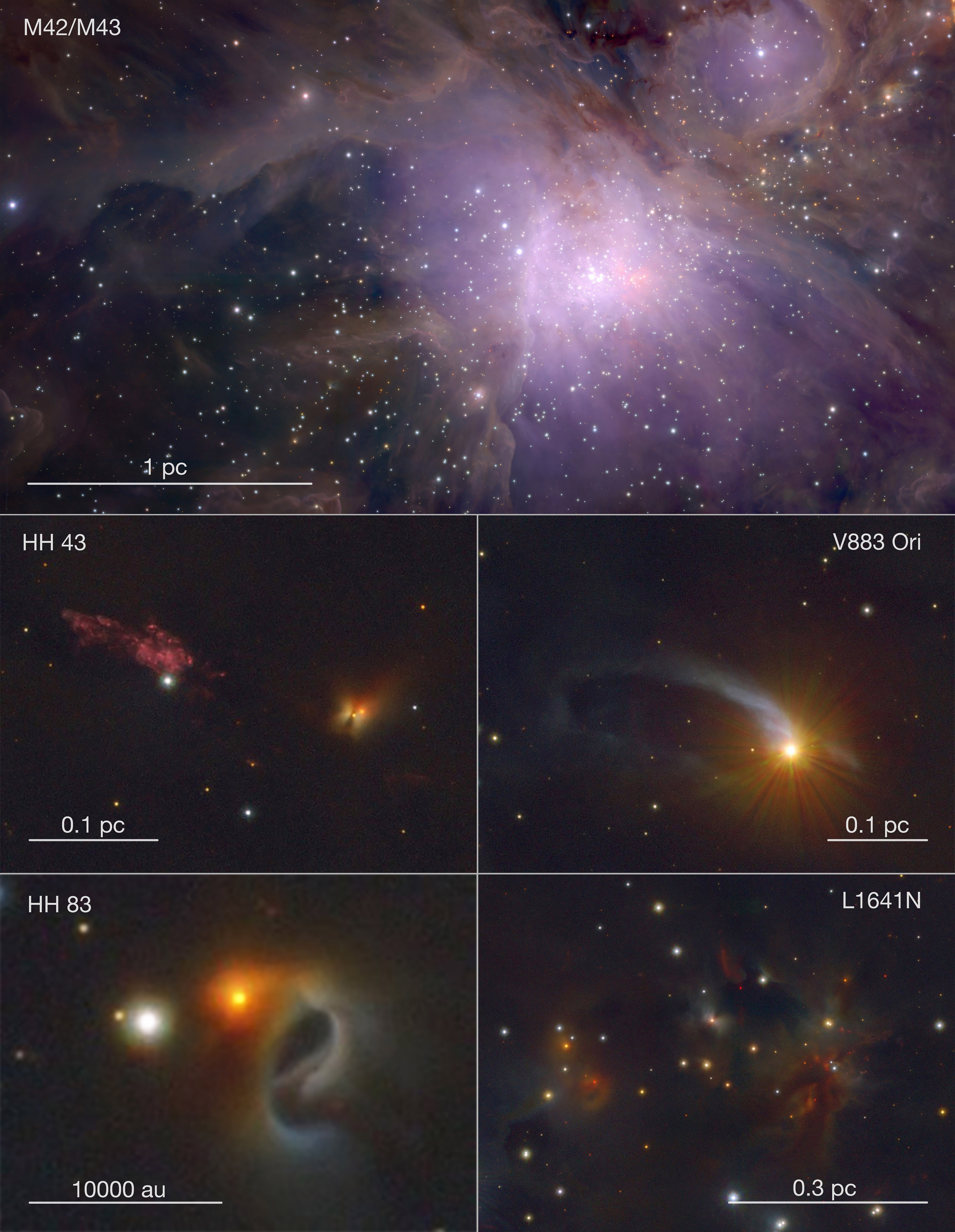}}
        \caption{Detailed view of some prominent objects in Orion~A as seen with VISTA. Here, the $J$, $H$, and $K_S$ bands were mapped to the blue, green, and red channels, respectively. All images are in a galactic projection (North is up, east is left). The physical length given in the scale bars was calculated with the adopted distance of 414 pc.}
        \label{img:vision_objects_1}
\end{figure*}

\begin{figure*}[p]
        \centering
    \resizebox{\hsize}{!}{\includegraphics[]{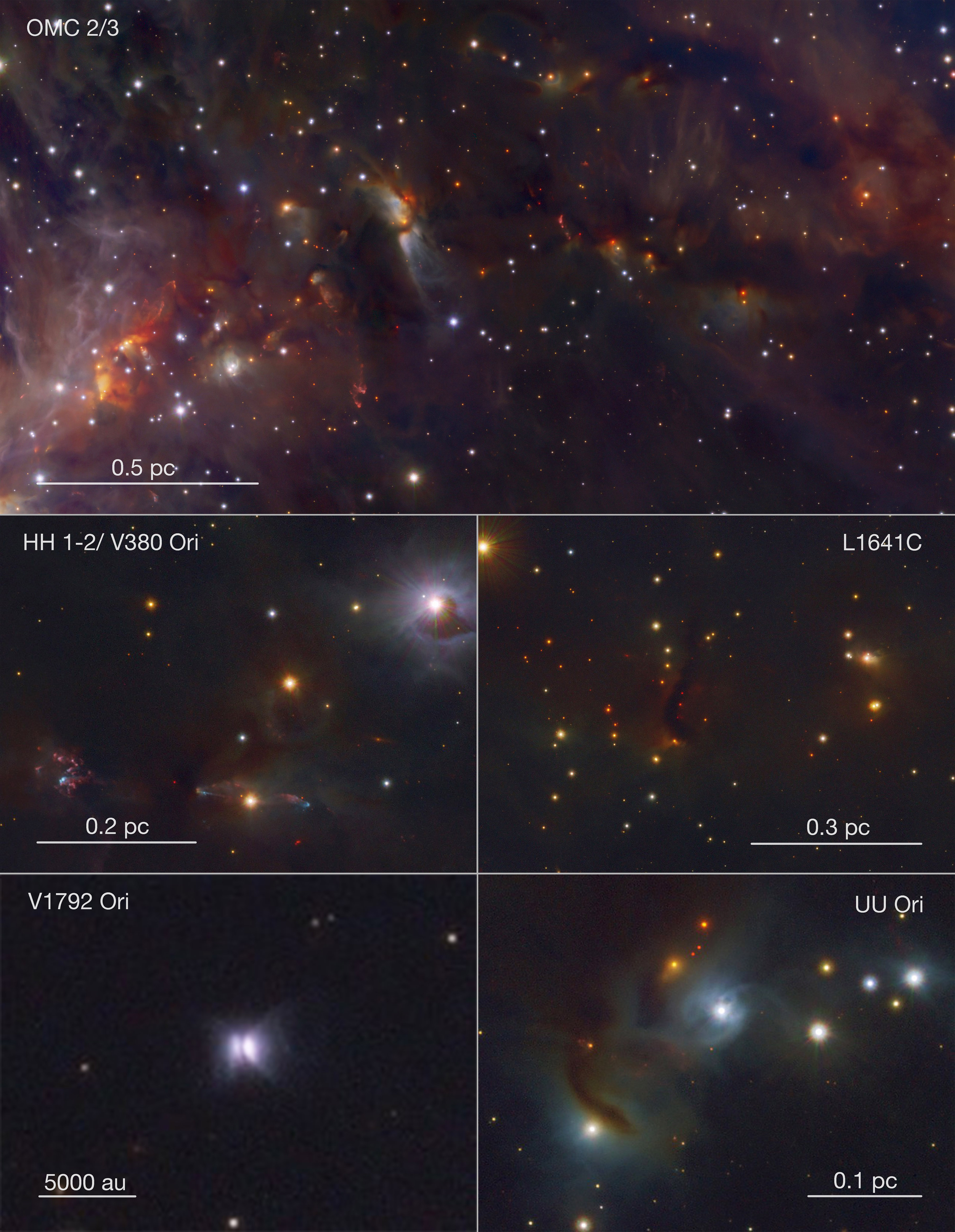}}
        \caption{Same as Fig. \ref{img:vision_objects_1}.}
        \label{img:vision_objects_2}
\end{figure*}

\begin{figure*}[tp]
        \centering
    \resizebox{\hsize}{!}{\includegraphics[]{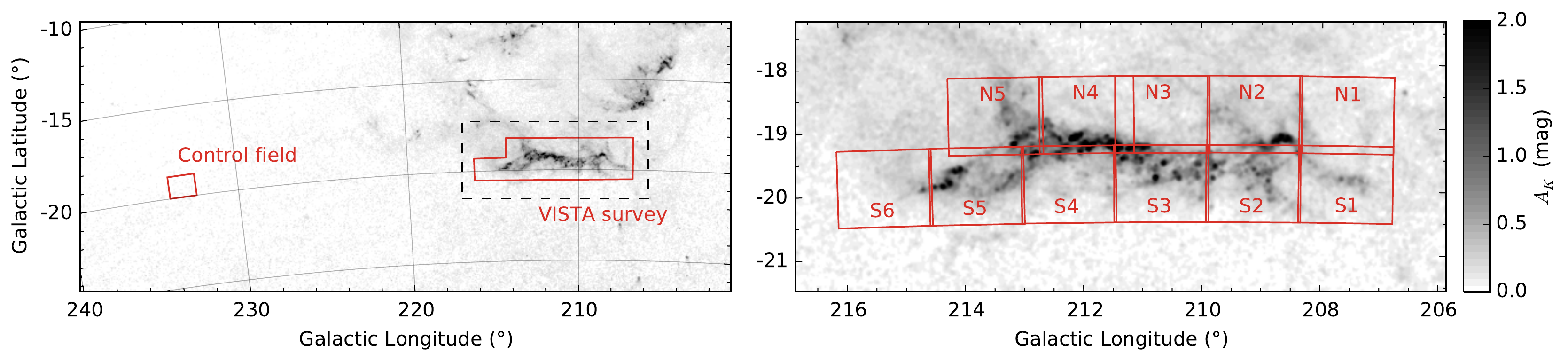}}
        \caption[]{VISTA survey coverage. The lefthand side plot shows the wide-field extinction map from \cite{2011A&A...535A..16L} with both the control field and Orion~A coverage marked as red boxes. The righthand side figure shows a close-up of Orion~A with the individual tiles labeled. The cutout region of this figure is marked with a black dashed box in the left plot.}
        \label{img:2mass_coverage}
\end{figure*}

\section{Introduction}

One of the major obstacles since the beginning of star formation studies in the late 1940s is that stars are embedded in molecular gas and dust during their formation and early evolution, inaccessible to optical imaging devices. The deployment of infrared imaging cameras on optical- and infrared-optimized telescopes during the past three decades has revolutionized the field, providing astronomers with the ability to detect, survey, and systematically study the earliest evolutionary phases of young stars within nearby molecular clouds. Since then, technological advancements have allowed us to constantly improve the sensitivity, resolution, and efficiency of infrared surveys culminating in the all-sky near-infrared (NIR) 2MASS survey  \citep{2006AJ....131.1163S} and in space-borne observatories, such as the \textit{Infrared Astronomical Satellite} \citep[\textit{IRAS},][]{1984ApJ...278L...1N}, the \textit{Spitzer Space Telescope} \citep{2004ApJS..154....1W}, the \textit{Wide-field Infrared Survey Explorer} \citep[\textit{WISE},][]{2010AJ....140.1868W}, and the \textit{Herschel Space Observatory} \citep{2010A&A...518L...1P}. The deployment of active- and adaptive-optics systems on 8--10m class telescopes, combined with advanced instrumentation, has allowed ground-based NIR observations to match the supreme sensitivity of space-borne observatories but reaching higher spatial resolutions thanks to the larger apertures, which is critical to star formation research.

Over the past 25 years, systematic NIR imaging surveys of molecular clouds, in particular of the Orion giant molecular clouds, have revealed much of what we currently know about the numbers and distributions of young stars in star-forming regions. For example, the early foundational NIR surveys \citep[e.g.,][]{1991ApJ...371..171L, 1993ApJ...412..233S, 1994ApJS...90..149C, 1994ApJS...94..615H, 1995AJ....109..709A, 1997ApJ...477..176P, 2000AJ....120.3139C, 2000ApJS..130..381C, 2009A&A...496..153D} revealed the importance of embedded clusters in the star-forming process. By  combining information on the distribution of young stars with surveys of the distribution and properties of molecular gas, important insight into how nature transforms gas into stars have been gained \citep[e.g.,][]{1992ApJ...393L..25L, 1995ApJ...450..201C, 1997ApJ...488..286L, 1997AJ....114.1106M, 2000ApJS..130..381C, 2006ApJ...636L..45T, 2008ApJ...672..410L, 2008ApJ...672..861R, 2009ApJS..181..321E, 2009ApJS..184...18G, 2011ApJ...739...84G}.

In the Orion star-forming complex, one finds a few of the best-studied testbeds for  star formation theories, such as the embedded clusters NGC 2024, NGC 2068, and NGC 2071, as well as the optically visible young clusters $\lambda$, $\sigma$ Ori, $\iota$ Ori, and NGC1981. However, none of these regions have drawn nearly as much attention as the famous Orion nebula cluster (ONC), embedded in the Orion~A molecular cloud. The ONC itself is the closest massive star factory and therefore a prime laboratory for addressing many open questions of current star formation research. Many fundamental quantities regarding the formation of stars have been tested against this benchmark cluster, but orders of magnitude fewer studies have been published about objects in other parts of Orion~A, and even fewer have addressed the molecular cloud as a whole, creating a biased view of the region. To help visualize this bias, we show in Fig. \ref{img:herschel_optical} a composite of an optical image overlaid on the \textit{Herschel-Planck} column density map from \cite{2014A&A...566A..45L}. Here we marked several objects and star-forming regions throughout Orion~A, which is mentioned later in this paper. On top of the image we plot a histogram of the number of articles referenced in the SIMBAD \citep{2000A&AS..143....9W} database\footnote{The references were extracted from the SIMBAD database on 2015 May 10.} for all objects at a given longitude slice in bins of 3 arcmin. 

While the ONC and its surroundings are subject to various studies in thousands of articles, objects in the eastern region of the cloud (in galactic frame\footnote{During the rest of this paper, we always refer to the galactic coordinate frame when using cardinal directions.}) receive considerably less attention with a few tens of published studies. We note that the SIMBAD database is not complete, nonetheless these numbers are a good indicator of the bias in the astronomical community for some regions of Orion~A. Figures \ref{img:vision_objects_1} and \ref{img:vision_objects_2} show  examples of prominent objects observed in our survey of Orion A. Together, these objects alone have more than 5000 bibliographic references listed in the SIMBAD database. 

While previous NIR surveys of Orion~A have given us important insights, they are limited in their depth and sensitivity and/or only cover a fraction of the entire molecular cloud. As a fundamental step toward a complete picture of the star formation processes in Orion~A, we present the most sensitive NIR survey of an entire massive star-forming molecular cloud yet. Table \ref{tab:vision_comparison} lists NIR surveys throughout the past two decades. Compared to the ONC surveys from the 1990s, our survey is about four times more sensitive (in terms of source counts) and covers a $\sim$50 times larger area at the same time. Moreover, we also increase source counts by about 40\% compared to the more recent dedicated NIR survey of the ONC by \cite{2010AJ....139..950R}. Compared to the Two Micron All Sky Survey \citep[2MASS,][]{2006AJ....131.1163S}, which obviously has a greater coverage, we gain almost a factor of 10 in sensitivity. For completeness we mention here that a similar survey has been conducted of the Orion~B molecular cloud. These results are presented in \cite{2015arXiv150504631S}.

\begin{table}
        \caption{On-sky coverage and relative gain in source counts for selected NIR surveys toward Orion~A. Our survey improves both coverage and sensitivity when compared to the literature.}
        \label{tab:vision_comparison}
        \begin{tabular}{lrrr}
        \hline\hline
        Reference                                                                       &       Coverage\tablefootmark{a}       &       Gain\tablefootmark{b}           &       Bands           \\
                                                                                        &       (arcmin$^2$)                            &                                                               &                               \\
        \hline                                                                                                                                  
        \cite{1993ApJ...412..233S}                                      &       2772                                            &       $\sim$4--6\tablefootmark{c}     &       $JHK$                   \\
        \cite{1995AJ....109..709A}                                      &       1472                                            &       4                                                       &       $K$                     \\
    \cite{2000AJ....120.3139C}\tablefootmark{d} &       65857                                           &       9.3                                                     &       $JHK_S$         \\
    \cite{2007MNRAS.379.1599L}\tablefootmark{e} &       $\sim$26500                                     &       1.4                                                     &       $ZYJHK_S$       \\
    \cite{2010AJ....139..950R}                                  &       1200                                            &       1.4                                                     &       $JHK_S$         \\
        \hline                                                  
    This work                                                                   &       65857                                           &                                                               &       $JHK_S$         \\
        \hline    
        \end{tabular}
\tablefoot{
\tablefoottext{a}{Refers to the common on-sky area of the given survey with our VISTA coverage.}
\tablefoottext{b}{Approximate gain in source counts when restricted to the same on-sky coverage.}
\tablefoottext{c}{Estimate based on completeness limits since no source catalog is available.}
\tablefoottext{d}{Study based on the second incremental 2MASS data release. Source counts in this table were taken from the final 2MASS all-sky data release.}
\tablefoottext{e}{Data from UKIDSS DR10. Because of the many spurious detections of nebulosity in the UKIDSS survey, we estimated the gain in source counts by selecting a ``clean'' subregion.}
}
\end{table}

The target of our survey, the Orion~A giant molecular cloud, extends for about 8 deg ($\sim$60 pc) and contains several well-studied objects and an extensive literature: we refer the reader to the review papers of \cite{2008hsf1.book..459B}, \cite{2008hsf1.book..838B}, \cite{2008hsf1.book..544O}, \cite{2008hsf1.book..621A},  \cite{2008hsf1.book..801A}, \cite{2008hsf1.book..483M}, and \cite{2008hsf1.book..590P}. Here we only list a selection of the many results for this important region, including studies of the ONC \citep{1998ApJ...492..540H, 2000AJ....120.3162L, 2002AAS...201.6002M, 2012ApJ...748...14D}, Herbig-Haro objects \citep[HH; for a historic overview, see, e.g.,][]{1997IAUS..182....3R}, such as HH 1-2 \citep[see, e.g.,][]{1983AJ.....88.1040H, 1985ARA&A..23..267L, 2010A&A...518L.122F} and HH 34 \citep[e.g.,][]{2002AJ....123..362R}, and variable FU Ori type pre-main-sequence stars such as V883 \citep[e.g.,][]{1993ApJ...412L..63S, 2013ApJ...768...99P}. Along the ``spine'' of Orion~A, there are also multiple noteworthy minor star-forming regions, such as L1641-N \citep[e.g.,][]{2008A&A...489.1409G, 2012ApJ...746...25N}, which are themselves, however, much less prominent than the Orion nebula and its surroundings. 

Studies referring to the entire cloud are rare. \cite{2012AJ....144..192M} present a Spitzer-based catalog of young stellar objects (YSO) for both Orion~A and Orion B. They identify 2446 pre-main-sequence stars with disks and 329 protostars in Orion~A. \cite{2013ApJ...768...99P} present an XMM-Newton survey of L1641 where they investigate clustering properties of Class II and Class III YSOs. They find an unequal spatial distribution in L1641, which suggests multiple star formation events along the line of sight, in agreement with the interpretation of \cite{2012A&A...547A..97A} and \cite{2014A&A...564A..29B}, and migration of older stars. More recently, \cite{2014A&A...566A..45L} have used a 2MASS dust extinction map \citep{2011A&A...535A..16L}, along with Planck dust emission measurements, to calibrate \textit{Herschel} data and construct higher angular-resolution and high dynamic range column-density and effective dust-temperature maps. \cite{2015A&A...577L...6S} investigate variations in the probability distribution functions of individual star-forming clouds in Orion~A and suggest a connection between the shape of the distribution functions and the evolutionary state of the gas.

Regarding the overall evolution of the Orion star-forming region and following \cite{1964ARA&A...2..213B}, \cite{1998AJ....115.1524G} speculated on the presence of multiple overlapping populations in the direction of the ONC with a possible triggered star formation scenario. As also mentioned by \cite{2008hsf1.book..459B}, recent studies by \cite{2012A&A...547A..97A} and \cite{2014A&A...564A..29B} reveal a slightly older foreground population associated with NGC 1980 with distance and age estimates of $\sim$380 pc and 5-10 Myrs, respectively. They find 2123 potential members for this foreground population, which, however, is an incomplete estimate, because they did not cover the entire Orion~A molecular cloud owing to lack of data in the eastern regions. Based on a shift in  X-ray luminosity functions across Orion~A, \citet{2013ApJ...768...99P} also find evidence of a more evolved foreground population near NGC 1980 at a distance of 300 -- 320 pc. Proposing an alternative view, \citet[][subm.]{2015arXiv151104147D} find that sources near NGC 1980 do not have significantly different kinematic properties from the embedded population, concluding that NGC 1980 is part of Orion~A's star formation history and is currently emerging from the cloud.

Early distance estimates from \cite{1931PASP...43..255T} placed the ONC at 540 pc. Subsequent studies find distances of $480 \pm 80$ pc \citep{1981ApJ...244..884G}, $437 \pm 19$ pc \citep{2007PASJ...59..897H}, $389^{+24}_{-21}$ pc \citep{2007ApJ...667.1161S}, $440 \pm 34$ pc \citep[$392 \pm 32$ pc with a different subset of target stars,][]{2007MNRAS.376.1109J}, and $371 \pm 10$ pc \citep{2011A&A...535A..16L}. Based on optical photometry and a \textit{Planck}-based dust screen model, \cite{2014ApJ...786...29S} find a distance of $420 \pm 42$ pc toward the ONC, while the eastern edge of Orion~A appears to be 70 pc more distant. For the remainder of this paper, we adopt the distance of $414 \pm 7$ pc from \cite{2007A&A...474..515M}.

The VISION (VIenna Survey In OrioN) data presented in this paper (Orion~A source catalog and three-color image mosaic) are made available to the community via CDS. In future publications the survey data will allow us and the community to refine, extend, and characterize several critical properties of Orion~A as a whole. This includes characterizing individual YSOs with the improved resolution and sensitivity, searching for HH objects and jets in a uniform manner, characterizing YSO clustering properties, determining IMFs down to the brown dwarf regime, and describing the gas mass distribution with respect to YSO positions. 

This article is structured as follows. In Sect. \ref{sec:observations} we present a survey overview, including its design and observing strategy. Section \ref{sec:DP} describes all our data-processing procedures from basic image reduction, co-addition, astrometric, and photometric calibration to catalog generation and cleaning. In Sect. \ref{sec:data_products} we review the main data products of our survey and provide a first look at the resulting photometry, including the possibilities for accessing both the generated source catalog and image data. In addition, we present a catalog of interesting objects that includes some new YSO candidates based on their morphological appearance and new candidate galaxy clusters. In Sect. \ref{sec:results} we present first results obtained from this new database, where we derive an estimate for the YSO population in Orion~A and investigate the foreground populations. Section \ref{sec:summary} contains a brief summary, and Appendices \ref{app:data_characteristics} and \ref{app:tables} contain additional information on the quality of the data products and supplementary data tables, respectively.

\section{Observations}
\label{sec:observations}

\subsection{Instrumentation}

The observations of the Orion~A molecular cloud have been carried out with the Visible and Infrared Survey Telescope for Astronomy \citep[VISTA,][]{2006Msngr.126...41E}, a 4m class telescope that is operated by the European Southern Observatory (ESO) as part of its Cerro Paranal facilities. A single instrument, the VISTA Infrared Camera \citep[VIRCAM,][]{2006SPIE.6269E..0XD}, is attached to the telescope's Cassegrain mount, which offers a range of broadband and narrowband filters in the NIR covering a wavelength range from about 0.85 $\mu$m to 2.4 $\mu$m. VIRCAM features a set of sixteen 2k $\times$ 2k Raytheon VIRGO detectors arranged in a sparse 4 $\times$ 4 pattern. Each detector covers about 11.6 $\times$ 11.6 arcmin on sky with gaps of 10.4 arcmin and 4.9 arcmin between them in the instrument's X/Y setup, respectively. Working at a mean pixel scale of 0.339 arcsec/pix in both axes, the instrument field of view in the telescope's beam is 1.292 $\times$ 1.017 deg.

The detectors offer a quantum efficiency above 90\% across the $J, H, \mathrm{and}$ $K_S$ bands but suffer from significant cosmetic deficiencies (e.g., bad pixel rows and columns, as well as bad readout channels) and nonlinearity effects, which need to be taken care of during data calibration. The gaps between the individual detectors make it necessary to observe multiple overlapping fields for a contiguous coverage. This is achieved by a six-step offset pattern that can be executed in several ways. As a consequence of this observing strategy, the effective coverage (hence exposure time) over a single field varies with position. The standard offset pattern offers a coverage of as little as just one frame on the edge of the field, two frames for most of the area and up to six overlapping exposures for only a tiny portion of the final frame. As is usual for NIR observations, a dither or jitter\footnote{Here we use the term dither for user-defined offset positions, whereas jitter refers to random telescope positioning.} pattern is usually executed at each offset position to mitigate saturation effects and to increase the total frame coverage to facilitate bad pixel rejection during co-addition.

For the rest of this paper, we use VISTA terminology to describe the telescope's data products and its parameters: a simultaneous integration from all sixteen detectors is called a ``pawprint'', and a fully sampled image resulting from the co-added frames of the six-step offset pattern is called a ``tile''. The integration time for a single readout from all detectors is referred to as DIT (detector integration time), whereas multiples of these single integrations can be stacked internally before readout. The number of integrations in such a stack is referred to as NDIT.

\begin{table*}[htbp]
        \centering
    \caption{Observing dates and basic parameters for the Orion~A VISTA survey.}
    \label{tab:observations}
    \begin{tabular*}{\textwidth}{c @{\extracolsep{\fill}} c c c c c c c c}
    \hline\hline
    Tile ID     &       Filter          &       Start time                      &       DIT     &       NDIT    &       NJitter &       Airmass range   &       Image quality\tablefootmark{a}  & 90\% Completeness\tablefootmark{b}            \\
                &                               &       UT                                      &       (s)     &       (\#)    &       (\#)    &                                       &       (arcsec)                &       (mag)           \\
        \hline
    S1          &       $J$                     &       2012/10/02 08:38:24     &       5       &       8               &       3               &       1.074 - 1.088 &       0.78 - 0.92             &       20.47           \\
    S1          &       $H$                     &       2013/02/27 02:09:44     &       2       &       27              &       5               &       1.308 - 1.657 &       0.68 - 0.83             &       20.06           \\
    S1          &       $K_S$           &       2013/01/20 00:23:40     &       2       &       20              &       5               &       1.110 - 1.197 &       0.63 - 1.02             &       18.91           \\
    S2          &       $J$                     &       2012/12/25 05:20:26     &       5       &       9               &       6               &       1.122 - 1.328 &       0.75 - 0.86             &       20.08           \\
    S2 (a)      &       $H$                     &       2013/02/08 01:17:15     &       2       &       17              &       5               &       1.058 - 1.136 &       0.72 - 0.89             &       19.58           \\
    S2 (b)      &       $H$                     &       2013/02/17 01:24:44     &       2       &       17              &       5               &       1.089 - 1.245 &       0.75 - 0.89             &       19.58           \\
    S2 (a)\tablefootmark{c}     &       $K_S$           &       2013/01/25 00:35:11        &       2       &       15              &       5               &       1.060 - 1.125 &       0.87 - 1.06             &       18.85           \\
    S2 (b)      &       $K_S$           &       2012/10/04 08:07:02     &       2       &       15              &       5               &       1.058 - 1.112 &       0.62 - 0.75             &       18.85           \\
    S3          &       $J$                     &       2012/10/05 08:39:35     &       5       &       8               &       3               &       1.054 - 1.064 &       0.61 - 0.69             &       20.87           \\
    S3          &       $H$                     &       2013/03/02 01:12:20     &       2       &       27              &       5               &       1.137 - 1.319 &       0.70 - 0.84             &       20.01           \\
    S3          &       $K_S$           &       2013/01/30 00:33:06     &       2       &       20              &       5               &       1.054 - 1.093 &       0.67 - 0.81             &       19.08           \\
    S4          &       $J$                     &       2012/11/13 08:09:40     &       5       &       8               &       3               &       1.103 - 1.143 &       0.70 - 0.92             &       20.48           \\
    S4          &       $H$                     &       2013/03/01 02:18:03     &       2       &       27              &       5               &       1.304 - 1.665 &       0.65 - 0.91             &       20.27           \\
    S4          &       $K_S$           &       2013/02/09 01:30:51     &       2       &       20              &       5               &       1.048 - 1.093 &       0.63 - 0.88             &       19.13           \\
    S5          &       $J$                     &       2013/02/24 02:53:52     &       5       &       8               &       3               &       1.352 - 1.461 &       0.66 - 0.92             &       20.41           \\
    S5          &       $H$                     &       2013/03/06 00:30:46     &       2       &       27              &       5               &       1.072 - 1.189 &       0.62 - 0.82             &       20.25           \\
    S5          &       $K_S$           &       2013/02/16 01:45:14     &       2       &       20              &       5               &       1.076 - 1.162 &       0.81 - 1.04             &       18.97           \\
    S6          &       $J$                     &       2013/02/24 03:35:28     &       5       &       8               &       3               &       1.573 - 1.750 &       0.76 - 0.96             &       20.23           \\
    S6          &       $H$                     &       2013/03/09 00:22:19     &       2       &       27              &       5               &       1.066 - 1.182 &       0.80 - 1.09             &       19.69           \\
    S6          &       $K_S$           &       2013/01/31 00:30:20     &       2       &       20              &       5               &       1.037 - 1.080 &       0.72 - 1.00             &       18.95           \\
    N1          &       $J$                     &       2012/11/13 07:48:48     &       5       &       8               &       3               &       1.108 - 1.141 &       0.72 - 0.86             &       20.55           \\
    N1          &       $H$                     &       2013/02/27 00:09:19     &       2       &       27              &       5               &       1.072 - 1.135 &       0.65 - 0.86             &       20.26           \\
    N1          &       $K_S$           &       2013/01/27 00:38:09     &       2       &       20              &       5               &       1.077 - 1.122 &       0.69 - 0.84             &       19.17           \\
    N2          &       $J$                     &       2013/02/25 02:23:36     &       5       &       9               &       6               &       1.298 - 1.753 &       0.78 - 0.97             &       20.14           \\
    N2 (a)      &       $H$                     &       2013/02/28 01:07:48     &       2       &       17              &       5               &       1.125 - 1.337 &       0.66 - 0.84             &       19.69           \\
    N2 (b)      &       $H$                     &       2013/03/01 00:56:10     &       2       &       17              &       5               &       1.112 - 1.306 &       0.64 - 0.78             &       19.69           \\
    N2 (a)      &       $K_S$           &       2013/01/28 00:40:27     &       2       &       15              &       5               &       1.060 - 1.107 &       0.81 - 0.93             &       18.86           \\
    N2 (b)      &       $K_S$           &       2013/01/29 00:45:10     &       2       &       15              &       5               &       1.059 - 1.096 &       0.84 - 0.98             &       18.86           \\
    N3          &       $J$                     &       2012/11/04 08:10:11     &       5       &       8               &       3               &       1.070 - 1.092 &       0.58 - 0.78             &       20.84           \\
    N3          &       $H$                     &       2013/03/03 01:58:40     &       2       &       27              &       5               &       1.270 - 1.588 &       0.70 - 0.89             &       20.16           \\
    N3          &       $K_S$           &       2013/02/01 00:31:35     &       2       &       20              &       5               &       1.056 - 1.093 &       0.66 - 0.88             &       19.20           \\
    N4          &       $J$                     &       2012/11/15 07:41:11     &       5       &       8               &       3               &       1.075 - 1.103 &       0.64 - 1.03             &       20.70           \\
    N4          &       $H$                     &       2013/03/03 23:58:43     &       2       &       27              &       5               &       1.050 - 1.111 &       0.60 - 0.79             &       20.16           \\
    N4          &       $K_S$           &       2013/02/15 00:19:06     &       2       &       20              &       5               &       1.045 - 1.050 &       0.80 - 1.18             &       18.83           \\
    N5          &       $J$                     &       2013/02/24 03:14:42     &       5       &       8               &       3               &       1.452 - 1.590 &       0.78 - 0.93             &       20.40           \\
    N5          &       $H$                     &       2013/03/07 01:16:29     &       2       &       27              &       5               &       1.158 - 1.369 &       0.62 - 0.84             &       20.18           \\
    N5          &       $K_S$           &       2013/02/16 00:33:10     &       2       &       20              &       5               &       1.038 - 1.073 &       0.88 - 1.02             &       18.92           \\
        \hline 
    CF          &       $J$                     &       2013/01/02 07:40:14     &       5       &       8               &       3               &       1.484 - 1.624 &       0.63 - 0.76             &       20.67           \\
    CF          &       $H$                     &       2013/02/15 03:48:50     &       2       &       27              &       5               &       1.216 - 1.489 &       0.65 - 0.95             &       19.78           \\
    CF          &       $K_S$           &       2013/02/18 04:08:35     &       2       &       20              &       5               &       1.339 - 1.609 &       0.67 - 0.87             &       18.99           \\
        \hline
        \end{tabular*}
        \tablefoot{
                \tablefoottext{a}{The image quality refers to measured FWHM estimates of point-like sources, which varies across each tile because of camera distortion and variable observing conditions.}
                \tablefoottext{b}{Completeness estimates are derived from the full combined Orion~A mosaics and are calculated on the basis of artificial star tests. Details on the method are described in \ref{app:phot}.}
                \tablefoottext{c}{Rejected in co-addition due to large differences in image quality with respect to Tile S2 (b).)}
        }
\end{table*}

\begin{figure*}[tp]
        \centering
        \resizebox{\hsize}{!}{\includegraphics[]{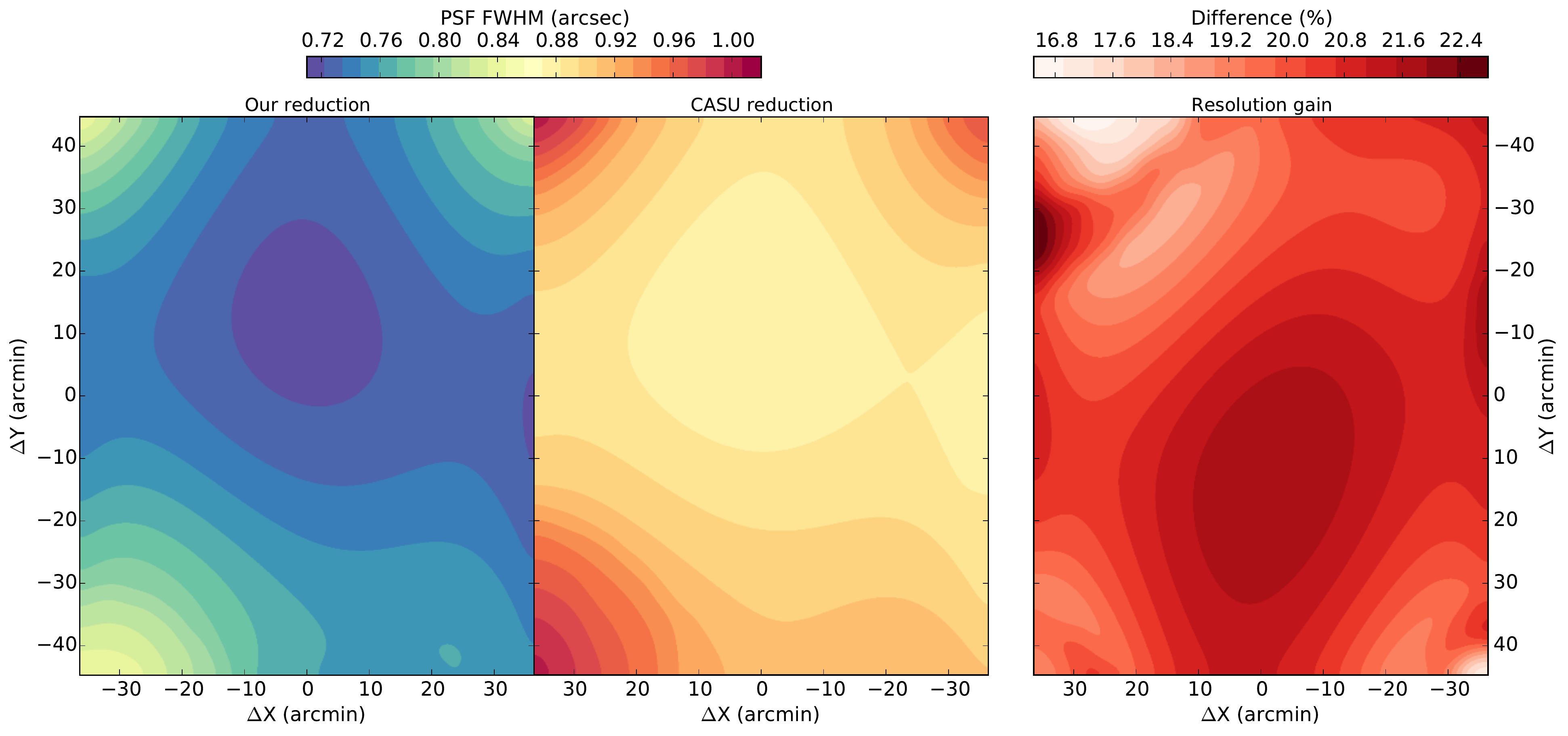}}
        \caption[]{FWHM maps for both our reduction and the standard CASU pipeline of tile S1 in $H$ band. Clearly a significant gain in image quality is achieved in our reduction.}
        \label{img:psf_gain}
\end{figure*}

\subsection{Survey design and strategy}

The Orion~A molecular cloud is centered at approximately $l = 210^{\circ}$, $b = -19^{\circ}$ and extends for about eight degrees, which is well aligned with the Galactic plane. The spine of the cloud (i.e., regions with high extinction) is very narrow with only about 0.3 deg at its widest point. However, shallower extinction levels are observed much more widely, which did not allow us to cover the entire cloud with only one series of pointings along the molecular ridge. Therefore we designed the survey to feature 11 individual pointings with two parallel sequences of tiles aligned with its spine. For each tile we also included overlaps with its neighboring field to ensure a contiguous coverage. Figure \ref{img:2mass_coverage} shows the final tile coverage on top of the extinction map from \cite{2011A&A...535A..16L}. Also shown are our designations for each tile labeled from N1, ..., N5, S1, ..., S6 indicating row (north/south) and column position (west to east). An observation of a tile in one of the three filters defined an observation block (OB). All 11 tiles were observed in $J, H, \mathrm{and}$ $K_S$, where all except the tiles N2 and S2 were executed with a standard jitter pattern with a maximum random throw within a 25 arcsec wide box centered on the initially acquired position. The N2 and S2 tiles include the ONC and therefore a large amount of extended emission. These positions were observed with a separate sky offset field centered at $l=209.272^{\circ}$, $b=-21.913^{\circ}$. Because the sky offset field had to be observed in addition to the science fields, the total duration of these sequences was greater than the maximum allowed OB length of one hour in the $H$ and $K_S$ bands. Therefore each of these four OBs was executed twice. Starting in October 2012 and spreading out over about six months until early March 2013, a total of 37 individual OBs were executed to complete the observations of Orion~A.

For statistical comparisons in the following analyses, we also observed a control field (CF) in the same filters in addition to the science field, centered at $l = 233.252^{\circ}$, $b = -19.399^{\circ}$ (see Fig. \ref{img:2mass_coverage}). Table \ref{tab:observations} gives a comprehensive overview of all collected data of Orion~A including basic parameters and statistics of each observing sequence. For the $H$ and $K_S$ bands, the DIT was chosen to be only 2~s owing to saturation issues and was compensated by increasing numbers of NDIT ranging from 15 to 27 for these bands. In the $J$ band we reached a more efficient duty cycle with larger DITs since saturation is less critical at this wavelength. The number of jitter positions at each of the six telescope offsets to form a tile was chosen to be a minimum of 3three for the $J$ band to allow for reliable bad pixel rejection. For the $H$ and $K_S$ bands, we observed five jittered positions for each pawprint. The total on-source exposure time is given by the product of the minimum exposure time (DIT $\times$ NDIT) and the number of observations taken at this position determined by Njitter and the six-step offset pattern. For the large majority of sources in a tile, this is given by DIT $\times$ NDIT $\times$ Njitter $\times$ 2.

In addition to the science and control fields, calibration frames were also needed to process the raw files into usable data products. Dark frames, sky flat fields for each band, and dome flats to measure detector nonlinearity were provided as part of ESO's standard calibration plan for VIRCAM.

\section{Data processing}
\label{sec:DP}

A total amount of $\sim$280 GB of science data, along with $\sim$680 GB of calibration frames, was obtained for the Orion~A VISTA survey. Together with the complex observing routine and camera setup, only a dedicated pipeline is able to handle the data reduction procedure. Calibrated science data products are available through the VISTA data flow system \citep{2004SPIE.5493..411I} provided by the Cambridge Astronomical Survey Unit (CASU). However, since the pipeline is designed for stability and is optimized for reducing a much larger amount of data under many different observing conditions, we identified several drawbacks in this system. To achieve the best possible data quality of the VISTA Orion~A survey, we decided to implement all key data reduction procedures ourselves. Details on the methods, including a mathematical description of the CASU pipeline modules, can be found in the VISTA data reduction library design and associated documents \citep[accessible through \url{http://casu.ast.cam.ac.uk},][]{2010ASPC..434...91L}.

\subsection{Motivation}
\label{sec:motivation}

Below we list the specific points that motivated us to develop our own customized reduction pipeline for the Orion~A VISTA data.

\begin{itemize}

        \item Owing to the observing strategy with VIRCAM, sources are sampled several times not only at different detector positions but also by different detectors. To optimally co-add all reduced paw prints, each input image needs to be resampled and aligned with a chosen final tile projection. The CASU pipeline uses a radial distortion model, together with fast bilinear interpolation, to remap the images for the final tiling step. Bilinear interpolation, however, has several drawbacks. Primarily it can introduce zero-point offsets and a significant dispersion in the measured fluxes. Typically a Moir\'e pattern is also seen on the background noise, and additionally it ``smudges'' the images, leading to lower output resolution (see \citealp{2010ascl.soft10068B} for details and examples). To test for the absolute gain in resolution over the bilinear resampling kernel, full width half maximum (FWHM) maps for each observed tile were calculated with PSFex \citep{2011ASPC..442..435B}. Figure \ref{img:psf_gain} shows a FWHM map for the tile S1 in $H$ band for both the CASU and our reduction, as well as the gain in resolution. We typically achieve 20\% higher resolution, i.e. a 20\% smaller FWHM, by simply using more suitable resampling kernels\footnote{We observed a dependency of the resolution gain on observing conditions. We get only 10-15\% for bad seeing conditions ($> 1$ arcsec) and up to almost 30\% for excellent conditions ($\sim0.6$ arcsec)} (for an interesting in-depth discussion of the importance of resampling methods, see, e.g., \citealp{2014AJ....147..108L}). We find that our resampling method recovers the image quality of the pawprint level for the combined tiles, which is not the case for the CASU reduction.

        \item The CASU pipeline only produces source catalogs for individual tiles. As can be seen in Fig. \ref{img:2mass_coverage}, it would be beneficial to co-add all input tiles to increase the effective coverage on the tile's edges and run the source extraction on the entire survey region. The spatially correlated noise (which is not traced by weight maps) and the necessity of yet another resampling pass make this step highly undesirable for the CASU tiles.
   
        \item Even with such short integration times as in our survey, stars brighter than $\sim$12th magnitude in $J$ (11.5 and 11 mag in $H$ and $K_S$, respectively) show saturation and residual nonlinearity effects when compared to the 2MASS catalog. Replacing these measurements with reliable photometry from 2MASS requires that both catalogs are calibrated toward the same photometric system. As already demonstrated by \cite{2011A&A...534A...3G}, among others, this is not the case, and a comparison with 2MASS requires the recalibration of the photometric zero point. Reliable color transformations can be found in \cite{2013A&A...552A.101S}. We also tested this by producing magnitudes from the CASU tile catalogs via
    
        \begin{equation}
                m_{i,j} = - 2.5 \log \left(\frac{F_{i,j}}{t_{j}}\right) - \mathrm{apcor}_{j} + \mathrm{ZP}_{j} 
        \end{equation}
    
        where $m_{i,j}$ are the calculated magnitudes for the $i$-th source measured on the $j$-th tile, $F_{i,j}$ the flux measurements given in the CASU tile catalogs, $t_{j}$ the exposure times, apcor$_{j}$ the aperture corrections, and ZP$_{j}$ the zero points as given in the tile headers. We also applied the appropriate zero-point transformation for the given atmospheric extinction and used the recommended aperture radius. We then concatenated all original tile catalogs and cross-matched the data with 2MASS. The comparison of the photometry is shown in Fig. \ref{img:casu_mag_turnoff} where magnitude differences between both data sets are displayed as a function of the 2MASS flux measurement. A systematic offset is visible with values around 0.1 mag in $H$ and 0.12 mag in $K_S$.
 
\begin{figure}[tp]
        \resizebox{\hsize}{!}{\includegraphics{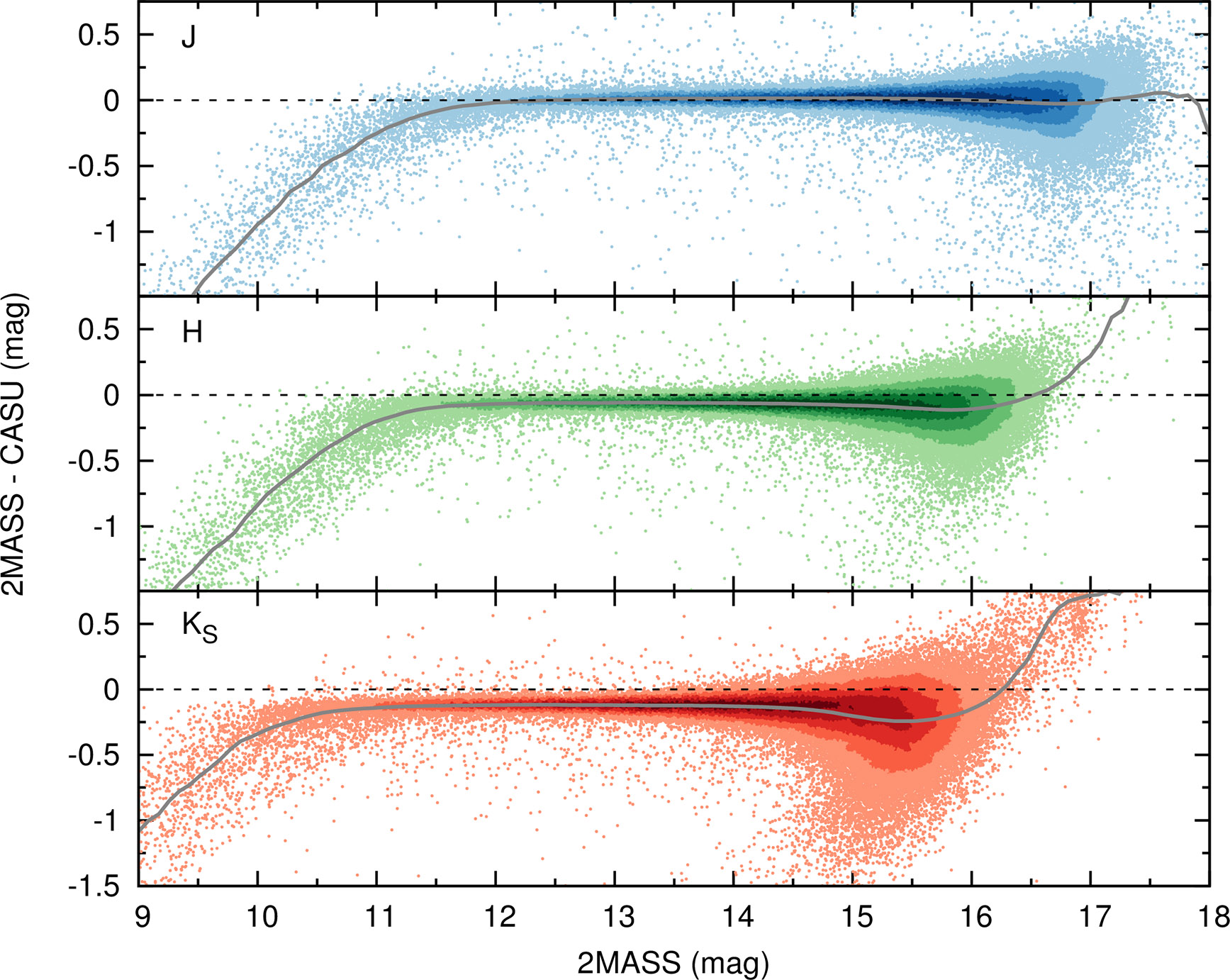}}
        \caption[]{Difference in the photometry between the CASU default reduction and the 2MASS catalog. The blue, green, and red data points show the $J$, $H$, and $K_S$ bands, respectively. The color shading represents source density in a 0.2 $\times$ 0.05 box in the given parameter space and the gray solid line a running median along the abscissa with a box width of 0.5 mag. Several processing steps contribute to the apparent offsets, which are all avoided in our data reduction.}
        \label{img:casu_mag_turnoff}
\end{figure}

        \item For the zero-point calculation of each observed field, the CASU pipeline applies a galactic extinction correction to all photometric measurements. To this end the pipeline uses the \cite{1998ApJ...500..525S} all-sky extinction maps (with a resolution of a few arc-minutes), together with the correction from \cite{2000AJ....120.2065B}. For each source, a bilinear interpolation yields the extinction correction factor for the zero point. This will also add systematic offsets with respect to photometric data for which no such correction was applied. More critically, for surveys covering multiple fields with variable extinction, systematic offsets are expected between the
tiles. For studies concerned with the intrinsic color of stars (e.g., extinction mapping), however, it is critical not to be biased in any way by such systematic offsets.

        \item The CASU pipeline by default stacks all frames of an entire set to build a single background model for one tile. This only works well if spatial sky variations across the detector array are constant for the entire duration of the observations. In the NIR this typically applies for small sets of data with relatively short total exposure times, such as for the VISTA survey products (e.g., VVV, \citealp{2010NewA...15..433M}). However, since our OBs were at the limit of the maximum allowed execution time of 1 h, significant changes in atmospheric conditions are expected for nonphotometric nights. This can lead to residual gradients across single detector frames, which can result in cosmetically imperfect reductions and difficult sky level estimates. This is especially the case for fields with separate offset sky positions with large gaps in the sky sampling. 

\begin{figure}
        \centering
    \resizebox{\hsize}{!}{\includegraphics{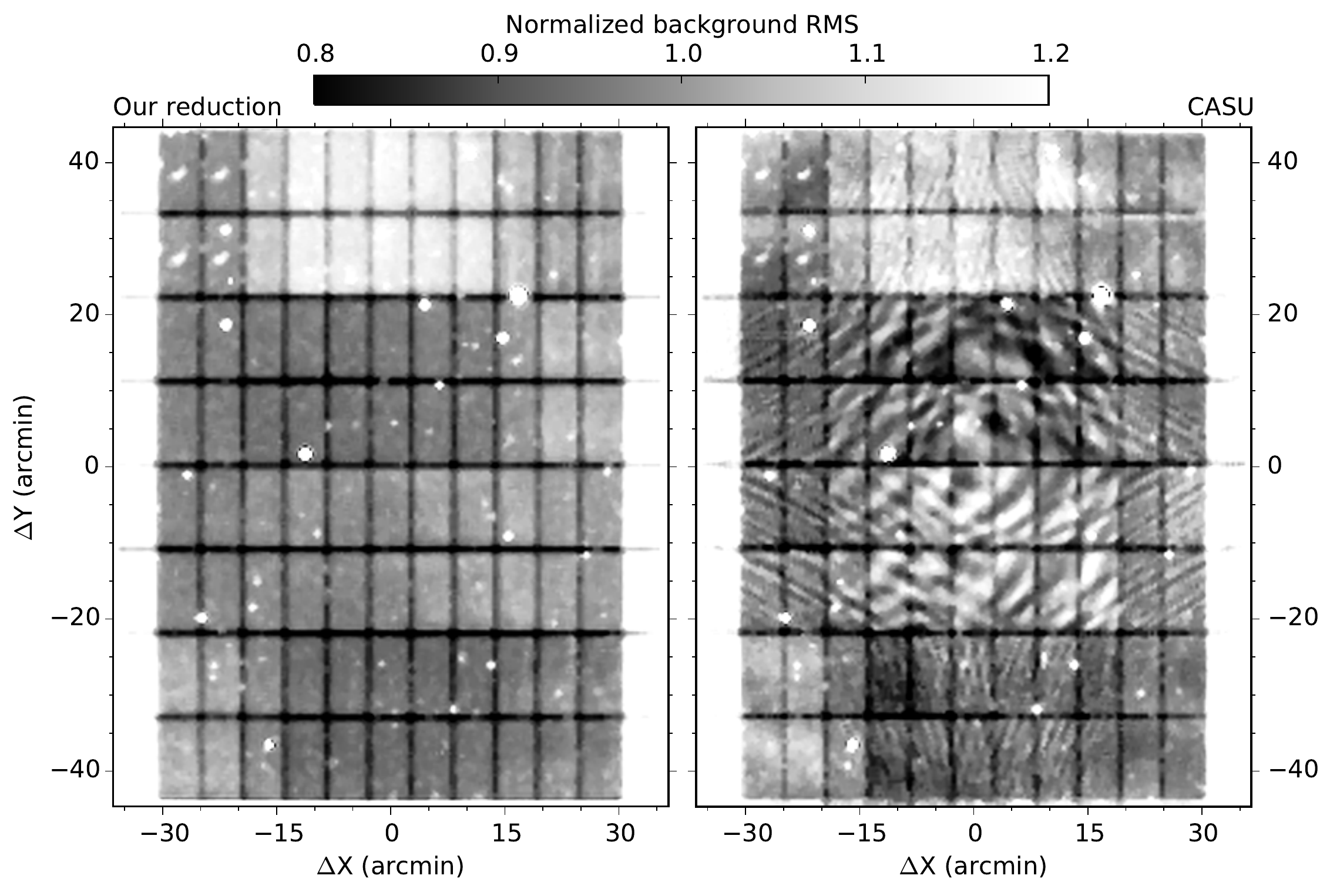}}
        \caption[]{Noise RMS maps of the CF in $K_S$ for our reduction and the CASU pipeline. The bilinear interpolation and the radial distortion model clearly leave spatially correlated noise in the tiled images. On the other hand, the variance in our reduction is only dominated by detector coverage and intrinsic detector characteristics. No sources were masked prior to noise calculations.}
        \label{img:sauron}
\end{figure}

    \item In total there are two interpolation steps employed by the CASU pipeline. The first generates stacks for each jitter sequence, and the second is used during the tiling procedure to correct for astrometric and photometric distortions (the latter to account for variable on-sky pixel size due to field distortion). Both of these steps use bilinear interpolation, which can introduce spatially correlated noise. The difference between the original bilinear interpolation and our data product, for which we use higher order resampling kernels, is illustrated in Fig. \ref{img:sauron}, where the background RMS maps of the CF in the $K_S$ band are shown. The radial distortion model, together with the fast interpolation, clearly leaves its mark. These RMS maps were generated with SExtractor \citep{1996A&AS..117..393B} and represent smoothed-noise RMS models with a background mesh size of 64 pixels. For reliable source detection, however, one has to keep track of the variable noise throughout an image. As a consequence this makes it difficult to reliably run external source detection packages on the output CASU tiles in cases the pipeline does not work satisfactorily, as in regions with extended emission such as the ONC.

\end{itemize}

From all these points, only the photometric offset relative to the 2MASS system can be corrected for via color transformations. Bias-free photometry and high resolution are both critical for all further studies with the Orion~A VISTA data. Therefore we have written a semi-automatic data-reduction package that is completely independent of the CASU pipeline. All functionalities of this package will be offered in open-source Python code in a future paper. The implemented reduction steps are discussed in detail during the following sections. In summary, the following capabilities have been implemented specifically for the VIRCAM reduction package:

\begin{itemize}
        \item Calculation of all required master calibration frames and parameters: bad pixel masks (BPM), dark frames, flat fields, and nonlinearity coefficients;
        \item Basic image calibration: nonlinearity correction, removal of the dark current, and first-order gain harmonization with the master flat;
        \item Accurate weight map generation for co-addition and source detection;
        \item Static and dynamic background modeling;
        \item Removal of cosmetic deficiencies (bad pixel masking, global background harmonization, etc.);
        \item Illumination correction (second-order gain harmonization) using external standards;
        \item Source detection, astrometric calibration, and co-addition via external packages;
        \item Robust aperture photometry using variable aperture corrections;
        \item Photometric calibration based on the 2MASS reference catalog.
\end{itemize}

Many of the techniques are similar to the methods used in the CASU pipeline. However, the problems listed above are carefully avoided. All sequential data reduction procedures are described in the following sections.

\subsection{Master calibration frames}

The basic image reduction steps include the generation of all required calibration frames and parameters and their application to the raw science data to remove the instrumental signature from VIRCAM.

\subsubsection{Bad pixel masking}

Before any other calibration step can be performed, a BPM is required to avoid introducing systematic offsets in, for instance, dark current calculations or linearity estimations. This step, however, has to be independent of any further calibration steps. Therefore we used a set of dome flats with constant exposure times that are first stacked at the detector level. The median of each detector served as a preliminary master flat and was then used to normalize each input image. Good pixels in each recorded flat field would then theoretically contain only values around unity due to the constant exposure time. Then, all pixels that deviated by more than 4\% with respect to the expected unity value were marked. Finally, if a single pixel was marked in this way in more than 20\% of all images in the sequence, it was propagated as a bad pixel to the final master BPM. Typical bad pixel-count fractions were found between 0.1\% and 0.2\% for the best detectors and around 2\% for the worst. 

\subsubsection{Nonlinearity correction}
\label{sec:nonlin}

To correct for detector nonlinearities, we used the same method as for the CASU pipeline. For details on this method, the reader is referred to the VISTA data reduction library design document\footnote{\url{http://casu.ast.cam.ac.uk/surveys-projects/vista/technical/data-processing/design.pdf/view}}. In principle a set of dome flat fields with increasing exposure time was first masked, i.e. with the BPM and pixels above the saturation level, and corrected for dark current with the accompanying dark frames. Then the flux was determined for the detector as the mode of each masked frame. The increasing exposure time should then provide a constant slope in the flux vs. exposure time relation for a completely linear detector with a given constant zero point (in double-correlated read mode used for our observations this offset should be close to 0). A least-squares fit to these data using a function of the form

\begin{equation}
        \Delta I = \sum_{m=0}^{3} b_m t_i^m \left[ \left( 1+k_i \right)^m - k_i^m \right]
\end{equation}

\noindent was performed, where $i$ indicates each detector, $m$ indicates the order of the function, $\Delta I$ are the measured nonlinear fluxes for the reset-corrected double-correlated read output, $b_m$ are the coefficients to be solved for, $t_i^m$ are the integration times, and $k_i$ are the ratios between the reset-read overhead and the integration times. All least-squares fits in our reduction package made use of the MPFIT IDL library described in \cite{2009ASPC..411..251M}. These nonlinearity coefficients were stored in look-up tables and were later applied to each input frame by a simple nonlinear inversion. We also tested nonlinearity corrections on the channel level (each of the 16 detectors of VIRCAM hosts 16 separate readout channels) and found no significant differences in the output data quality.

\subsubsection{Dark current estimation}

To estimate the dark current, a set of dark frames with the same exposure time parameters (i.e., DIT and NDIT) as the science frames was stacked at the detector level. The output of this procedure was a master dark pawprint that was calculated as the average of the pixel stack with a simple rejection of the minimum and maximum pixel value. We favored this method over a median because of the small number of available dark frames (typically five per unique DIT/NDIT combination for the VIRCAM calibration plan).

\subsubsection{First-order gain harmonization}

For photometric consistency across all detectors, one has to calibrate all pixels to the same gain level. We used a series of twilight flats to correct for pixel-to-pixel gain variations. For camera arrays, however, it is usually not enough to create master flats for each detector separately since the detectors themselves also need to be brought to the same gain value with respect to each other. In a first step, all input flats were linearized, the dark current was removed, and bad pixels were masked. We then normalized all input pawprints by the median flux over all channels to account for the variable illumination, preserving detector-to-detector differences. The master flat field was then simply calculated as the median of the stacked, calibrated, and scaled input pawprints.

\subsubsection{Weightmaps}

To accurately trace variable noise and bad pixels across the frames, we used weight maps initially generated from the master flat field. The normalized master flat field already accounted for variable sensitivity across the focal plane introduced by vignetting from filter holders and other detector/camera characteristics. We simply added bad pixels and rejected pixels with an unusually low/high response. These weights were later used for source detection to trace the spatial variations in background noise and during co-addition for an optimized weighting scheme.

\subsubsection{Saturation levels, read-noise, and gain}

The saturation levels, read-noise, and gain of each detector are important parameters for any source detection method and error calculations. We  determined the read-noise and gain following Janesick's method \citep[e.g.,][]{2001sccd.book.....J} on a set of dedicated calibration frames. Initial values for the saturation levels of each detector were taken from the VIRCAM user manual\footnote{\url{https://www.eso.org/sci/facilities/paranal/instruments/vircam/doc.html}}. These were then checked during the calculation of the coefficients for the nonlinear inversion as described in Sec. \ref{sec:nonlin}. The saturation level of Detector 6 had to be refined since we still saw significant nonlinearity below the given threshold. We lowered the original value of 36000 ADU to 24000 ADU. All calculated parameters were stored in look-up tables for later processing steps.

\subsection{Science data calibration}

After producing all the necessary calibration master files and parameters as described above, we consecutively applied the nonlinearity correction, dark current subtraction, gain harmonization, and bad pixel masking. In addition to these standard data reduction procedures, NIR data typically benefit from the removal of the (highly variable) background signature and detector-dependent cosmetic corrections.

\subsubsection{Background model}

Additional additive background signatures (e.g., atmospheric emission, residual scattered light) can be removed by creating a background model. For our observing sequences with only a few individual exposures, we masked any contaminating sources prior to calculating the residual background. As a first step, we therefore created a static background model, calculated from a simple median of all stacked data, to allow for a rough first-pass source detection. 

These temporary background models were applied to the science data, which in turn were used to create source masks with SExtractor. Very bright sources produced large halo structures on the VIRCAM detectors, which were simply masked by placing a circular mask with a radius proportional to a preliminary calculated magnitude:

\begin{equation}
        \mathrm{m_{preliminary}} = -2.5\,\mathrm{log}\left(\frac{F}{t}\right) + \mathrm{ZP_{VISTA}}
,\end{equation}

\noindent where $\mathrm{m_{preliminary}}$ is the preliminary adopted magnitude, $F$ is the measured flux from SExtractor, $\mathrm{ZP_{VISTA}}$ is the zero point for each band from the VISTA user manual, and $t$ is the integration time (DIT $\times$ NDIT). By manually comparing some sources to 2MASS, we typically found errors of only a few 10~\% for these estimates ,which was sufficient for source masking. A star of magnitude eight received a mask with a 50 arcsec radius $r$. All other masks were calculated with $\Delta r/\Delta \mathrm{mag} = -10$ relative to this value. In addition, we also manually produced masks to cover regions of extended emission throughout Orion~A.

Subsequently, dynamic background models with variable window sizes, $w$, were calculated, where $w$ corresponds to the number of frames to include for each model. For our data, a compromise between accurate sky sampling and acceptable noise in the background model was found for values of $w$ around 15 to 20 in the $H$ and $K_S$ bands, and $w \approx 10$ for $J$. To this end we first normalized the input data by subtracting the mode of each frame and then calculated the median from the $w$ closest input pawprints in time. In all cases with separate offset sky observations, the background models were calculated from the offset observations alone.

\begin{figure*}
        \centering
        \resizebox{\hsize}{!}{\includegraphics[]{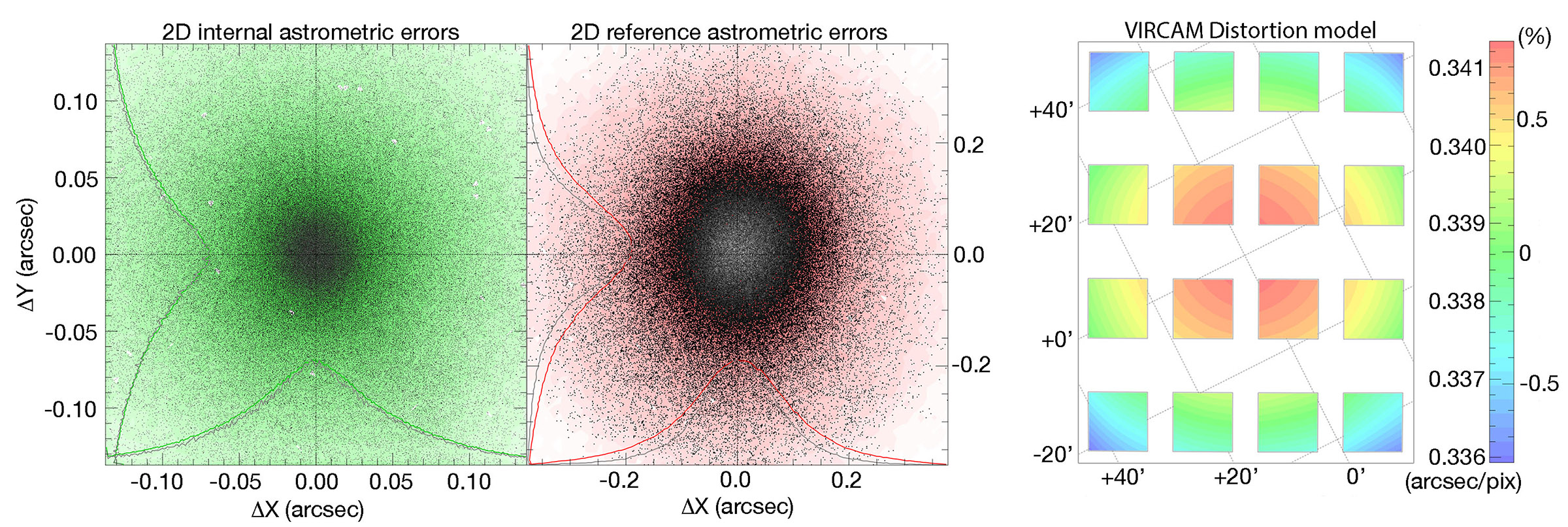}}
        \caption{Verification plots created by Scamp for all data on Orion~A in the $K_S$ band. The two figures on the left show the 2-dimensional internal and external dispersions in the astrometric matching procedure. The symmetry of the Gaussian-shaped distributions indicates that the remaining systematic errors are negligible compared to other noise terms. The typical global errors (RMS) were found around 40 mas internally and 70 mas with respect to the reference catalog. The plot on the righthand side shows the derived distortion model of VIRCAM, where the color indicates the variation in pixel scale across the focal plane. Only very small distortion levels are seen with an amplitude of $\sim$2\% from the center to the edge.}
        \label{img:scamp}
\end{figure*}

\subsubsection{Cosmetics}

As a last step in the basic reduction, some cosmetic flaws, which were still visible after the preceding calibration stages, had to be taken care of. Mainly, a residual horizontal pattern could be seen on the background across all detectors. This, however, can easily be removed by just subtracting the median of each detector row from all input frames and is referred to as ``de-striping'' in the CASU reduction. Since de-striping only works for images where regions of extended emission (if present) are smaller than a detector, we skipped this step for all frames that included the Orion nebula.

As a final step in the data reduction, bad pixels, as given in the BPM, were interpolated. This was necessary for successfully deploying our high-order resampling kernels owing to the large number of bad pixels on the VIRCAM detectors. Running such kernels on regions with bad pixels produces ``holes'' of the size of the kernel in the resampled frames, which were more difficult to reject in the final pixel stack and, in general, increased the overall noise level. Bilinear interpolation kernels, such as those used by the CASU pipeline, suffer considerably less from this problem, however at the cost of introducing more systematic errors. Bad pixel interpolation in general is not desirable since these should naturally be rejected during co-addition. To keep the impact at a minimum, we used a nonlinear bi-cubic spline interpolation code where only pixels that have fewer than 20~\% ``bad neighbors'' within a radius of four pixels were interpolated. For an overview of the impact of different interpolation methods see \cite{2013PASP..125.1119P}, among others.

\subsubsection{Remarks}

A first inspection of the data did not reveal any strong contamination by cosmic ray events. Also subsequent visual inspection of the reduced combined images showed only very few artifacts that might have originated in cosmic rays. Therefore, no attempt to identify and mask those was made. Also, no fringe correction was applied during any stage of the data processing. Fringes can occur for various reasons, such as interference effects in the detector or scattered light. If these patterns are not highly variable on spatial scales, they can be mistaken for sky background emission. They differ from them by variable amplitudes and different time scales and therefore, if present, must be removed in a separate reduction step. After inspecting many of the science frames in our survey, only very localized and low amplitude fringe patterns could be found, which were mostly taken care of during the background modeling and/or co-addition. Any attempts to correct for those small effects would have undoubtedly introduced more systematic errors, so they were neglected.

\subsection{Astrometric calibration}
\label{sec:astromcal}

Before any co-addition could be performed, the frames needed to be registered to a common reference frame. To calculate astrometric solutions, we used Scamp \citep[v2.0.1,][]{2006ASPC..351..112B}. This software package performs pattern matching of arbitrary source catalogs with any available reference catalog, and subsequently determines accurate astrometric (and to some extent also photometric) solutions. From the reduced pawprints, relatively shallow source catalogs were generated with SExtractor to match the 2MASS dynamic range, which served as the basis for the astrometric calibration. 
Scamp offers several options for treating multi-extension Flexible Image Transport System \citep[FITS,][]{2010A&A...524A..42P} files, in our case focal plane arrays, depending on the reliability and completeness of the initial input parameters in the FITS headers. We tried several combinations of the available modes in Scamp, but only running the software in LOOSE\footnote{In this mode each detector is treated individually without a global focal plane model.} mode offered an unbiased global astrometric solution without any systematics across the focal plane. In principle, the FIX\_FOCALPLANE\footnote{Here Scamp attempts to derive a common WCS projection followed by computing the median of the detector positions with respect to the focal plane.} mode was also employed successfully, but in that case we observed systematic source clipping toward the outer detectors, resulting in mismatches between individual tiles. 

All three bands were calibrated separately with a third-order distortion model over the focal plane. Figure \ref{img:scamp} shows the internal (VISTA source-to-source scatter) and external (VISTA-to-2MASS scatter) astrometric errors along with the derived VIRCAM$@$VISTA distortion model as generated by Scamp. The dispersion (RMS) for the global astrometric solution in all bands was between 40 and 45 mas with respect to internal source matches and about 70 mas with respect to external (2MASS) matches. This compares very well to the CASU mean RMS value of 70 mas as given in the headers of the assembled tiles, which also uses 2MASS as an astrometric reference. 

In addition to the astrometric solutions, Scamp can be used to derive photometric scaling factors to calibrate all input data to the same zero point. This method, however, has two major drawbacks: (a) Scamp only calculates zero-point offsets between entire pawprints and does not take residual detector-to-detector differences into account, and (b) Scamp requires single sources to be visible in all input data. The observing strategy, together the sparse focal plane coverage, provides only a tiny overlapping field for all telescope pointings. For our jitter box width, we found overlaps smaller than 1 arcmin across. Therefore in most cases there would be no sources available in these overlaps. For these reasons Scamp cannot be used for a global fine-tuned gain harmonization based on relative internal source measurements alone. We therefore adjusted the relative zero points by comparing the source catalogs for each detector with 2MASS reference stars (see Sect. \ref{sec:gainharm} for details). The typical internal photometric scatter at this stage was around 0.01 mag. For more details on the astrometric properties of our VISTA survey, see Appendix \ref{app:astr}.

\subsection{Tile and Orion~A mosaic assembly}
\label{sec:coadd}

Prior to assembling the final mosaics, additional processing steps were required to produce science-ready data. These included resampling onto a common reference frame, global background modeling, and the fine-tuned gain harmonization. For quality control and computational reasons, we chose to co-add each single tile before assembling the final Orion~A mosaic. 

For all co-addition tasks during the data processing, we used the method of \cite{2014PASP..126..158G}, who implemented an algorithm for optimized artifact removal while retaining superior noise characteristics in the co-added frame. In principle this method works in a similar way to a $\kappa - \sigma$ clipping technique, but allows for an additional degree of freedom to account for variable point spread function (PSF) shapes.  Not only did we observe excellent artifact removal, but also the standard deviation in the background was found to typically be 10 - 20\% lower than a median-combined mosaic. The photometric calibrations referred to in the following sections are described in Sect. \ref{sec:photcal}.

\subsubsection{Resampling}

With the focal plane model and astrometrically calibrated science frames in place, the images were resampled onto a common reference frame using SWarp \citep[v2.38.0,][]{2002ASPC..281..228B} using a third-order Lanczos kernel \citep{1979JApMe..18.1016D}.  To avoid complex flux-scaling applications across the tiles due to variable on-sky pixel sizes, we chose a conic equal area projection \citep[COE,][]{2002A&A...395.1077C} in equatorial coordinates with one standard parallel at $\delta = -3^\circ$ as the projection type, the field center to be aligned with the center of each tile, and a pixel scale of 1/3 arcsec/pix. Choosing an equal area projection over the standard gnomonic tangential projection assured that every pixel covered the same area on-sky, which avoids further flux adjustments for the subsequent photometry.  

\subsubsection{Global background modeling}

During resampling, SWarp can fit a user-defined background model to each frame, but it neglects overlaps between the individual images. In cases of very crowded fields, particularly in the presence of extended emission, this method introduces discontinuities across the full tiles. To correct for these last remaining offsets between overlapping images, we calculated a global background model with the Montage software package (\url{http://montage.ipac.caltech.edu/}) while using very large mesh sizes (a constant offset for the tiles N2 and S2 and 1/3 of a detector for all other fields) with SWarp. It is important to note here that despite our efforts to apply as little spatial filtering as possible in the background correction, minor residuals are still visible throughout the assembled tiles. For this reason we do not encourage measurements of nebulous emission on our mosaics. We estimate that structures of few arcminutes in size should mostly be preserved in our reduction.

\subsubsection{Illumination correction}
\label{sec:gainharm}

Up to this point any zero-point offsets between the detectors were only corrected for during the calibration with the master flat field. This, however, proved to be mostly insufficient. Unaccounted-for scattered light in the optical train or imperfect flat fields are two examples of effects that can create variable photometric zero points over the field of view. As already mentioned above, it is not possible to attempt a gain harmonization based on internal photometric measurements from the science fields with VIRCAM owing to the nonexistent overlaps in the offset pattern. The calibration plan offers standard field observations specifically for this correction, but since these measurements could not be performed simultaneously with the Orion~A field and are carried out only once per night or upon user request, it was safer to rely on external standard catalogs.

To this end, we defined subsets of the data for which we assumed stable photometric conditions with respect to the zero point and the PSF shape. Each of these subsets comprised one detector for each jitter sequence (5 frames for the $H$ and $K_S$ bands, 3 or 6 frames for the $J$ band; compare with Table \ref{tab:observations}). Thus each tile was split into 96 subsets (16 detectors, 6 offset positions). The jitter box width and the execution time for each of these sequences were in a range where this assumption should hold. This assumption only breaks down for the tiles with offset sky fields (S2 and N2), where one of the six-step offset patterns was completed before any jitter was executed. 

The images in each of these subsets were co-added, and for the resulting data we performed source extraction with SExtractor to calculate zero-point offsets relative to 2MASS. We then used the 96 determined zero points to calculate relative flux scaling factors. 

\subsubsection{Observing parameters}
\label{sec:obspar}

For quality control purposes, we also calculated several observing parameters for each of the given subsets as defined in Sect. \ref{sec:gainharm}. These include the local seeing conditions (FWHM; estimated with PSFEx), effective exposure time, frame coverage, and the local effective observing time (MJD). Most important, aperture correction maps were also generated for apertures with discrete radii of 2/3, 1, 2, 3, and 4 arcsec. The fluxes were corrected to an aperture of 5 arcsec for which no variation due to changing seeing conditions was expected. Only point-like sources (as classified by SExtractor) with a high signal-to-noise ratio (S/N) were included in the calculation of the aperture corrections. Examples of the quality control parameters are shown in Appendix \ref{app:qc}.

\subsubsection{Co-addition}

Once the photometric flux scaling was adjusted with the correct zero-point offsets, the original resampled frames were co-added to the final tiles with SWarp. We then again created shallow source catalogs and calculated relative zero-point offsets for each tile, and we finally merged all tiles into the Orion~A mosaic. The final mosaic constructed from all data for each filter also features a COE projection with the same standard parallel and pixel scale as the individual tiles. For easier data access and three-color image assembly, we used the same projection for all filters.

Unfortunately, the two separate observations in $K_S$ of tile S2 featured one of the best and one of the worst observing conditions in terms of image quality (FWHM), respectively. As a consequence, when co-adding these tiles, we saw a significant drop in S/N after source extraction. For this reason we decided to only include the data set taken during the better ambient conditions.

\subsection{Photometric calibration}
\label{sec:photcal}

The recipes described in this section apply to all stages throughout the data processing where photometric calibration was performed.

\subsubsection{Source detection and extraction}

Source detection and extraction was performed with SExtractor where we tested several different detection thresholds with respect to the background noise level, $\sigma$. For the final source catalog we chose a threshold of 1.5$\sigma$, requiring at least three connected pixels above this level, while lowering the default deblending threshold by two orders of magnitude to also detect sources in high-contrast regions. This combination proved to be optimal because a visual inspection of multiple regions in the mosaic showed only a few misdetections (< 1\%) of nebulosity and residual artifacts. We interpreted the low threshold as a validation of the methods for creating the weight maps and co-added the data. For zero-point determinations and astrometric matching, the threshold was typically set to 7$\sigma$.

The resulting source catalogs were cleaned by removing all bad measurements, i.e. sources with negative fluxes or a SExtractor flag larger than or equal to four (essentially saturated or truncated objects). For tiles and the final Orion~A mosaic catalog, we applied the previously determined aperture corrections to all the extracted sources. In addition, each source was also assigned an effective observing MJD, exposure time, frame coverage, and local seeing value using the quality control data as described in Sect. \ref{sec:obspar}. 

\subsubsection{Photometric zero point}

For reliably determining the photometric zero point, only a limited dynamic range was used since bright stars still showed signs of nonlinearity, and for fainter stars we found a large dispersion relative to 2MASS owing to low S/N in the reference catalog. For the $J, H$, and $K_S$ bands, we used ranges of $[12,15], [11.5, 14.5]$, and $[11,14]$ mag, respectively. Also, we required an \textit{A} quality flag in 2MASS for a reference source to be used in the zero-point determination, a SExtractor flag of 0 (i.e., no blending, truncation, or incomplete or corrupted data), and an error below 0.1 mag in our catalog. These requirements offer both good S/N values in our survey and 2MASS, and typically also several thousand available reference stars for a single tile. For the co-added subset described in Sect. \ref{sec:gainharm}, we typically find several tens of sources matching these criteria. For cross-correlation between the catalogs, we searched for matches within a radius of 1 arcsec where, in cases of multiple possibilities, the nearest match was always selected. 

The zero point was determined by applying a one-pass 2-$\sigma$ clipping in the 2MASS - VISTA parameter space and by fitting a simple linear function with a forced slope of 1 to the data, weighted by the sum of the inverse measurement errors. Typical errors for the zero point were found to be around 0.01 mag. We also decided not to include color terms in the photometric calibration since (a) we did not see any significant dependency on those within the measurement errors, and (b) we aimed for separate calibrations for each individual filter without the need for detections in multiple bands. 

\subsubsection{Catalog magnitudes}

In summary, magnitudes and errors were calibrated onto the 2MASS photometric system (in contrast to the CASU pipeline) and calculated using equations of the form
\begin{align}
        \mathrm{m}_{i,r} &= -2.5\log\left(\frac{F_{i,r}}{t_i}\right) + \mathrm{apcor}_{i,r} + \mathrm{ZP}_{i,r} \\
        \Delta \mathrm{m}_{i,r} &= 1.0857 \times \frac{\sqrt{A_{i,r}\sigma_{i,r} + F_{i,r}/g}}{F_{i,r}}
\end{align}

\noindent where $r$ refers to each aperture size, $F_i$ are the measured fluxes, $t_i$ the exposure times, $\mathrm{apcor}_{i}$ the aperture corrections,  $\mathrm{ZP}_{i}$ the determined zero point, $\Delta \mathrm{m}_{i}$ the calculated errors, $A_i$ the area of the aperture, $\sigma_i$ the standard deviation of the noise, and $g$ the gain. We note here that magnitude errors calculated in this way should only be taken as lower limits because (a) SExtractor does not include a term describing the absolute background flux in the aperture typically found in CCD S/N equations and (b) we do not include systematic errors. Furthermore, we do not include uncertainties from the determination of the zero point (typically $\Delta \mathrm{m}_{ZP} \approx 0.01$ mag.) for individual source magnitude errors. Since the errors are calculated independently for each source based on photon statistics alone, they can be considered random and may only show spatial correlations due to variable observing conditions and changes in the image quality across the focal plane array.

Finally, the adopted catalog magnitude for each source was chosen so that the selected aperture maximized the S/N among all measurements. After extensive tests, we found that the best overall measurement to represent fluxes for all sources can be achieved by selecting the catalog magnitude from only the two smallest apertures (2/3 and 1 arcsec).

\subsection{Final catalog assembly}

For the final source catalog, we applied the aforementioned source extraction procedures to the entire Orion~A mosaic. In this way, we avoided the issue of multiple detections of the same sources and at the same time increased the S/N in the overlapping regions. In contrast to all intermediate catalogs, we included additional processing steps for assembling the final catalog of the full Orion~A mosaic. We focused on four remaining issues: 

\begin{enumerate}
        \item Morphological classification to distinguish between extended and point-like objects;
    \item Cleaning of spurious detections;
    \item Sources in the residual nonlinearity and saturation range;
    \item Source detection near the Orion Nebula due to significant and highly variable extended emission.
\end{enumerate}

During the following sections we address these supplementary processing steps individually.

\subsubsection{Morphological classification}

SExtractor itself is able to distinguish between extended and point-like objects on the basis of neural networks. The successful application of this method, however, critically depends on the input parameters; in particular, a correct guess of local seeing conditions (FWHM) is crucial. Since the FWHM varies by more than a factor of 2 across the entire Orion~A mosaic, the classification with SExtractor shows residual systematics correlating with the variable PSF sizes. To mitigate this situation, we ran SExtractor several times on subsets of the mosaic with similar seeing (see Sect. \ref{app:qc}).

\begin{figure}
        \resizebox{\hsize}{!}{\includegraphics{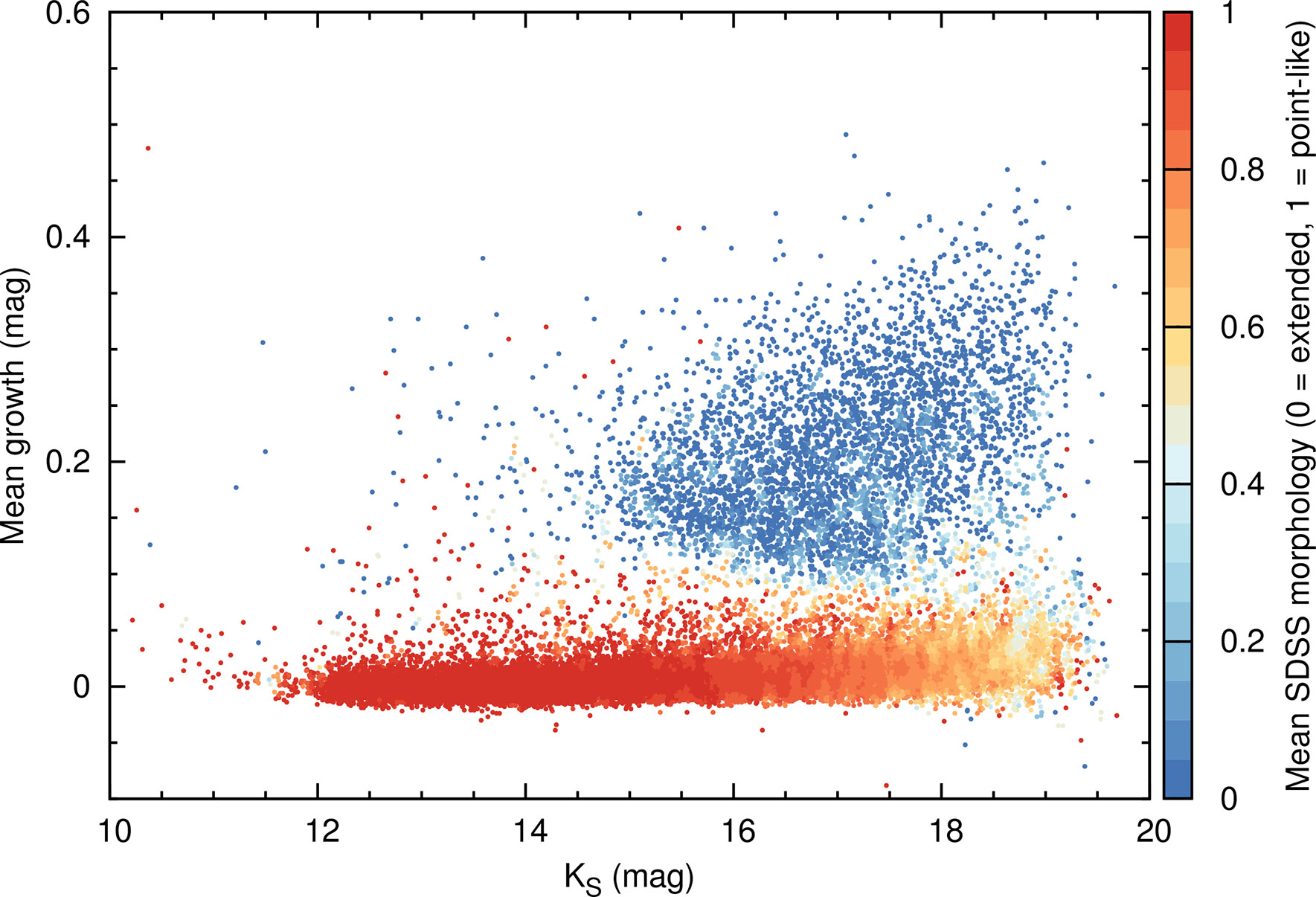}}
        \caption[]{Magnitude growth when increasing photometric apertures from 2/3 to 1 arcsec as a function of $K_S$ band magnitude. The color indicates the mean SDSS morphology in a box of $\Delta K_S$=0.1 mag, $\Delta$growth = 0.02 mag. In this parameter space, galaxies are well separated from point-like objects, which was used for a refined morphological classification.}
        \label{img:sdss}
\end{figure}

The morphological classification for each source was chosen among all three bands to minimize the effect of the local seeing. Despite all these efforts, systematic trends are still visible with the SExtractor classification that correlates with observing conditions. Nevertheless, for sources well above the detection threshold, this method produced reliable results. 

These issues led us to decide to implement an independent method for distinguishing sources with point-like or extended morphology. We found that a very robust parameter for describing the shape of a source was provided by the curve-of-growth analysis calculated earlier. Since our aperture corrections are only valid for point-like sources (only point-like sources were allowed in its calculation), any elliptically or irregularly shaped source should show a growth value different from 0 in our aperture-corrected magnitudes. Among all available apertures, the best parameter for the classification was the difference between aperture corrected magnitudes for the 1 and 2/3 arcsec apertures. 

We then cross-matched (1 arcsec radius) all detected sources with the Sloan Digital Sky Survey (SDSS) catalog \citep[DR7,][]{2009ApJS..182..543A}, which includes one of the most reliable galaxy classifications in this field down to very faint magnitudes. From the cross-matched sample, we constructed a relatively clean subset by selecting only those sources with magnitudes brighter than 23 mag in all available bands $\left( u,g,r,i,z \right)$, which seemed to be a good compromise between acceptable S/N and source counts. This subset contained about 47\,000 objects, among which about 80 \% were classified as stars. 

Figure \ref{img:sdss} shows the curve-of-growth parameter as a function of $K_S$ magnitude, with  color indicating the corresponding SDSS morphology. Clearly, galaxies separate very well from stars. We then used the cross-matched SDSS subset as a training sample for a $k$-nearest-neighbor analysis ($k$NN, $k=30$) applied to the entire survey catalog. For details on the classification method and the Python implementation we used, see \cite{scikit-learn}. In reference to Fig. \ref{img:sdss}, this method tends to favor point-like sources for faint objects simply because the training set included about four times more stars than galaxies.

For completeness, we mention here that we also attempted a multivariate classification based on the available colors, $(J - H)$ and $(H - K_S)$. Unfortunately, the limited available color parameters were not enough to separate galaxies from point sources reliably with these methods. Supplementary optical data of equivalent completeness would aid tremendously in identifying background galaxies. Unfortunately, SDSS covers only a portion of our Orion~A field. 

\begin{figure}[t]
        \resizebox{\hsize}{!}{\includegraphics{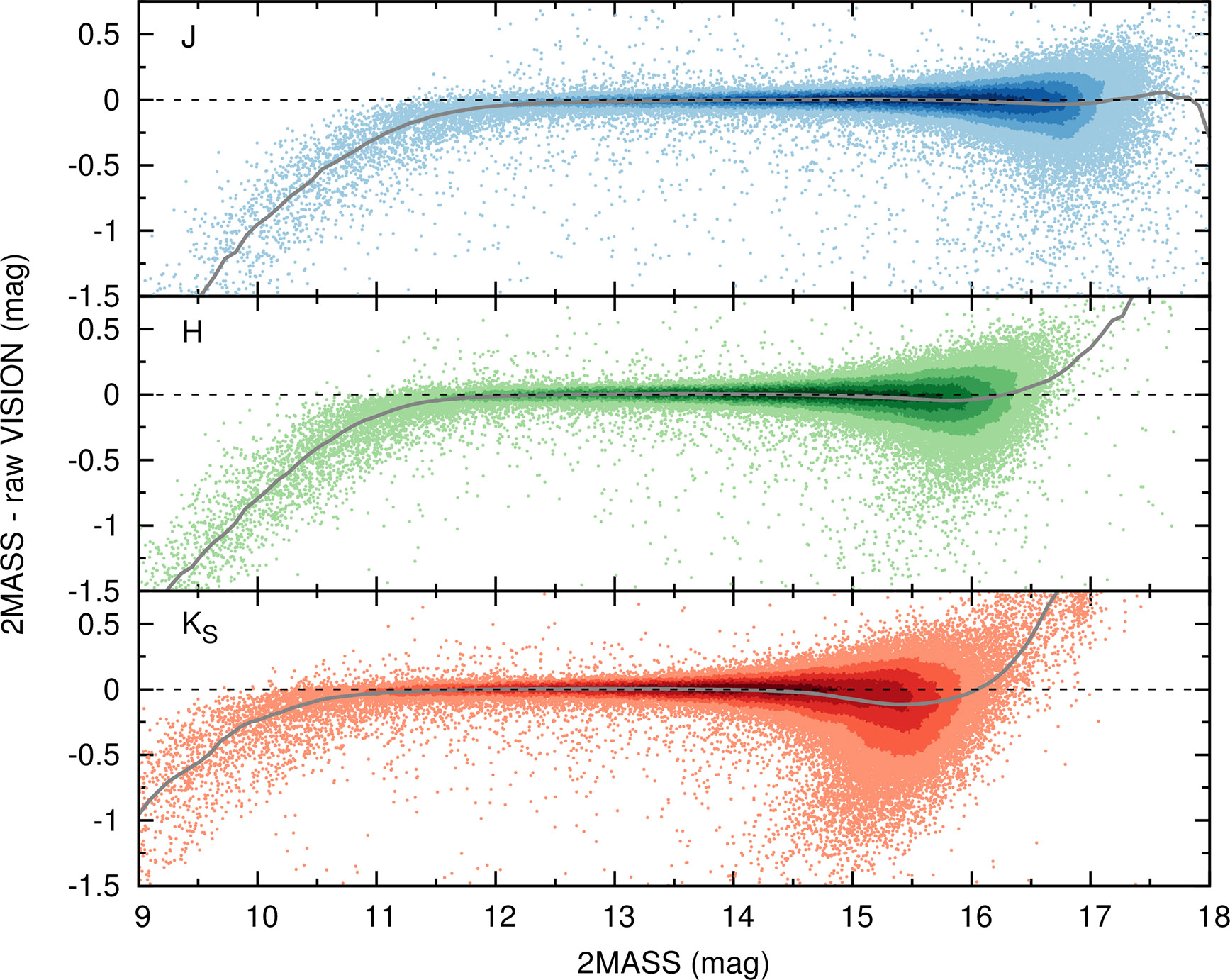}}
        \caption[]{2MASS vs. VISTA photometry. Clearly residual nonlinearity and saturation effects have an impact on the bright end of the measured magnitudes. For this reason we decided to replace VISION photometry with 2MASS photometry for sources brighter than $(13,12,11.5)$ in $\left( J, H, K_S \right)$, respectively. The shading indicates source density in a $0.2 \times 0.05$ box in this parameter space. The gray line indicates a running median with a box width of 0.5 mag, and the dotted horizontal line marks the reference value of zero mag difference between the 2MASS and VISTA catalogs.}
        \label{img:mag_turnoff}
\end{figure}

\subsubsection{Spurious detections}
\label{sec:spurdet}

SExtractor has its own, quite robust, implementation for cleaning spurious detections by assessing local detection thresholds for each individual source. These were mostly picked up in the vicinity of bright stars  ($\la$ 10 mag) and could be identified well by their morphological classification. To remove these, we used a cleaning radius proportional to the 2MASS magnitude to approximately fit the halo structures and simply removed all extended objects. In addition to this cleaning iteration, we also found a few hundred detections associated with extended emission. This subset, however, was easily identified by large curve-of-growth values, extended SExtractor morphology, and proximity to the detection limit. A visual inspection of the remaining sources only revealed very few spurious detections. We did not attempt to remove those manually, since such a task could not have been applied in a consistent manner for the entire mosaic.

\subsubsection{Residual nonlinearity and saturation}
\label{sec:cat_clean_2mass}

Comparing the resulting catalog with 2MASS, as displayed in Fig. \ref{img:mag_turnoff}, one can see that stars brighter than a wavelength-dependent magnitude limit were either affected by residual nonlinearity problems or were saturated. We therefore replaced all the sources in our catalog that were located in the immediate vicinity of stars brighter than 13, 12, 11.5 mag in $J$, $H$, $K_S$ in 2MASS, respectively, with the corresponding single clean reference catalog measurement from 2MASS. All clean, high S/N detections (quality flag \textit{A}) were propagated from the 2MASS catalog. A few remaining very bright sources (quality flag \textit{B} or worse) can therefore only be found in the 2MASS catalog. Most of them are detections of nebulosity near the ONC and only very few ($\sim$15) are associated with saturated sources. In addition we found about 20 sources in 2MASS with \textit{A} quality flags across all three bands that are fainter than the above-mentioned limits and were not detected in the VISTA images. These sources were not added to the final catalog.

\subsubsection{Sources near the Orion nebula}

Common source detection techniques unfortunately do not provide satisfactory results in regions where the background varies significantly on very small scales (i.e., on scales smaller than a few times the size of the core of the PSF). In this case the modeling of the background fails even for advanced methods. The SExtractor method (and also the CASU pipeline) fails for the regions around the ONC where we find background variation on sub-arcsec levels, even when using specialized filtering kernels. As a result some localized emission peaks get easily picked up as sources, producing a relatively large number of false detections. We therefore decided to make a 2000 $\times$ 2000 pix ($\sim$11 $\times$ 11 arcmin) cutout around the ONC for which we manually cleaned the SExtractor catalogs, while also adding missed sources. All sources in this subset were recentered by calculating a Gaussian least-squares fit at the input coordinates with IRAF \citep{1986SPIE..627..733T}. From this new and cleaned coordinate list, we created an artificial image with Skymaker \citep[v3.10.5,][]{2009MmSAI..80..422B} and used this image as input for SExtractor in double imaging mode, while extracting the sources from the original cutout. We note here, that this, of course, does not avoid the problems in the photometry associated with such highly variable background (e.g., flux over- or underestimations depending on aperture radius owing to imperfect removal of the extended emission and systematic offsets in measured source positions). Compared to automated 2MASS photometry that shows many misdetections, we are confident that our source catalog in this region is among the most reliable ones.

\begin{table}
        \caption{Accessible data columns for our VISION survey. Apart from astrometric and photometric measurements, we also include several quality control parameters and two morphological classification schemes.}
        \label{tab:vision_columns}
        \begin{tabular}{lcc}
        \hline\hline
        Column name 				& Description 				& Unit \\
        \hline
        VISION\_ID					& internal numbering 		& \# \\
        RAJ2000						& Right ascension (J2000) 	& hh:mm:ss \\
        DEJ2000						& Declination (J2000) 		& dd:mm:ss\\
        ($J/H/K_S$) 				& corrected magnitude 		& mag \\
        ($J/H/K_S$)\_err 			& magnitude error 			& mag \\
        Class\_cog\tablefootmark{a} & $k$NN-morphology 			& \{0,1\} \\
    Class\_sex\tablefootmark{b}  	& SExtractor-morphology 	& [0,1] \\
        ($J/H/K_S$)\_mjd 			& effective MJD 			& d \\
        ($J/H/K_S$)\_exptime 		& eff. exposure time 		& s \\
        ($J/H/K_S$)\_fwhm 			& source FWHM 				& arcsec \\
        ($J/H/K_S$)\_seeing 		& local seeing 				& arcsec \\ 
        ($J/H/K_S$)\_coverage 		& frame coverage 			& \# exposures \\
        ($J/H/K_S$)\_aper 			& aperture radius 			& arcsec \\
        ($J/H/K_S$)\_2mass\_id 		& 2MASS identifier 			& \\
        ($J/H/K_S$)\_origin 		& original catalog 			& \\
    \hline
        \end{tabular}
\tablefoot{
        \tablefoottext{a}{Based on curve-of-growth (cog) characteristics and a subsequent $k$NN analysis with SDSS classifications in the training sample.}
    \tablefoottext{b}{Based on neural networks and delivered with SExtractor; the published value is the one across all three bands for which the best seeing conditions were measured.}
}
\end{table}

\begin{figure*}[tp]
        \centering
        \resizebox{\hsize}{!}{\includegraphics[]{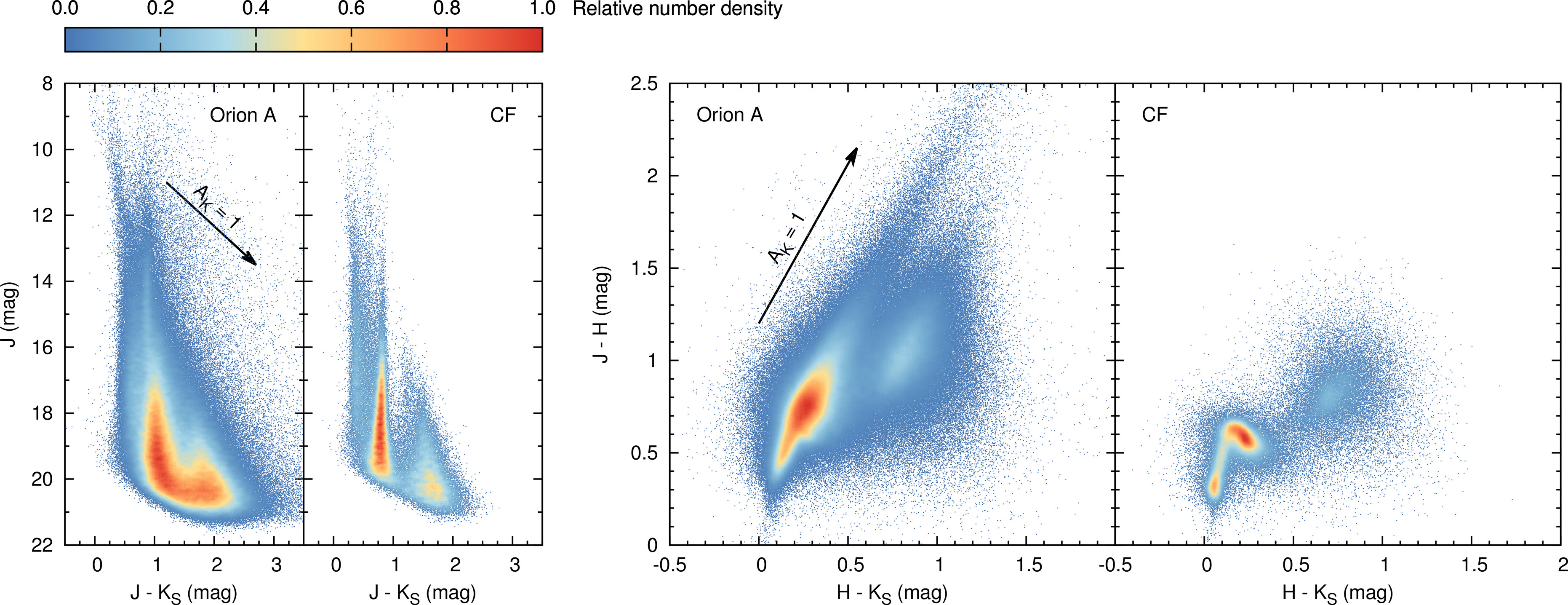}}
        \caption{Color-magnitude and color-color diagrams for both the Orion~A and the CF data. The colors indicate normalized source density within a $0.1 \times 0.1$ mag box in the $J$ vs. $J - K_S$ parameter space and $0.02 \times 0.02$ mag in $J - H$ vs. $H - K_S$. The black arrows indicate the effect of an extinction of 1 magnitude in $K_S$. The presence of heavy dust extinction in Orion~A pushes many sources toward redder colors when compared to the CF, which itself has a clearly defined main-sequence ($0.2 \lesssim J-K_S \lesssim 1.2$, $0 \lesssim H-K_S \lesssim 0.4$) and galaxy locus ($1.2 \lesssim J-K_S \lesssim 2.2$, $0.5 \lesssim H-K_S \lesssim 1.2$) in both diagrams.}
        \label{img:cmd_ccd_dens}
\end{figure*}

\section{VISTA Orion~A survey data products}
\label{sec:data_products}

In this section we describe the main data products of the VISTA Orion~A survey: Section \ref{sec:sourcecat} contains a concise overview on the source catalog, together with the presentation of the resulting color-magnitude and color-color diagrams for both the Orion~A and CF data. In Sect. \ref{sec:result_lrgb} we present an L-RGB version of the entire mosaic and a catalog of interesting objects extracted from the provided image. This includes a selection of prominent, already known YSOs and, based on morphology criteria, an identification of five new YSO candidates. We also used the catalog to identify probable new galaxy clusters. Details on photometric, astrometric, and quality control properties can be found in Appendix \ref{app:data_characteristics}. Appendix \ref{app:tables} contains supplementary data tables.

\subsection{VISTA Orion~A source catalog}
\label{sec:sourcecat}

After all calibration, merging, and cleaning steps, the final source catalog contained 161 parameters across all three bands. Since most of the calculated parameters were only used for calibration purposes and can only be fully understood with access to all details of the data reduction recipes, we decided to reduce the load by including the 35 most important columns in the published catalog. An overview of the available data is given in Table \ref{tab:vision_columns}, and a sample of the VISION data is shown in Table \ref{tab:vision_sample}. 

For a total on-sky coverage (including the jitter sequences and pixel rejection during co-addition) of 18.2935 deg$^2$, we detected a total of 799\,995 individual sources across all three observed bands; 505\,339 of these were detected in all three filters, 653\,888 in at least in two bands. For the individual $J$, $H$, and $K_S$ bands, we detect 571\,458, 747\,290, and 640\,474 sources, respectively. In contrast, the 2MASS point source catalog contains 86\,460 sources in the area covered by our survey, and the 2MASS Extended Catalog \citep{2006AJ....131.1163S} only a few hundred. For the CF we found a total of 93\,909 sources with 65\,665 detected sources in all three bands, and 80\,526 sources were detected in at least two filters. For both the Orion~A observations and the CF, we classified about 30\% as extended objects with our curve-of-growth analysis. Above the residual nonlinearity and saturation limits, we added or replaced 7788, 6298, and 5355 sources from the 2MASS catalog in $J$, $H$, and $K_S$, respectively. The mean image quality (FWHM) for the Orion~A data is 0.78, 0.75, and 0.8 arcsec with a standard deviation of 0.07, 0.08, and 0.1 arcsec in $J$, $H$, and $K_S$. The survey catalog for all sources will be made available through the CDS.

\subsubsection{VISTA photometry}
\label{sec:results_phot}

We estimate the completeness to be 20.3 mag for $J$, 19.7 mag for $H$, and 18.7 mag for $K_S$ as determined by the histogram peaks in the three luminosity distributions. This, however, is highly variable throughout the mosaic because of unequal coverage, variable extinction, and extended emission in the region (for details see Appendix \ref{app:phot}). The median absolute deviations of the photometric errors are 0.056, 0.048, and 0.048 mag in $J$, $H$, and $K_S$, respectively.

As an example of the calibration and photometric properties of the survey, we show color-color and color-magnitude diagrams for both Orion~A and the CF in Fig. \ref{img:cmd_ccd_dens}. The CF in the color-magnitude diagram exhibits the typical early-type dwarf sequence ($J - K_S \sim 0.4$ mag) and the fainter but very visible branch of M dwarfs ($J - K_S \sim 0.8$ mag), together with the fainter locus of galaxies at $J - K_S \sim 1.5$ mag. The color-color diagram for the CF is very clean with a well-defined main sequence that is not obviously affected by extinction and a good separation between the main sequence and the somewhat extended locus of galaxies. The clump in the CF near $H - K_S \sim 0.1$ mag, $J - H \sim 0.3$ mag shows bright early-type dwarf stars and appear in overabundance here since these stars are visible at much further distances than the fainter late-type stars \citep{1998PhDT........24A}. Since Orion~A is not projected against the galactic plane or the bulge, the CF lacks the typical giant sequence, which separates from the dwarf sequence at $H - K_S \sim 0.15$ mag, $J - H \sim 0.7$ mag \citep{1988PASP..100.1134B}. In contrast to the CF, stars and galaxies toward the Orion~A cloud can be substantially affected by dust extinction, introducing a color excess due to reddening. 

\subsection{VISTA Orion~A mosaic L-RGB}
\label{sec:result_lrgb}

Supplementarily to the Orion~A source catalog, we also created a full resolution L-RGB image optimized for displaying the whole dynamic range of the data. To this end we created an artificial luminosity channel by co-adding all tiles including all filters at once. Since the total mosaic size exceeded the data limits for ordinary FITS conversion software, we converted each individual band from FITS to TIFF via STIFF \citep[v2.4.0,][]{2012ASPC..461..263B}. We then mapped the $J, H, K_S$ bands to the blue, green, and red channels, respectively, and created a luminance channel from the combined image in Photoshop\texttrademark. Residual image defects (saturated stars, imperfect background subtraction, etc.) were modestly rectified for better artistic impressions. We then extracted the three RGB channels and subsequently built a Hierarchical Progressive Sky (HiPS) with the Aladin Sky Atlas \citep{2000A&AS..143...33B, 2014ASPC..485..277B}. In this way we were able to preserve the optimized nonlinear curve stretch, together with the cosmetic corrections from Photoshop. Aladin resamples the input data onto a HEALPix grid \citep{2005ApJ...622..759G}, which only allows for discrete tile orders (and therefore pixel scales). Even though the original pixel scale of the reduced image was set to $1/3$ arcsec, we chose a grid with a slightly coarser scale at 402.6 mas to reduce loading times and disk use. Given that the typical seeing of our survey is mostly around 0.8 arcsec, we only undersample our data in a few cases (see Fig. \ref{img:qc_quality}). The full L-RGB progressive sky map will be made available through the CDS at \url{http://alasky.u-strasbg.fr/VISTA/VISTA-Orion-A-Colored}. Until the integration of the data, the HiPS will also be available through \url{http://homepage.univie.ac.at/stefan.meingast/Orion_A_VISTA_RGB/}. It will not be possible to perform photometry on these data.

\begin{figure*}[p]
        \centering
    \resizebox{\hsize}{!}{\includegraphics[]{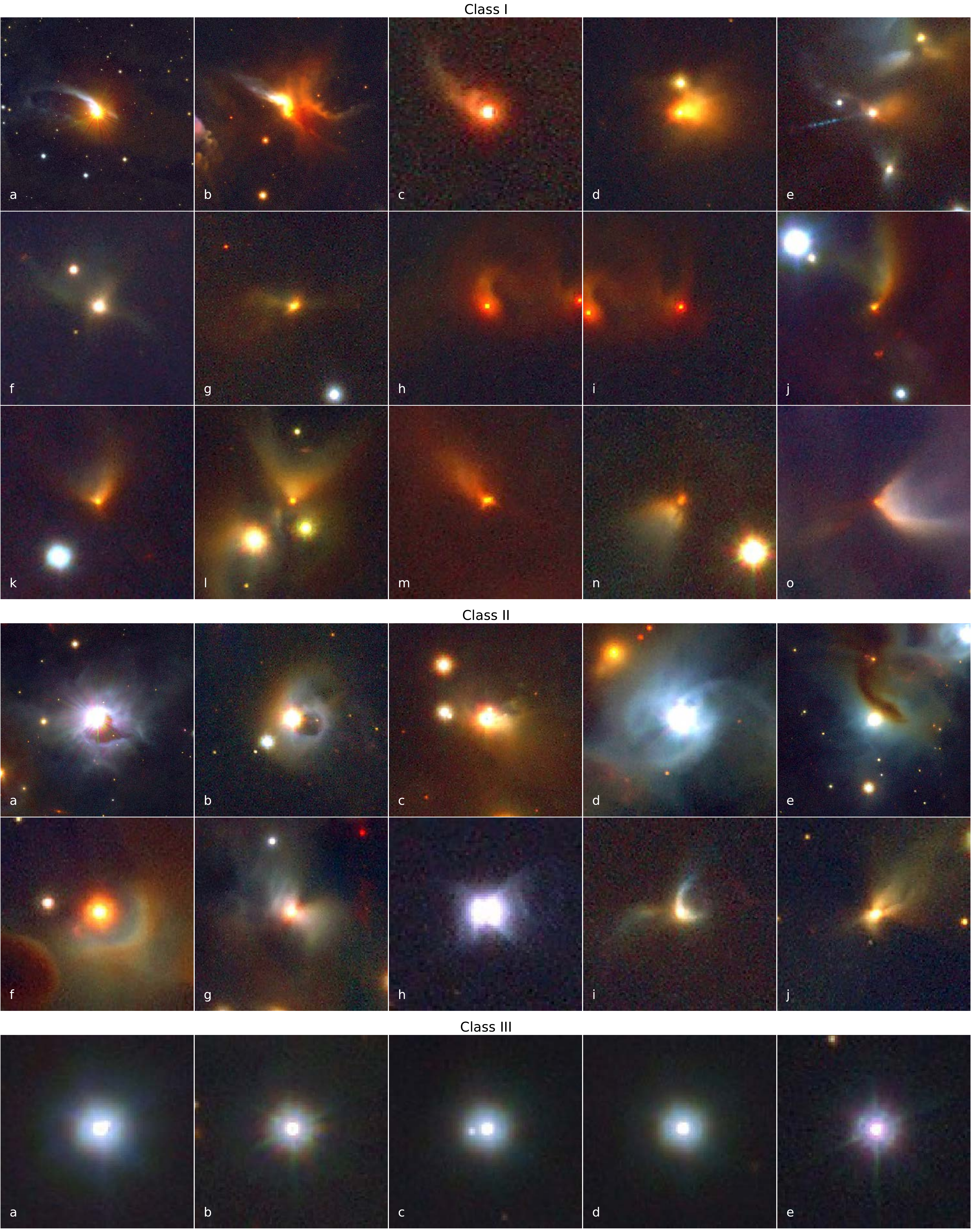}}
        \caption{Selected YSOs in our VISTA Orion~A survey. Class I and Class II sources were taken from \cite{2012AJ....144..192M}, Class III objects from \cite{2013ApJ...768...99P}. Class III sources are virtually indistinguishable from more evolved stars in NIR colors, while Class I and Class II sources very often show characteristic structures of scattered light. The labels in the bottom left corners refer to Tab. \ref{tab:newobj}.}
        \label{img:matrix_ysos}
\end{figure*}

\begin{figure*}[tp]
        \centering
    \resizebox{\hsize}{!}{\includegraphics[]{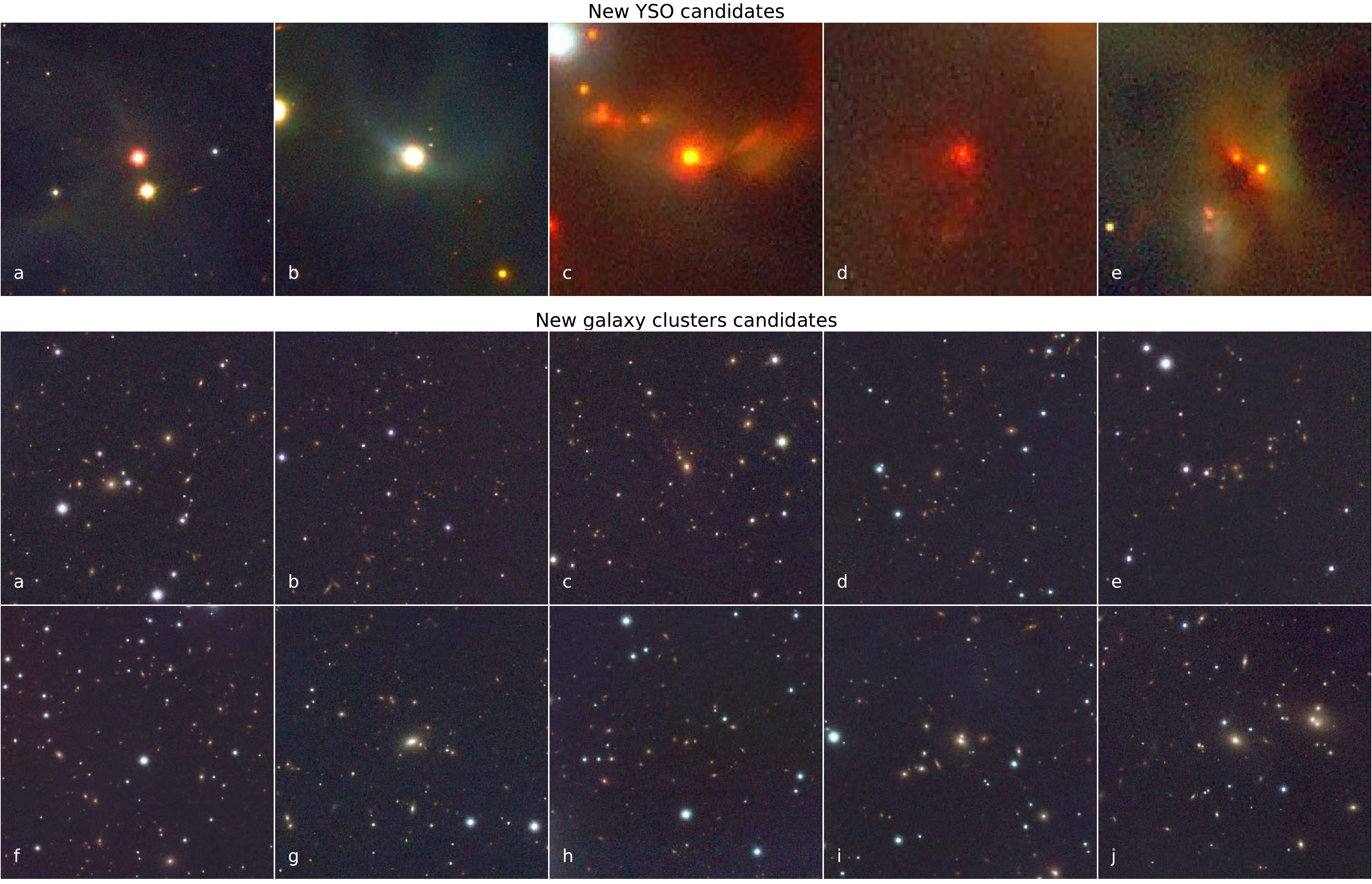}}
        \caption{Newly identified objects in the VISTA Orion~A survey. The top row shows potential new YSOs based on similar morphology compared to sources listed in the literature (see Fig. \ref{img:matrix_ysos}). The bottom matrix shows new candidate galaxy clusters near Orion~A, identified as over-densities in the galaxy distribution in the presented catalog. The labels in the bottom left corners refer to Tab. \ref{tab:newobj}.}
        \label{img:matrix_new_ysos}
\end{figure*}

\subsubsection{Catalog of interesting objects}
\label{sec:results_newobjects}

With the superior resolution and sensitivity of the Orion~A VISTA survey, it becomes possible to investigate the morphology of some already known YSOs and even extend this view to identify new candidates. We have compiled a representative list of young stars associated with prominent features of scattered light and outflows in different evolutionary stages as given in the literature \citep[e.g., ][]{1987IAUS..115....1L, 2000prpl.conf...59A, 2009ApJS..181..321E}. Figure \ref{img:matrix_ysos} shows examples of YSOs at different ages extracted from the provided Orion~A L-RGB. 
Class I and Class II identifications were taken from \cite{2012AJ....144..192M}, and Class III identifications from \cite{2013ApJ...768...99P}. While many Class I and II sources show spectacular nebulous structure in their vicinity, Class III sources are virtually indistinguishable from stars on the main sequence in NIR colors. Based on a visual inspection of the morphology of these known YSOs, we were able to identify five new YSO candidates not mentioned in previous studies (e.g., due to the restricted coverage of the \textit{Spitzer} survey). These candidate YSOs are shown in the top row of Fig. \ref{img:matrix_new_ysos}, and all together they show the typical morphological characteristics of our test sample.

In addition to the investigation of YSO morphology, we also used the VISTA Orion~A catalog to identify overdensities in background galaxies, which are potential previously undiscovered galaxy clusters. Here we evaluated the spatial density distribution of extended sources as given in our catalog with a 2 arcmin wide Epanechnikov kernel on a 1 $\times$ 1 arcmin grid. The sample was restricted to class\_sex  $\leq 0.3$, class\_cog = 0, $J/H/K_S > 15$ mag, and $K_S < 18$ mag. The last requirement filters most misclassifications for faint unresolved sources. Significant overdensities ($>5 \sigma$) were inspected visually. With this method we selected ten outstanding overdensities and classify them as potential new galaxy clusters. Postage stamps of these regions are displayed in the bottom matrix of Fig. \ref{img:matrix_new_ysos}. Cross identifications, coordinates and magnitudes for all objects shown in Figs. \ref{img:matrix_ysos} and \ref{img:matrix_new_ysos} are listed in Table \ref{tab:newobj}.

\section{Young stellar populations toward Orion~A}
\label{sec:results}

To discuss the stellar populations seen toward Orion~A, we begin by deriving an estimate of the young stellar population associated with the molecular cloud via the $K_S$ band luminosity function (KLF). Here we use the term KLF for the general $K_S$ band magnitude distribution of sources in our survey and do not explicitly refer to the monochromatic luminosity distribution of a given sample (such as the YSOs associated with the molecular cloud). In the second part, we discuss the foreground populations.

\begin{figure*}[tp]
        \centering
        \resizebox{\hsize}{!}{\includegraphics[]{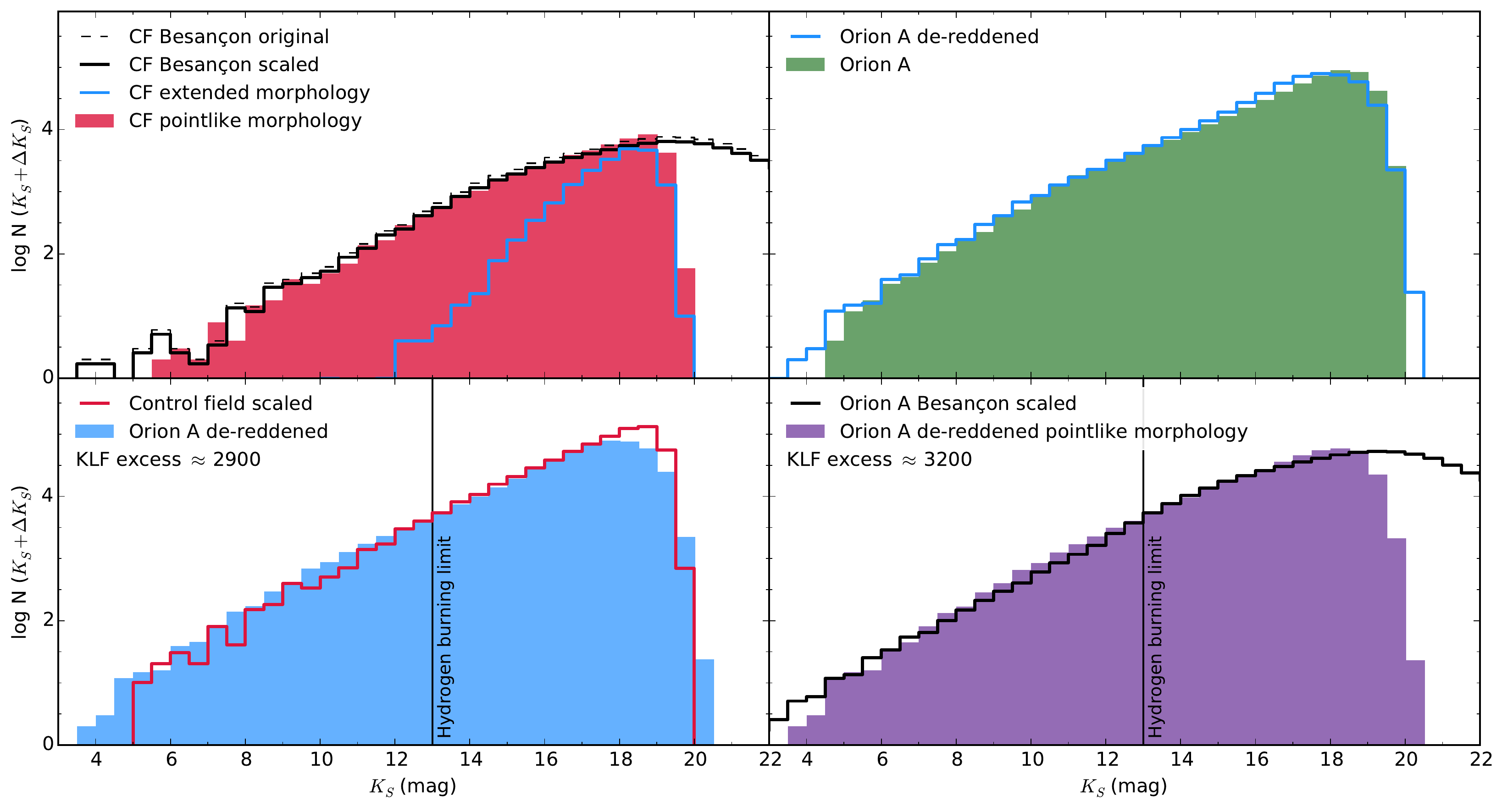}}
        \caption[]{$K_S$ band luminosity functions (histograms; 0.5 mag bin width) for the various analysis steps to estimate both young foreground and the total Orion~A population. Top left: Comparison of the CF KLF with the Besan\c{c}on model of the Galaxy. We find that a slight adjustment to all source counts in the model was necessary to fit the observations (see text for details). Top right: KLFs for the final survey catalog and the de-reddened data. Sources are mostly pushed to brighter magnitudes, however, we also allow (for statistical reasons) negative values of $A_K$ and therefore find a minor population at fainter magnitudes than in the original histogram (the last bin starting at $K_S = 20$ in the de-reddened histogram). Bottom left: Comparison of the de-reddened data with the scaled CF KLF. We clearly see an excess of bright sources, as expected from the young populations we find towards Orion~A. Note here the good agreement of the CF and Orion~A KLF for magnitudes fainter than $K_S = 13$ mag up to the completeness limit. Bottom right: Comparison of the de-reddened Orion~A KLF for point sources to the Besan\c{c}on model. Also here we find a clear excess of sources caused by the young populations seen towards the molecular cloud. Both comparisons deliver similar numbers which is interpreted as a validation for our statistical approach.}
        \label{img:klf}
\end{figure*}

\begin{figure*}[tp]
        \centering
        \resizebox{\hsize}{!}{\includegraphics[]{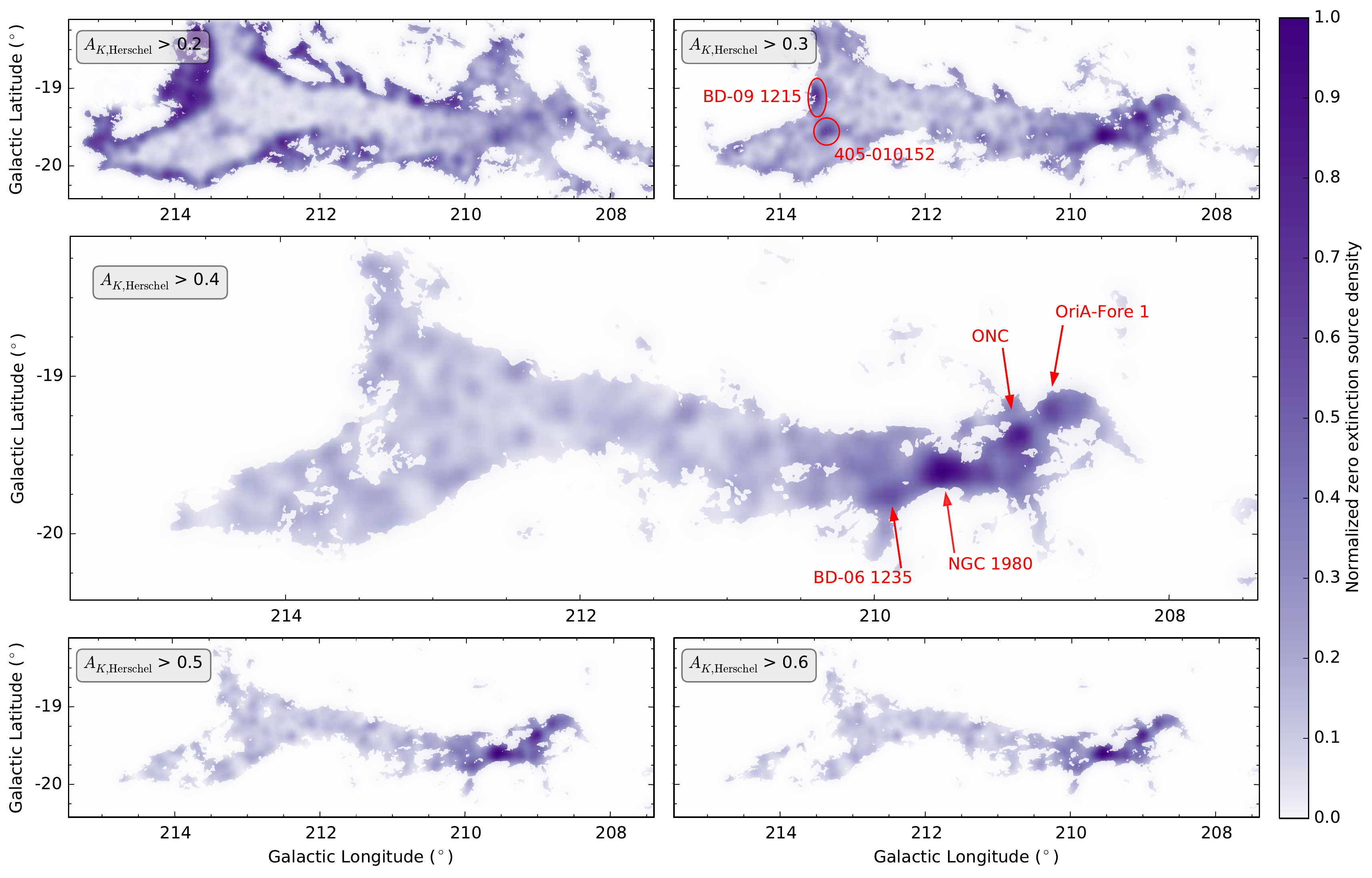}}
        \caption[]{Density maps of zero-extinction sources restricted to an area above a given column density threshold taken from the \textit{Planck-Herschel} map. Kernel densities were evaluated on a grid with 1 arcmin resolution and a gaussian kernel with 3 arcmin FWHM. Here we used the cloud to effectively block background sources to assess the population size of true foreground stars. For low column density, clearly the background contaminates the sample. Our final threshold of $A_{K,\mathrm{Herschel}}$ is a trade-off between reliable filtering and area coverage. Also note here the preferred clustering of sources towards NGC 1980 and the ONC.}
        \label{img:kde_foreground}
\end{figure*}

\subsection{Orion~A population}
\label{sec:oriapop}

Orion~A shows a significant gradient in star formation activity along its spine coinciding with the number of published articles (cf. Fig. \ref{img:herschel_optical}) with about twice as many YSOs near the integral-shaped filament compared to the rest of the cloud \citep{2012AJ....144..192M}. To the west we find - among others - the ONC, the BN/KL region, and the OMC 2-3 complex, which are well-studied regions. Toward the east of the ONC, however, the cloud has drawn much less attention to itself. Here we attempt to statistically derive a complete census of the entire Orion~A population by means of the KLF. To rule out critical systematic errors, we compared our survey data not only to the CF, but also to source counts given in the Besan\c{c}on model of the Galaxy \citep{2003A&A...409..523R}. Our analysis was split into the following consecutive steps that will be discussed in more detail individually:

\begin{enumerate}
        \item Comparing the source counts given in the Besan\c{c}on model with our CF;
    \item Scaling the CF to fit the coverage of the Orion~A survey;
    \item Dereddening the Orion~A data;
    \item Calculating the excess population toward Orion~A with respect to the CF, as well as the  Besan\c{c}on model. This excess includes both the young populations associated with Orion~A, as well as the young foreground population unrelated to the galactic field;
    \item Estimating the total foreground population toward Orion~A (galactic field + young foreground);
    \item Estimating the young stellar population in the foreground to Orion~A unrelated to the galactic field;
    \item Estimating the total young stellar population of the Orion~A molecular cloud.
\end{enumerate}

\textbf{Ad 1.} The Besan\c{c}on model of the Milky Way plays a crucial role in estimating the young populations Orion~A and the foreground because we can approximate the number of expected galactic field stars up to the adopted distance of Orion~A of 414 pc \citep{2007A&A...474..515M}. For this reason we wanted to verify that the source counts measured in our CF and the ones taken from the model match. The top lefthand plot in Fig. \ref{img:klf} shows the results of our comparison. Looking at the Besan\c{c}on data we see an offset in source counts relative to the measurements in the CF. The CF histogram shows only about 85\% of the source counts in the model when considering the bins with $7 \leq K_S \leq 15$ mag for which we have a reasonable number of source counts and, at the same time, do not expect any problems from a morphological misclassification. For this reason we decided to scale all model data with the scaling factor derived from the difference between the observed CF KLF and the Besan\c{c}on model in our subsequent analysis. We split the data from the CF into extended and point-like morphology since the Besan\c{c}on model does not include background galaxies. 

Comparing the histograms we note a discrepancy in source counts between the CF KLF and the model KLF at the faint end ($K_S = 19$ mag) where mostly galaxies are situated. This is readily explained by our morphological classification, which tends to classify sources as point-like for very noisy measurements. The (scaled) background model also overpredicts sources at the bright end of the spectrum ($K_S \lesssim 10$ mag) relative to our measurements. This can be explained by the absence of very bright sources in our catalog since only sources with clean measurements (quality flag \textit{A}) were adopted from the 2MASS catalog. For the further analysis we note that these potentially missing bright sources do not critically influence our population estimates because there are so few of them. Besides this seemingly constant offset and the aforementioned excesses, we find good agreement with respect to the shape of the two histograms.

\textbf{Ad 2.} To make statistical comparisons between the CF and the Orion~A surveys, we needed to scale the source counts of the CF to match the survey coverage. We decided to scale based on field coverage where we find that the Orion~A survey covers an area 10.149 times larger than the CF. Other scaling methods (e.g., total source counts) were ruled out because of dissimilar stellar populations and the presence of Orion~A, which blocks many background sources. 

\textbf{Ad 3.} The Orion~A molecular cloud covers large parts of our survey and therefore many sources in the background will exhibit non-negligible NIR excess. Also, stars embedded within the cloud will naturally show redder colors (compare Fig. \ref{img:cmd_ccd_dens}). \cite{2015ApJ...799..116S} do not find significant amounts of dust up to 300 pc in this region, therefore all stars showing NIR color excess should lie in or behind the cloud. For statistical comparisons we therefore needed to estimate the NIR excess for each source to get intrinsic $K_S$-band magnitudes. To this end we used the NICER method \citep{2001A&A...377.1023L} which de-projects the measured colors using both measurement errors and the color distribution of sources in the CF to derive line-of-sight extinctions. This method, together with its extension, NICEST, described in \cite{2009A&A...493..735L}, has found many successful applications showing its robustness (see, e.g., \citealp{2006A&A...454..781L, 2014A&A...565A..18A} for examples and \citealp{2009ApJ...692...91G} for an independent comparison of different column density tracers). We note here that NICEST has no effect on  extinction measurements for point sources since the method only attempts to correct for cloud substructure in the subsequent construction of extinction maps. The reliable determination of color-excesses with NICER depends not only on the photometric quality, but also on the intrinsic color distribution of the sources as measured from the CF. Typically for stars in the NIR, this distribution is very narrow (compare Fig. \ref{img:cmd_ccd_dens}), however, when also including galaxies the estimate of the color-excesses can be significantly biased\footnote{NICER includes the intrinsic color distribution in the calculation of the errors.} \citep{2008ApJ...674..831F}. This, however, is no problem here since galaxies were not included in the determination of excess source counts in Orion~A, assured by a subsequent magnitude cut at $K_S = 13$ mag (see step 4). Also, the method itself allows negative values for color excesses. These are sources lying below the adopted color zero point and comprise hot stars with low foreground extinction. For this application of NICER, we used the extinction coefficients from \cite{2005ApJ...619..931I}.  A comparison of the raw Orion~A data and its dereddened KLF is shown in the top righthand plot of Fig. \ref{img:klf}.

\textbf{Ad 4.} We estimated the excess of young stars over the Galactic field by calculating the differences between (a) the dereddened Orion~A KLF and the CF KLF and (b) the Orion~A KLF only for point-like sources and the Besan\c{c}on KLF. For a distance modulus of $\mu = 8.1$ mag and an age of 1 Myr, we find the assumed hydrogen burning limit of 0.08 M$_{\odot}$ at $K_S \approx 13$ mag \citep{1998A&A...337..403B, 2002A&A...382..563B}. Both methods clearly show an excess that is also readily visible in the histograms in the two lower panels of Fig. \ref{img:klf}. The Orion~A KLF shows an excess of $\sim$2900 sources compared to the CF and $\sim$3200 sources compared to the scaled Besan\c{c}on KLF. To test for the statistical significance of these results, we determined the error of the excess assuming Poisson statistics for the histograms. In both cases the errors amount to $\sim$170 sources or approximately 5--6\%. Including the absolute difference between the two methods, we assume an error of 10\% on these estimates for the following analysis. The comparison with the Besan\c{c}on KLF also produces an excess of bright sources when compared to the scaled CF data. While the difference on the bright end between the VISION KLF and the CF KLF can be explained by low number statistics (smaller on-sky coverage of the CF), the VISION KLF overall is not complete for these bright sources since only clean unsaturated measurements from 2MASS were adopted (which leads to a lack of bright sources when comparing to the Besan\c{c}on model). This difference, however, only has a negligible effect on our number estimates. For the subsequent population estimates we adopt a KLF excess of 3000 sources up to $K_S = 13$ mag.

\begin{figure}[tp]
        \resizebox{\hsize}{!}{\includegraphics{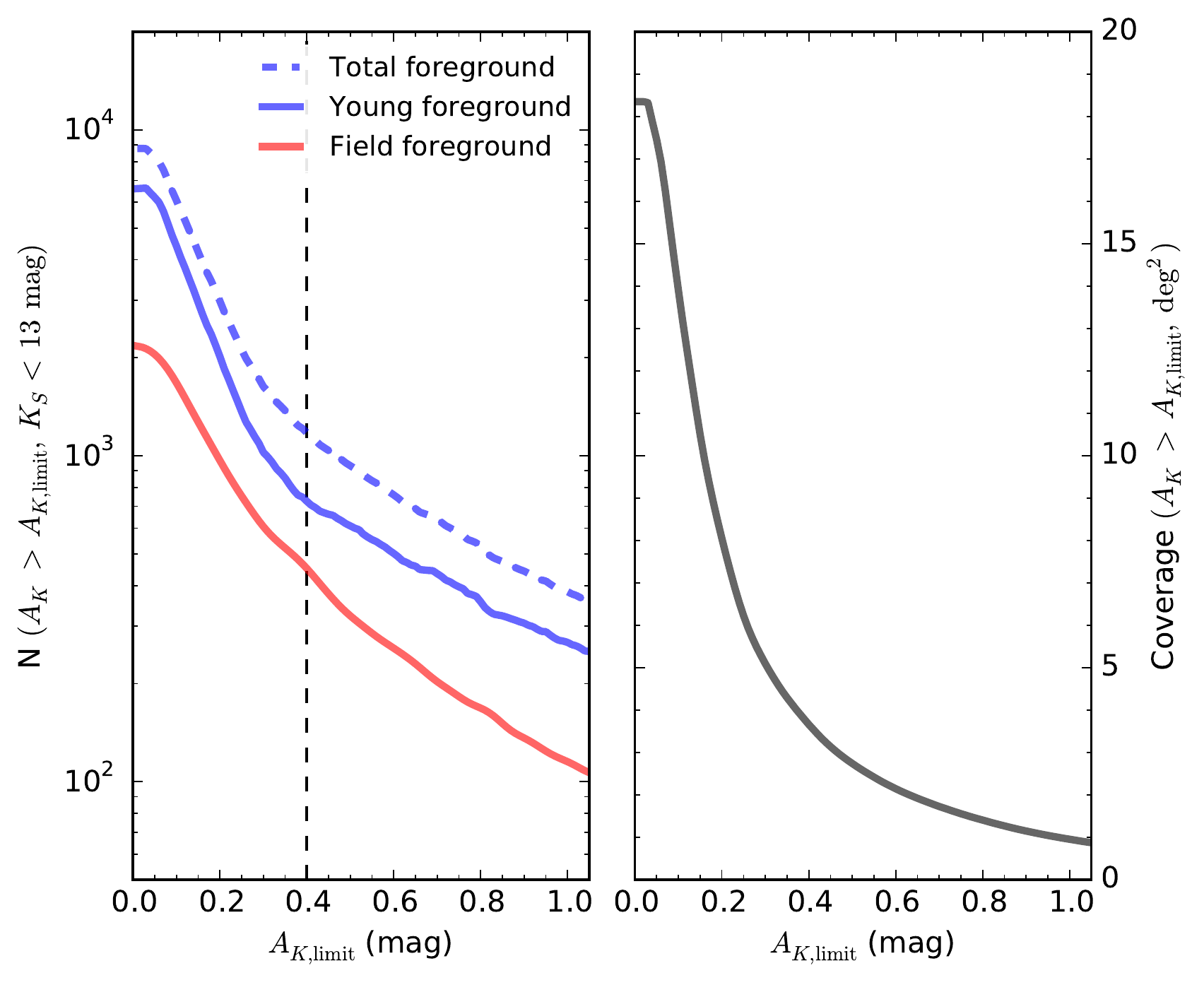}}
        \caption[]{Left: Number of statistically derived foreground stars as a function of \textit{Herschel} column density expressed in $A_K$. The total foreground was identified as stars seen in projection against areas above a given threshold while showing negligible extinction. The field foreground was extracted from the Besan\c{c}on model of the galaxy. The difference between those denotes the young foreground stars unrelated to the galactic field. The vertical dashed line indicates the optimal column density threshold for which background contamination is still negligible. Right: The given area on sky as a function of column density.}
        \label{img:ak_vs_nfore}
\end{figure}

\textbf{Ad 5.} The Orion~A KLF excess of 3000 sources includes not only the population of Orion~A, but also parts of the young populations mainly associated with NGC 1980, which is claimed to be a foreground population not emerging from the cloud \citep{2012A&A...547A..97A, 2013ApJ...768...99P, 2014A&A...564A..29B}. To estimate the number of foreground stars, we used the molecular cloud to effectively block background sources. In principle, sources seen in projection toward Orion~A that at the same time show negligible extinction as determined via NICER (i.e., $A_{K,\mathrm{NICER}} \leq 0$ mag), can only lie in the foreground of the cloud. In Fig. \ref{img:kde_foreground} we show zero-extinction source densities\footnote{The source densities were estimated with a symmetric 3 arcmin-wide Gaussian kernel} for different column density thresholds as given in the \textit{Herschel-Planck} map from \cite{2014A&A...566A..45L}. Here we note some key characteristics: 

\begin{enumerate}
        \item For low column densities, background sources contaminate the sample, whereas for large thresholds the background disappears, however at the cost of reduced coverage. 
    
        \item At about $A_{K, \mathrm{Herschel}} \ga$ 0.3 -- 0.4 mag, the background contamination starts to become negligible because we cannot see steep gradients toward the edges of the covered area. 
    
    \item We find significant substructure in the distribution of foreground sources, which are discussed in Sect. \ref{sec:foreground}.
\end{enumerate}

Surely, not all the entire foreground population is captured with this method. However, the bulk of the young foreground stars are thought to be associated with NGC 1980, which mostly falls on top of heavily extincted regions. For $A_{K, \mathrm{Herschel}} > 0.4$ mag, we find a total zero extinction, i.e., a foreground population (galactic field + young stars) of about 1200 sources with $K_S \leq 13$ mag. 

\textbf{Ad 6.} To split the estimate from step (5) into young foreground sources and galactic field we compared our findings to the Besan\c{c}on model of the Galaxy. For the area given by $A_{K, \mathrm{Herschel}} > 0.4$ mag we counted all stars up to the adopted distance of 414 pc. Here we find about 500 stars brighter than $K_S = 13$ mag. Given a total foreground population of 1200 sources (within our constraints), we therefore estimate the young foreground to comprise about 700 stars below the magnitude limit at $K_S = 13$ mag within our limited field. 

\textbf{Ad 7.} For a column density threshold of $A_{K, \mathrm{Herschel}} > 0.4$ mag and taking the 3000 KLF excess sources in the KLF together with the 700 sources associated with the young foreground, we estimated $\sim$2300 sources comprising the Orion~A stellar population. We consider this number a lower limit for two reasons: 

\begin{enumerate}
        \item Our survey is not complete with respect to protostars. Using \textit{Spitzer} observations covering large parts of Orion~A (and also Orion~B), \cite{2012AJ....144..192M} find a total of 2818 protostars and disk sources as defined in their paper in the region of Orion~A. Overall, we find 2751 out of these 2818 (98\%) sources to have a detection in at least one band when using a 2 arcsec cross-matching radius; here we are 99.6\% complete with respect to disks (10 out of 2446 missing) and 84\% complete regarding protostars (53 out of 329 missing). We checked the ten remaining undetected disks in the \textit{Spitzer} data, and about half of them turned out to be misdetections of nebulosity. As a result, we are essentially complete with respect to disks. Requiring a simultaneous detection in $J$, $H$, and $K_S$ (as for our dereddening step), the \cite{2012AJ....144..192M} YSO number count detected in our survey decreases by about 10\% to 2500 sources.
    \item The estimate of 700 young foreground sources with $K_S < 13$ mag can be biased by a population emerging from Orion~A. Such emerging stars would show negligible extinction and thus be included in our foreground sample. An indicator of the magnitude of this potential contamination, together with statistical errors, is derived in Sect. \ref{sec:foreground} and amounts to $\sim$10\%. We derive an additional indicator for contamination in our foreground sample by comparing the zero-extinction sample to the YSO catalog of \cite{2012AJ....144..192M}. We find that about 9\% of our total foreground sample are associated with \textit{Spitzer}--identified YSOs within our selection criteria, which matches the above-mentioned error estimate well.
\end{enumerate}

\noindent In light of these arguments and including the 10\% error indicator of the KLF excess determination, we find an upper limit for the young population toward Orion~A of about 3000 YSOs (excluding unresolved binaries).

The most critical problem with this deduction is the dependency of our population estimates on the column density threshold. In Fig. \ref{img:ak_vs_nfore} we plot the dependency of the foreground estimates as a function of the chosen threshold in the \textit{Herschel-Planck} column density map. Here, the total foreground refers to step (5), the young foreground and galactic field foreground to step (6). At about $A_{K} \sim 0.35$ mag, we see a sudden increase in the slope in the foreground estimates. At about this point, the background contamination becomes non-negligible, and estimates significantly below this level would be biased. This compares well to the appearance of the source density maps in Fig. \ref{img:kde_foreground}. Scaling our results to the size of the cloud is  difficult owing to the anisotropic distribution of young foreground stars \citep{2014A&A...564A..29B}. Linearly extrapolating to the cloud size would therefore not yield realistic results. We therefore settled for the number estimates obtained with a column density threshold of $A_{K, \mathrm{Herschel}} = 0.4$ mag. This threshold provides significant shielding from background contamination and at the same time encompasses most of NGC 1980 and therefore the bulk of the young foreground sources.

\begin{table}
        \caption{Identified groups of young foreground populations in this work. The list below the separator indicates potential groups and require further analysis. The names of these potential groups refer to the brightest star (optical) in the region.}
        \label{tab:foreground_groups}
    \begin{tabular}{l c c c}   
        \hline\hline
        Group name                                      & RA (J2000)    & Dec (J2000)     & Reference             \\
                                                        & (hh:mm)               & (dd:mm)         &                               \\
        \hline
    NGC 1980                                    & 05:35.4               & -05:54.9                & 1, 2, 3, 4, 5 \\
    OriA-Fore 1                                 & 05:35.3               & -05:10.3                & 1, 5                  \\
    BD-06 1235                                  & 05:35.07              & -06:17.12               & 1, 5                  \\
    \hline
    405-010152\tablefootmark{a} & 05:41.93              & -09:11.84             & 1                               \\
    BD-09 1215                                  & 05:43.9               & -09:01.99               & 1                             \\
    \hline
        \end{tabular}
    \tablebib{  (1) This work;
                        (2) \citet{2008hsf1.book..459B};
                        (3) \citet{2012A&A...547A..97A}; 
                (4) \citet{2013ApJ...768...99P};
                (5)     \citet{2014A&A...564A..29B}.
             }
  \tablefoot{
      \tablefoottext{a}{UCAC4 identifier.}
  }

\end{table}

\subsection{Foreground population}
\label{sec:foreground}

Figure \ref{img:kde_foreground} shows substantial substructure in the distribution of foreground sources. The peak associated with NGC 1980 confirms previous results from \cite{2012A&A...547A..97A} and \cite{2014A&A...564A..29B}. The latter study also finds an overdensity near OMC 2/3, called OriA-Fore 1, which is visible in our maps as well. We find one additional overdensity to the southeast of NGC 1980 near HH 322 and HH 323. Here we can also spot a few very bright and blue sources in the VISTA data, where BD-06 1235 is the brightest star in this region. Only a few arcminutes south, \cite{2014A&A...564A..29B} speculate on a possible foreground population in the vicinity of an overdensity in X-ray sources they called L1641W. The other prominent peak associated with the ONC and roughly centered on the Trapezium cluster does not have a straightforward interpretation. It appears in \cite{2012A&A...547A..97A}, which uses NIR photometry, but it is barely present in \cite{2014A&A...564A..29B}, which uses mostly optical data. Certainly the error associated with the density estimations in this region is by far the largest across the survey simply because of the extreme surface density of stars toward the Trapezium cluster. If the enhancement is not due to statistical errors alone, the next interpretation would be that the enhancement represents the Trapezium young population emerging from the dusty molecular cloud. We can estimate the global bias due to statistical errors and potentially emerging stars by comparing the source counts in our zero-extinction sample to the total number of sources in the region near the ONC. Here about 10\% of the total VISTA source catalog comprise the zero-extinction subsample. As a third alternative explanation for the ONC enhancement  one can consider the possibility of a real, yet undiscovered, foreground population. But in light of the previous scenarios and since no other study found any evidence of such a population, this third explanation seems unlikely. In addition to these significant overdensities in the western parts of Orion~A we can identify two new groups to the east when lowering the threshold to $A_{K,\mathrm{Herschel}} > 0.3$. This map might already show minor background contamination, so we refer to the identified overdensities as potential groups that require further analysis. Both regions are marked in the top right plot of Fig. \ref{img:kde_foreground}). The labels refer to the brightest stars in the vicinity of the peaks, which are BD-09 1215 and UCAC4 405-010152 \citep{2013AJ....145...44Z}. Since SDSS does not cover the eastern parts of Orion~A, newly acquired deep optical data supplemented by our NIR survey would greatly aid in the confirmation of these potential new foreground populations. 

By comparing the KLF excess with expected Galactic field stars in Sect. \ref{sec:oriapop}, we found that about 700 sources are related to the young foreground population. When limiting the $\sim$2100 foreground sources identified in \cite{2014A&A...564A..29B} to our constraints (field and magnitude limit), their sample decreases to $\sim$650, which agrees very well with our results.

\cite{2014A&A...564A..29B} propose a star formation scenario toward Orion~A where the foreground populations in this region formed 5 - 10 Myrs ago. Subsequent supernovae then could have triggered episodic star formation leading to the formation of the embedded clusters we see today. With our new results, we can extend this view. Both number estimates of the young foreground from previous studies \citep[$\sim$2100,][]{2014A&A...564A..29B} and the embedded population ($\sim$2300) are incomplete. The number of foreground stars suffers from incomplete data coverage, while our estimate of the embedded population misses some of the deeply embedded sources. The numbers, however, are similar, suggesting that a comparable star formation event took place a few million years ago. The formation of NGC 1980 with its massive stars might even have been similar to what we see in the Orion Nebula today. Furthermore, our data offers the first view of the foreground population toward all parts of Orion~A. That we only see a substantial amount of young foreground stars in the direction of the ONC and the integral-shaped filament, where the bulk of all star formation in Orion~A occurs today, suggests a causal connection. The star formation events that produced NGC 1980 and the other foreground groups are then not responsible for the formation of the molecular cloud as a whole, but indeed seem to have had an enhancing effect with respect to the formation of new stars on the western parts of the cloud, which is the only part where massive stars are forming in Orion~A.

\section{Summary}
\label{sec:summary}

The VISTA Orion~A survey provides the most detailed view of this massive star-forming region in the NIR yet. In this paper we presented survey strategy, data calibration, catalog generation, the main data products, and first results with this rich data set. Here, we summarize our main results.

\begin{enumerate}
        \item Our survey of the Orion~A molecular cloud in the NIR bands $J$, $H$, and $K_S$  covered in total $\sim$18.3 deg$^2$ on a pixel scale of 1/3 arcsec/pix.
        \item We implemented independent data reduction procedures that avoid some disadvantages in the standard CASU VIRCAM pipeline. Most notably we improved the resolution on average by about 20\% over the pipeline processed data. 
    \item The generated source catalog contains 799\,995 sources, a gain of almost an order of magnitude compared to 2MASS, translating into a gain of three to four magnitudes. The 90\% completeness levels (i.e., 90\% of the sources are detected with our source extraction) are $20.4, 19.9,$ and $19.0$ mag in $J, H$, and $K_S$, respectively. We also improved depth and coverage of all previously available ONC catalogs in the NIR. In contrast to the pipeline, our photometry is calibrated toward the 2MASS photometric system.
    The source catalog will be made available through the CDS.
    \item In addition to the source catalog, we also provided optimized three-color image data in HEALPix format, also available through the CDS in the future.
    \item Cross-matching with the YSO catalog from \cite{2012AJ....144..192M} reveals that we are essentially complete (99.6\%) with respect to disks and 84\% complete regarding protostars as classified in their paper. 
    \item From these data we identified several notable YSOs associated with characteristic nebulosity. Based on the morphology of this test set, we identified five new YSO candidates.
    \item Based on the surface density of extended sources in the catalog we identified ten new galaxy cluster candidates.
    \item We estimated the entire young stellar population in Orion~A by means of the $K_S$ band luminosity function. We find lower and upper limits of 2300 and 3000 sources, respectively, which compares well to results from earlier studies.
        \item Separated from the young population in Orion~A, we can confirm previous results regarding the young foreground population toward Orion~A. Here we find the same complex pattern of foreground groups mostly toward the integral-shaped filament, including the Orion nebula. Toward the eastern parts of the cloud we could identify two new potential small foreground groups.
        \item Given the asymmetric east-west projected distribution of foreground sources it is unlikely that this population played an important role in assembling the Orion~A cloud. Nevertheless, given the good correlation with the enhanced star formation activity in the integral-shaped filament, it is likely that the foreground population is responsible instead for compressing the western part of the cloud via feedback processes (winds, supernovas). 
\end{enumerate}

\begin{acknowledgements}
Stefan Meingast is a recipient of a DOC Fellowship of the Austrian Academy of Sciences at the Institute for Astrophysics, University of Vienna.
We gratefully acknowledge the referee, Tom Megeath, for carefully reading the manuscript and the useful comments that served to improve both the clarity and quality of this study.
This research made use of Montage, funded by the National Aeronautics and Space Administration's Earth Science Technology Office, Computation Technologies Project, under Cooperative Agreement Number NCC5-626 between NASA and the California Institute of Technology. Montage is maintained by the NASA/IPAC Infrared Science Archive;
The SIMBAD database, operated at the CDS, Strasbourg, France.
"Aladin sky atlas" developed at the CDS, Strasbourg Observatory, France. The VizieR catalogue access tool, CDS, Strasbourg, France.
Astropy is a community-developed core Python package for Astronomy \citep{2013A&A...558A..33A}.
H. Bouy is supported by the the Ram\'on y Cajal fellowship program number RYC-2009-04497 and by the Spanish Grant AYA2012-38897-C02-01.
\end{acknowledgements}

\bibliography{aanda}

\begin{thebibliography}{111}
\expandafter\ifx\csname natexlab\endcsname\relax\def\natexlab#1{#1}\fi

\bibitem[{{Abazajian} {et~al.}(2009){Abazajian}, {Adelman-McCarthy},
  {Ag{\"u}eros}, {Allam}, {Allende Prieto}, {An}, {Anderson}, {Anderson},
  {Annis}, {Bahcall}, \& et~al.}]{2009ApJS..182..543A}
{Abazajian}, K.~N., {Adelman-McCarthy}, J.~K., {Ag{\"u}eros}, M.~A., {et~al.}
  2009, \apjs, 182, 543

\bibitem[{{Alcal{\'a}} {et~al.}(2008){Alcal{\'a}}, {Covino}, \&
  {Leccia}}]{2008hsf1.book..801A}
{Alcal{\'a}}, J.~M., {Covino}, E., \& {Leccia}, S. 2008, {Orion Outlying
  Clouds}, ed. B.~{Reipurth}, 801

\bibitem[{{Ali} \& {Depoy}(1995)}]{1995AJ....109..709A}
{Ali}, B. \& {Depoy}, D.~L. 1995, \aj, 109, 709

\bibitem[{{Allen} \& {Davis}(2008)}]{2008hsf1.book..621A}
{Allen}, L.~E. \& {Davis}, C.~J. 2008, {Low Mass Star Formation in the Lynds
  1641 Molecular Cloud}, ed. B.~{Reipurth}, 621

\bibitem[{{Alves} \& {Bouy}(2012)}]{2012A&A...547A..97A}
{Alves}, J. \& {Bouy}, H. 2012, \aap, 547, A97

\bibitem[{{Alves} {et~al.}(2014){Alves}, {Lombardi}, \&
  {Lada}}]{2014A&A...565A..18A}
{Alves}, J., {Lombardi}, M., \& {Lada}, C.~J. 2014, \aap, 565, A18

\bibitem[{{Alves}(1998)}]{1998PhDT........24A}
{Alves}, J.~F. 1998, PhD thesis, European Southern Observatory

\bibitem[{{Andre} {et~al.}(2000){Andre}, {Ward-Thompson}, \&
  {Barsony}}]{2000prpl.conf...59A}
{Andre}, P., {Ward-Thompson}, D., \& {Barsony}, M. 2000, Protostars and Planets
  IV, 59

\bibitem[{{Astropy Collaboration} {et~al.}(2013){Astropy Collaboration},
  {Robitaille}, {Tollerud}, {Greenfield}, {Droettboom}, {Bray}, {Aldcroft},
  {Davis}, {Ginsburg}, {Price-Whelan}, {Kerzendorf}, {Conley}, {Crighton},
  {Barbary}, {Muna}, {Ferguson}, {Grollier}, {Parikh}, {Nair}, {Unther},
  {Deil}, {Woillez}, {Conseil}, {Kramer}, {Turner}, {Singer}, {Fox}, {Weaver},
  {Zabalza}, {Edwards}, {Azalee Bostroem}, {Burke}, {Casey}, {Crawford},
  {Dencheva}, {Ely}, {Jenness}, {Labrie}, {Lim}, {Pierfederici}, {Pontzen},
  {Ptak}, {Refsdal}, {Servillat}, \& {Streicher}}]{2013A&A...558A..33A}
{Astropy Collaboration}, {Robitaille}, T.~P., {Tollerud}, E.~J., {et~al.} 2013,
  \aap, 558, A33

\bibitem[{{Bally}(2008)}]{2008hsf1.book..459B}
{Bally}, J. 2008, {Overview of the Orion Complex}, ed. B.~{Reipurth}, 459

\bibitem[{{Baraffe} {et~al.}(1998){Baraffe}, {Chabrier}, {Allard}, \&
  {Hauschildt}}]{1998A&A...337..403B}
{Baraffe}, I., {Chabrier}, G., {Allard}, F., \& {Hauschildt}, P.~H. 1998, \aap,
  337, 403

\bibitem[{{Baraffe} {et~al.}(2002){Baraffe}, {Chabrier}, {Allard}, \&
  {Hauschildt}}]{2002A&A...382..563B}
{Baraffe}, I., {Chabrier}, G., {Allard}, F., \& {Hauschildt}, P.~H. 2002, \aap,
  382, 563

\bibitem[{{Bertin}(2006)}]{2006ASPC..351..112B}
{Bertin}, E. 2006, in Astronomical Society of the Pacific Conference Series,
  Vol. 351, Astronomical Data Analysis Software and Systems XV, ed.
  C.~{Gabriel}, C.~{Arviset}, D.~{Ponz}, \& S.~{Enrique}, 112

\bibitem[{{Bertin}(2009)}]{2009MmSAI..80..422B}
{Bertin}, E. 2009, \memsai, 80, 422

\bibitem[{{Bertin}(2010)}]{2010ascl.soft10068B}
{Bertin}, E. 2010, {SWarp: Resampling and Co-adding FITS Images Together},
  astrophysics Source Code Library

\bibitem[{{Bertin}(2011)}]{2011ASPC..442..435B}
{Bertin}, E. 2011, in Astronomical Society of the Pacific Conference Series,
  Vol. 442, Astronomical Data Analysis Software and Systems XX, ed. I.~N.
  {Evans}, A.~{Accomazzi}, D.~J. {Mink}, \& A.~H. {Rots}, 435

\bibitem[{{Bertin}(2012)}]{2012ASPC..461..263B}
{Bertin}, E. 2012, in Astronomical Society of the Pacific Conference Series,
  Vol. 461, Astronomical Data Analysis Software and Systems XXI, ed.
  P.~{Ballester}, D.~{Egret}, \& N.~P.~F. {Lorente}, 263

\bibitem[{{Bertin} \& {Arnouts}(1996)}]{1996A&AS..117..393B}
{Bertin}, E. \& {Arnouts}, S. 1996, \aaps, 117, 393

\bibitem[{{Bertin} {et~al.}(2002){Bertin}, {Mellier}, {Radovich}, {Missonnier},
  {Didelon}, \& {Morin}}]{2002ASPC..281..228B}
{Bertin}, E., {Mellier}, Y., {Radovich}, M., {et~al.} 2002, in Astronomical
  Society of the Pacific Conference Series, Vol. 281, Astronomical Data
  Analysis Software and Systems XI, ed. D.~A. {Bohlender}, D.~{Durand}, \&
  T.~H. {Handley}, 228

\bibitem[{{Bessell} \& {Brett}(1988)}]{1988PASP..100.1134B}
{Bessell}, M.~S. \& {Brett}, J.~M. 1988, \pasp, 100, 1134

\bibitem[{{Blaauw}(1964)}]{1964ARA&A...2..213B}
{Blaauw}, A. 1964, \araa, 2, 213

\bibitem[{{Boch} \& {Fernique}(2014)}]{2014ASPC..485..277B}
{Boch}, T. \& {Fernique}, P. 2014, in Astronomical Society of the Pacific
  Conference Series, Vol. 485, Astronomical Data Analysis Software and Systems
  XXIII, ed. N.~{Manset} \& P.~{Forshay}, 277

\bibitem[{{Bonifacio} {et~al.}(2000){Bonifacio}, {Monai}, \&
  {Beers}}]{2000AJ....120.2065B}
{Bonifacio}, P., {Monai}, S., \& {Beers}, T.~C. 2000, \aj, 120, 2065

\bibitem[{{Bonnarel} {et~al.}(2000){Bonnarel}, {Fernique}, {Bienaym{\'e}},
  {Egret}, {Genova}, {Louys}, {Ochsenbein}, {Wenger}, \&
  {Bartlett}}]{2000A&AS..143...33B}
{Bonnarel}, F., {Fernique}, P., {Bienaym{\'e}}, O., {et~al.} 2000, \aaps, 143,
  33

\bibitem[{{Bouy} {et~al.}(2014){Bouy}, {Alves}, {Bertin}, {Sarro}, \&
  {Barrado}}]{2014A&A...564A..29B}
{Bouy}, H., {Alves}, J., {Bertin}, E., {Sarro}, L.~M., \& {Barrado}, D. 2014,
  \aap, 564, A29

\bibitem[{{Briceno}(2008)}]{2008hsf1.book..838B}
{Briceno}, C. 2008, {The Dispersed Young Population in Orion}, ed.
  B.~{Reipurth}, 838

\bibitem[{{Calabretta} \& {Greisen}(2002)}]{2002A&A...395.1077C}
{Calabretta}, M.~R. \& {Greisen}, E.~W. 2002, \aap, 395, 1077

\bibitem[{{Carpenter}(2000)}]{2000AJ....120.3139C}
{Carpenter}, J.~M. 2000, \aj, 120, 3139

\bibitem[{{Carpenter} {et~al.}(2000){Carpenter}, {Heyer}, \&
  {Snell}}]{2000ApJS..130..381C}
{Carpenter}, J.~M., {Heyer}, M.~H., \& {Snell}, R.~L. 2000, \apjs, 130, 381

\bibitem[{{Carpenter} {et~al.}(1995){Carpenter}, {Snell}, \&
  {Schloerb}}]{1995ApJ...450..201C}
{Carpenter}, J.~M., {Snell}, R.~L., \& {Schloerb}, F.~P. 1995, \apj, 450, 201

\bibitem[{{Chen} \& {Tokunaga}(1994)}]{1994ApJS...90..149C}
{Chen}, H. \& {Tokunaga}, A.~T. 1994, \apjs, 90, 149

\bibitem[{{Da Rio} {et~al.}(2012){Da Rio}, {Robberto}, {Hillenbrand},
  {Henning}, \& {Stassun}}]{2012ApJ...748...14D}
{Da Rio}, N., {Robberto}, M., {Hillenbrand}, L.~A., {Henning}, T., \&
  {Stassun}, K.~G. 2012, \apj, 748, 14

\bibitem[{{Da Rio} {et~al.}(2015){Da Rio}, {Tan}, {Covey}, {Cottaar}, {Foster},
  {Cullen}, {Tobin}, {Kim}, {Meyer}, {Nidever}, {Stassun}, {Chojnowski},
  {Flaherty}, {Majewski}, {Skrutskie}, {Zasowski}, \&
  {Pan}}]{2015arXiv151104147D}
{Da Rio}, N., {Tan}, J.~C., {Covey}, K.~R., {et~al.} 2015, ArXiv e-prints
  [\eprint[arXiv]{1511.04147}]

\bibitem[{{Dalton} {et~al.}(2006){Dalton}, {Caldwell}, {Ward}, {Whalley},
  {Woodhouse}, {Edeson}, {Clark}, {Beard}, {Gallie}, {Todd}, {Strachan},
  {Bezawada}, {Sutherland}, \& {Emerson}}]{2006SPIE.6269E..0XD}
{Dalton}, G.~B., {Caldwell}, M., {Ward}, A.~K., {et~al.} 2006, in Society of
  Photo-Optical Instrumentation Engineers (SPIE) Conference Series, Vol. 6269,
  Society of Photo-Optical Instrumentation Engineers (SPIE) Conference Series

\bibitem[{{Davis} {et~al.}(2009){Davis}, {Froebrich}, {Stanke}, {Megeath},
  {Kumar}, {Adamson}, {Eisl{\"o}ffel}, {Gredel}, {Khanzadyan}, {Lucas},
  {Smith}, \& {Varricatt}}]{2009A&A...496..153D}
{Davis}, C.~J., {Froebrich}, D., {Stanke}, T., {et~al.} 2009, \aap, 496, 153

\bibitem[{{Duchon}(1979)}]{1979JApMe..18.1016D}
{Duchon}, C.~E. 1979, Journal of Applied Meteorology, 18, 1016

\bibitem[{{Emerson} {et~al.}(2006){Emerson}, {McPherson}, \&
  {Sutherland}}]{2006Msngr.126...41E}
{Emerson}, J., {McPherson}, A., \& {Sutherland}, W. 2006, The Messenger, 126,
  41

\bibitem[{{Evans} {et~al.}(2009){Evans}, {Dunham}, {J{\o}rgensen}, {Enoch},
  {Mer{\'{\i}}n}, {van Dishoeck}, {Alcal{\'a}}, {Myers}, {Stapelfeldt},
  {Huard}, {Allen}, {Harvey}, {van Kempen}, {Blake}, {Koerner}, {Mundy},
  {Padgett}, \& {Sargent}}]{2009ApJS..181..321E}
{Evans}, II, N.~J., {Dunham}, M.~M., {J{\o}rgensen}, J.~K., {et~al.} 2009,
  \apjs, 181, 321

\bibitem[{{Fischer} {et~al.}(2010){Fischer}, {Megeath}, {Ali}, {Tobin},
  {Osorio}, {Allen}, {Kryukova}, {Stanke}, {Stutz}, {Bergin}, {Calvet}, {di
  Francesco}, {Furlan}, {Hartmann}, {Henning}, {Krause}, {Manoj}, {Maret},
  {Muzerolle}, {Myers}, {Neufeld}, {Pontoppidan}, {Poteet}, {Watson}, \&
  {Wilson}}]{2010A&A...518L.122F}
{Fischer}, W.~J., {Megeath}, S.~T., {Ali}, B., {et~al.} 2010, \aap, 518, L122

\bibitem[{{Foster} {et~al.}(2008){Foster}, {Rom{\'a}n-Z{\'u}{\~n}iga},
  {Goodman}, {Lada}, \& {Alves}}]{2008ApJ...674..831F}
{Foster}, J.~B., {Rom{\'a}n-Z{\'u}{\~n}iga}, C.~G., {Goodman}, A.~A., {Lada},
  E.~A., \& {Alves}, J. 2008, \apj, 674, 831

\bibitem[{{G{\^a}lfalk} \& {Olofsson}(2008)}]{2008A&A...489.1409G}
{G{\^a}lfalk}, M. \& {Olofsson}, G. 2008, \aap, 489, 1409

\bibitem[{{Genzel} {et~al.}(1981){Genzel}, {Reid}, {Moran}, \&
  {Downes}}]{1981ApJ...244..884G}
{Genzel}, R., {Reid}, M.~J., {Moran}, J.~M., \& {Downes}, D. 1981, \apj, 244,
  884

\bibitem[{{Gomez} \& {Lada}(1998)}]{1998AJ....115.1524G}
{Gomez}, M. \& {Lada}, C.~J. 1998, \aj, 115, 1524

\bibitem[{{Gonzalez} {et~al.}(2011){Gonzalez}, {Rejkuba}, {Zoccali}, {Valenti},
  \& {Minniti}}]{2011A&A...534A...3G}
{Gonzalez}, O.~A., {Rejkuba}, M., {Zoccali}, M., {Valenti}, E., \& {Minniti},
  D. 2011, \aap, 534, A3

\bibitem[{{Goodman} {et~al.}(2009){Goodman}, {Pineda}, \&
  {Schnee}}]{2009ApJ...692...91G}
{Goodman}, A.~A., {Pineda}, J.~E., \& {Schnee}, S.~L. 2009, \apj, 692, 91

\bibitem[{{G{\'o}rski} {et~al.}(2005){G{\'o}rski}, {Hivon}, {Banday},
  {Wandelt}, {Hansen}, {Reinecke}, \& {Bartelmann}}]{2005ApJ...622..759G}
{G{\'o}rski}, K.~M., {Hivon}, E., {Banday}, A.~J., {et~al.} 2005, \apj, 622,
  759

\bibitem[{{Gruen} {et~al.}(2014){Gruen}, {Seitz}, \&
  {Bernstein}}]{2014PASP..126..158G}
{Gruen}, D., {Seitz}, S., \& {Bernstein}, G.~M. 2014, \pasp, 126, 158

\bibitem[{{Gutermuth} {et~al.}(2009){Gutermuth}, {Megeath}, {Myers}, {Allen},
  {Pipher}, \& {Fazio}}]{2009ApJS..184...18G}
{Gutermuth}, R.~A., {Megeath}, S.~T., {Myers}, P.~C., {et~al.} 2009, \apjs,
  184, 18

\bibitem[{{Gutermuth} {et~al.}(2011){Gutermuth}, {Pipher}, {Megeath}, {Myers},
  {Allen}, \& {Allen}}]{2011ApJ...739...84G}
{Gutermuth}, R.~A., {Pipher}, J.~L., {Megeath}, S.~T., {et~al.} 2011, \apj,
  739, 84

\bibitem[{{Herbig} \& {Jones}(1983)}]{1983AJ.....88.1040H}
{Herbig}, G.~H. \& {Jones}, B.~F. 1983, \aj, 88, 1040

\bibitem[{{Hillenbrand} \& {Hartmann}(1998)}]{1998ApJ...492..540H}
{Hillenbrand}, L.~A. \& {Hartmann}, L.~W. 1998, \apj, 492, 540

\bibitem[{{Hirota} {et~al.}(2007){Hirota}, {Bushimata}, {Choi}, {Honma},
  {Imai}, {Iwadate}, {Jike}, {Kameno}, {Kameya}, {Kamohara}, {Kan-Ya},
  {Kawaguchi}, {Kijima}, {Kim}, {Kobayashi}, {Kuji}, {Kurayama}, {Manabe},
  {Maruyama}, {Matsui}, {Matsumoto}, {Miyaji}, {Nagayama}, {Nakagawa},
  {Nakamura}, {Oh}, {Omodaka}, {Oyama}, {Sakai}, {Sasao}, {Sato}, {Sato},
  {Shibata}, {Shintani}, {Tamura}, {Tsushima}, \&
  {Yamashita}}]{2007PASJ...59..897H}
{Hirota}, T., {Bushimata}, T., {Choi}, Y.~K., {et~al.} 2007, \pasj, 59, 897

\bibitem[{{Hodapp}(1994)}]{1994ApJS...94..615H}
{Hodapp}, K.-W. 1994, \apjs, 94, 615

\bibitem[{{Indebetouw} {et~al.}(2005){Indebetouw}, {Mathis}, {Babler}, {Meade},
  {Watson}, {Whitney}, {Wolff}, {Wolfire}, {Cohen}, {Bania}, {Benjamin},
  {Clemens}, {Dickey}, {Jackson}, {Kobulnicky}, {Marston}, {Mercer},
  {Stauffer}, {Stolovy}, \& {Churchwell}}]{2005ApJ...619..931I}
{Indebetouw}, R., {Mathis}, J.~S., {Babler}, B.~L., {et~al.} 2005, \apj, 619,
  931

\bibitem[{{Irwin} {et~al.}(2004){Irwin}, {Lewis}, {Hodgkin}, {Bunclark},
  {Evans}, {McMahon}, {Emerson}, {Stewart}, \& {Beard}}]{2004SPIE.5493..411I}
{Irwin}, M.~J., {Lewis}, J., {Hodgkin}, S., {et~al.} 2004, in Society of
  Photo-Optical Instrumentation Engineers (SPIE) Conference Series, Vol. 5493,
  Optimizing Scientific Return for Astronomy through Information Technologies,
  ed. P.~J. {Quinn} \& A.~{Bridger}, 411--422

\bibitem[{{Janesick}(2001)}]{2001sccd.book.....J}
{Janesick}, J.~R. 2001, {Scientific charge-coupled devices}, Press Monographs
  (Society of Photo Optical)

\bibitem[{{Jeffries}(2007)}]{2007MNRAS.376.1109J}
{Jeffries}, R.~D. 2007, \mnras, 376, 1109

\bibitem[{{Lada}(1985)}]{1985ARA&A..23..267L}
{Lada}, C.~J. 1985, \araa, 23, 267

\bibitem[{{Lada}(1987)}]{1987IAUS..115....1L}
{Lada}, C.~J. 1987, in IAU Symposium, Vol. 115, Star Forming Regions, ed.
  M.~{Peimbert} \& J.~{Jugaku}, 1--17

\bibitem[{{Lada} {et~al.}(2000){Lada}, {Muench}, {Haisch}, {Lada}, {Alves},
  {Tollestrup}, \& {Willner}}]{2000AJ....120.3162L}
{Lada}, C.~J., {Muench}, A.~A., {Haisch}, Jr., K.~E., {et~al.} 2000, \aj, 120,
  3162

\bibitem[{{Lada} {et~al.}(2008){Lada}, {Muench}, {Rathborne}, {Alves}, \&
  {Lombardi}}]{2008ApJ...672..410L}
{Lada}, C.~J., {Muench}, A.~A., {Rathborne}, J., {Alves}, J.~F., \& {Lombardi},
  M. 2008, \apj, 672, 410

\bibitem[{{Lada}(1992)}]{1992ApJ...393L..25L}
{Lada}, E.~A. 1992, \apjl, 393, L25

\bibitem[{{Lada} {et~al.}(1991){Lada}, {Depoy}, {Evans}, \&
  {Gatley}}]{1991ApJ...371..171L}
{Lada}, E.~A., {Depoy}, D.~L., {Evans}, II, N.~J., \& {Gatley}, I. 1991, \apj,
  371, 171

\bibitem[{{Lada} {et~al.}(1997){Lada}, {Evans}, \&
  {Falgarone}}]{1997ApJ...488..286L}
{Lada}, E.~A., {Evans}, II, N.~J., \& {Falgarone}, E. 1997, \apj, 488, 286

\bibitem[{{Lang}(2014)}]{2014AJ....147..108L}
{Lang}, D. 2014, \aj, 147, 108

\bibitem[{{Lawrence} {et~al.}(2007){Lawrence}, {Warren}, {Almaini}, {Edge},
  {Hambly}, {Jameson}, {Lucas}, {Casali}, {Adamson}, {Dye}, {Emerson},
  {Foucaud}, {Hewett}, {Hirst}, {Hodgkin}, {Irwin}, {Lodieu}, {McMahon},
  {Simpson}, {Smail}, {Mortlock}, \& {Folger}}]{2007MNRAS.379.1599L}
{Lawrence}, A., {Warren}, S.~J., {Almaini}, O., {et~al.} 2007, \mnras, 379,
  1599

\bibitem[{{Lewis} {et~al.}(2010){Lewis}, {Irwin}, \&
  {Bunclark}}]{2010ASPC..434...91L}
{Lewis}, J.~R., {Irwin}, M., \& {Bunclark}, P. 2010, in Astronomical Society of
  the Pacific Conference Series, Vol. 434, Astronomical Data Analysis Software
  and Systems XIX, ed. Y.~{Mizumoto}, K.-I. {Morita}, \& M.~{Ohishi}, 91

\bibitem[{{Lombardi}(2009)}]{2009A&A...493..735L}
{Lombardi}, M. 2009, \aap, 493, 735

\bibitem[{{Lombardi} \& {Alves}(2001)}]{2001A&A...377.1023L}
{Lombardi}, M. \& {Alves}, J. 2001, \aap, 377, 1023

\bibitem[{{Lombardi} {et~al.}(2006){Lombardi}, {Alves}, \&
  {Lada}}]{2006A&A...454..781L}
{Lombardi}, M., {Alves}, J., \& {Lada}, C.~J. 2006, \aap, 454, 781

\bibitem[{{Lombardi} {et~al.}(2011){Lombardi}, {Alves}, \&
  {Lada}}]{2011A&A...535A..16L}
{Lombardi}, M., {Alves}, J., \& {Lada}, C.~J. 2011, \aap, 535, A16

\bibitem[{{Lombardi} {et~al.}(2014){Lombardi}, {Bouy}, {Alves}, \&
  {Lada}}]{2014A&A...566A..45L}
{Lombardi}, M., {Bouy}, H., {Alves}, J., \& {Lada}, C.~J. 2014, \aap, 566, A45

\bibitem[{{Markwardt}(2009)}]{2009ASPC..411..251M}
{Markwardt}, C.~B. 2009, in Astronomical Society of the Pacific Conference
  Series, Vol. 411, Astronomical Data Analysis Software and Systems XVIII, ed.
  D.~A. {Bohlender}, D.~{Durand}, \& P.~{Dowler}, 251

\bibitem[{{Megeath} {et~al.}(2012){Megeath}, {Gutermuth}, {Muzerolle},
  {Kryukova}, {Flaherty}, {Hora}, {Allen}, {Hartmann}, {Myers}, {Pipher},
  {Stauffer}, {Young}, \& {Fazio}}]{2012AJ....144..192M}
{Megeath}, S.~T., {Gutermuth}, R., {Muzerolle}, J., {et~al.} 2012, \aj, 144,
  192

\bibitem[{{Megeath} \& {Wilson}(1997)}]{1997AJ....114.1106M}
{Megeath}, S.~T. \& {Wilson}, T.~L. 1997, \aj, 114, 1106

\bibitem[{{Menten} {et~al.}(2007){Menten}, {Reid}, {Forbrich}, \&
  {Brunthaler}}]{2007A&A...474..515M}
{Menten}, K.~M., {Reid}, M.~J., {Forbrich}, J., \& {Brunthaler}, A. 2007, \aap,
  474, 515

\bibitem[{{Minniti} {et~al.}(2010){Minniti}, {Lucas}, {Emerson}, {Saito},
  {Hempel}, {Pietrukowicz}, {Ahumada}, {Alonso}, {Alonso-Garcia}, {Arias},
  {Bandyopadhyay}, {Barb{\'a}}, {Barbuy}, {Bedin}, {Bica}, {Borissova},
  {Bronfman}, {Carraro}, {Catelan}, {Clari{\'a}}, {Cross}, {de Grijs},
  {D{\'e}k{\'a}ny}, {Drew}, {Fari{\~n}a}, {Feinstein}, {Fern{\'a}ndez
  Laj{\'u}s}, {Gamen}, {Geisler}, {Gieren}, {Goldman}, {Gonzalez}, {Gunthardt},
  {Gurovich}, {Hambly}, {Irwin}, {Ivanov}, {Jord{\'a}n}, {Kerins}, {Kinemuchi},
  {Kurtev}, {L{\'o}pez-Corredoira}, {Maccarone}, {Masetti}, {Merlo},
  {Messineo}, {Mirabel}, {Monaco}, {Morelli}, {Padilla}, {Palma}, {Parisi},
  {Pignata}, {Rejkuba}, {Roman-Lopes}, {Sale}, {Schreiber}, {Schr{\"o}der},
  {Smith}, {}, {Soto}, {Tamura}, {Tappert}, {Thompson}, {Toledo}, {Zoccali}, \&
  {Pietrzynski}}]{2010NewA...15..433M}
{Minniti}, D., {Lucas}, P.~W., {Emerson}, J.~P., {et~al.} 2010, \na, 15, 433

\bibitem[{{Muench} {et~al.}(2008){Muench}, {Getman}, {Hillenbrand}, \&
  {Preibisch}}]{2008hsf1.book..483M}
{Muench}, A., {Getman}, K., {Hillenbrand}, L., \& {Preibisch}, T. 2008, {Star
  Formation in the Orion Nebula I: Stellar Content}, ed. B.~{Reipurth}, 483

\bibitem[{{Muench} {et~al.}(2002){Muench}, {Lada}, \&
  {Lada}}]{2002AAS...201.6002M}
{Muench}, A.~A., {Lada}, E.~A., \& {Lada}, C.~J. 2002, in Bulletin of the
  American Astronomical Society, Vol.~34, American Astronomical Society Meeting
  Abstracts, 1210

\bibitem[{{Nakamura} {et~al.}(2012){Nakamura}, {Miura}, {Kitamura},
  {Shimajiri}, {Kawabe}, {Akashi}, {Ikeda}, {Tsukagoshi}, {Momose}, {Nishi}, \&
  {Li}}]{2012ApJ...746...25N}
{Nakamura}, F., {Miura}, T., {Kitamura}, Y., {et~al.} 2012, \apj, 746, 25

\bibitem[{{Neugebauer} {et~al.}(1984){Neugebauer}, {Habing}, {van Duinen},
  {Aumann}, {Baud}, {Beichman}, {Beintema}, {Boggess}, {Clegg}, {de Jong},
  {Emerson}, {Gautier}, {Gillett}, {Harris}, {Hauser}, {Houck}, {Jennings},
  {Low}, {Marsden}, {Miley}, {Olnon}, {Pottasch}, {Raimond}, {Rowan-Robinson},
  {Soifer}, {Walker}, {Wesselius}, \& {Young}}]{1984ApJ...278L...1N}
{Neugebauer}, G., {Habing}, H.~J., {van Duinen}, R., {et~al.} 1984, \apjl, 278,
  L1

\bibitem[{{O'Dell} {et~al.}(2008){O'Dell}, {Muench}, {Smith}, \&
  {Zapata}}]{2008hsf1.book..544O}
{O'Dell}, C.~R., {Muench}, A., {Smith}, N., \& {Zapata}, L. 2008, {Star
  Formation in the Orion Nebula II: Gas, Dust, Proplyds and Outflows}, ed.
  B.~{Reipurth}, 544

\bibitem[{Pedregosa {et~al.}(2011)Pedregosa, Varoquaux, Gramfort, Michel,
  Thirion, Grisel, Blondel, Prettenhofer, Weiss, Dubourg, Vanderplas, Passos,
  Cournapeau, Brucher, Perrot, \& Duchesnay}]{scikit-learn}
Pedregosa, F., Varoquaux, G., Gramfort, A., {et~al.} 2011, Journal of Machine
  Learning Research, 12, 2825

\bibitem[{{Pence} {et~al.}(2010){Pence}, {Chiappetti}, {Page}, {Shaw}, \&
  {Stobie}}]{2010A&A...524A..42P}
{Pence}, W.~D., {Chiappetti}, L., {Page}, C.~G., {Shaw}, R.~A., \& {Stobie}, E.
  2010, \aap, 524, A42

\bibitem[{{Peterson} \& {Megeath}(2008)}]{2008hsf1.book..590P}
{Peterson}, D.~E. \& {Megeath}, S.~T. 2008, {The Orion Molecular Cloud 2/3 and
  NGC 1977 Regions}, ed. B.~{Reipurth}, 590

\bibitem[{{Phelps} \& {Lada}(1997)}]{1997ApJ...477..176P}
{Phelps}, R.~L. \& {Lada}, E.~A. 1997, \apj, 477, 176

\bibitem[{{Pilbratt} {et~al.}(2010){Pilbratt}, {Riedinger}, {Passvogel},
  {Crone}, {Doyle}, {Gageur}, {Heras}, {Jewell}, {Metcalfe}, {Ott}, \&
  {Schmidt}}]{2010A&A...518L...1P}
{Pilbratt}, G.~L., {Riedinger}, J.~R., {Passvogel}, T., {et~al.} 2010, \aap,
  518, L1

\bibitem[{{Pillitteri} {et~al.}(2013){Pillitteri}, {Wolk}, { eath}, {Allen},
  {Bally}, {Gagn{\'e}}, {Gutermuth}, {Hartman}, {Micela}, {Myers}, {Oliveira},
  {Sciortino}, {Walter}, {Rebull}, \& {Stauffer}}]{2013ApJ...768...99P}
{Pillitteri}, I., {Wolk}, S.~J., { eath}, S.~T., {et~al.} 2013, \apj, 768, 99

\bibitem[{{Popowicz} {et~al.}(2013){Popowicz}, {Kurek}, \&
  {Filus}}]{2013PASP..125.1119P}
{Popowicz}, A., {Kurek}, A.~R., \& {Filus}, Z. 2013, \pasp, 125, 1119

\bibitem[{{Reipurth} \& {Heathcote}(1997)}]{1997IAUS..182....3R}
{Reipurth}, B. \& {Heathcote}, S. 1997, in IAU Symposium, Vol. 182, Herbig-Haro
  Flows and the Birth of Stars, ed. B.~{Reipurth} \& C.~{Bertout}, 3--18

\bibitem[{{Reipurth} {et~al.}(2002){Reipurth}, {Heathcote}, {Morse},
  {Hartigan}, \& {Bally}}]{2002AJ....123..362R}
{Reipurth}, B., {Heathcote}, S., {Morse}, J., {Hartigan}, P., \& {Bally}, J.
  2002, \aj, 123, 362

\bibitem[{{Robberto} {et~al.}(2010){Robberto}, {Soderblom}, {Scandariato},
  {Smith}, {Da Rio}, {Pagano}, \& {Spezzi}}]{2010AJ....139..950R}
{Robberto}, M., {Soderblom}, D.~R., {Scandariato}, G., {et~al.} 2010, \aj, 139,
  950

\bibitem[{{Robin} {et~al.}(2003){Robin}, {Reyl{\'e}}, {Derri{\`e}re}, \&
  {Picaud}}]{2003A&A...409..523R}
{Robin}, A.~C., {Reyl{\'e}}, C., {Derri{\`e}re}, S., \& {Picaud}, S. 2003,
  \aap, 409, 523

\bibitem[{{Rom{\'a}n-Z{\'u}{\~n}iga} {et~al.}(2008){Rom{\'a}n-Z{\'u}{\~n}iga},
  {Elston}, {Ferreira}, \& {Lada}}]{2008ApJ...672..861R}
{Rom{\'a}n-Z{\'u}{\~n}iga}, C.~G., {Elston}, R., {Ferreira}, B., \& {Lada},
  E.~A. 2008, \apj, 672, 861

\bibitem[{{Sandstrom} {et~al.}(2007){Sandstrom}, {Peek}, {Bower}, {Bolatto}, \&
  {Plambeck}}]{2007ApJ...667.1161S}
{Sandstrom}, K.~M., {Peek}, J.~E.~G., {Bower}, G.~C., {Bolatto}, A.~D., \&
  {Plambeck}, R.~L. 2007, \apj, 667, 1161

\bibitem[{{Schlafly} {et~al.}(2014){Schlafly}, {Green}, {Finkbeiner}, {Rix},
  {Bell}, {Burgett}, {Chambers}, {Draper}, {Hodapp}, {Kaiser}, {Magnier},
  {Martin}, {Metcalfe}, {Price}, \& {Tonry}}]{2014ApJ...786...29S}
{Schlafly}, E.~F., {Green}, G., {Finkbeiner}, D.~P., {et~al.} 2014, \apj, 786,
  29

\bibitem[{{Schlafly} {et~al.}(2015){Schlafly}, {Green}, {Finkbeiner}, {Rix},
  {Burgett}, {Chambers}, {Draper}, {Kaiser}, {Martin}, {Metcalfe}, {Morgan},
  {Price}, {Tonry}, {Wainscoat}, \& {Waters}}]{2015ApJ...799..116S}
{Schlafly}, E.~F., {Green}, G., {Finkbeiner}, D.~P., {et~al.} 2015, \apj, 799,
  116

\bibitem[{{Schlegel} {et~al.}(1998){Schlegel}, {Finkbeiner}, \&
  {Davis}}]{1998ApJ...500..525S}
{Schlegel}, D.~J., {Finkbeiner}, D.~P., \& {Davis}, M. 1998, \apj, 500, 525

\bibitem[{{Skrutskie} {et~al.}(2006){Skrutskie}, {Cutri}, {Stiening},
  {Weinberg}, {Schneider}, {Carpenter}, {Beichman}, {Capps}, {Chester},
  {Elias}, {Huchra}, {Liebert}, {Lonsdale}, {Monet}, {Price}, {Seitzer},
  {Jarrett}, {Kirkpatrick}, {Gizis}, {Howard}, {Evans}, {Fowler}, {Fullmer},
  {Hurt}, {Light}, {Kopan}, {Marsh}, {McCallon}, {Tam}, {Van Dyk}, \&
  {Wheelock}}]{2006AJ....131.1163S}
{Skrutskie}, M.~F., {Cutri}, R.~M., {Stiening}, R., {et~al.} 2006, \aj, 131,
  1163

\bibitem[{{Soto} {et~al.}(2013){Soto}, {Barb{\'a}}, {Gunthardt}, {Minniti},
  {Lucas}, {Majaess}, {Irwin}, {Emerson}, {Gonzalez-Solares}, {Hempel},
  {Saito}, {Gurovich}, {Roman-Lopes}, {Moni-Bidin}, {Santucho}, {Borissova},
  {Kurtev}, {Toledo}, {Geisler}, {Dominguez}, \&
  {Beamin}}]{2013A&A...552A.101S}
{Soto}, M., {Barb{\'a}}, R., {Gunthardt}, G., {et~al.} 2013, \aap, 552, A101

\bibitem[{{Spezzi} {et~al.}(2015){Spezzi}, {Petr-Gotzens}, {Alcal{\'a}},
  {J{\o}rgensen}, {Stanke}, {Lombardi}, \& {Alves}}]{2015arXiv150504631S}
{Spezzi}, L., {Petr-Gotzens}, M.~G., {Alcal{\'a}}, J.~M., {et~al.} 2015, ArXiv
  e-prints [\eprint[arXiv]{1505.04631}]

\bibitem[{{Strom} \& {Strom}(1993)}]{1993ApJ...412L..63S}
{Strom}, K.~M. \& {Strom}, S.~E. 1993, \apjl, 412, L63

\bibitem[{{Strom} {et~al.}(1993){Strom}, {Strom}, \&
  {Merrill}}]{1993ApJ...412..233S}
{Strom}, K.~M., {Strom}, S.~E., \& {Merrill}, K.~M. 1993, \apj, 412, 233

\bibitem[{{Stutz} \& {Kainulainen}(2015)}]{2015A&A...577L...6S}
{Stutz}, A.~M. \& {Kainulainen}, J. 2015, \aap, 577, L6

\bibitem[{{Teixeira} {et~al.}(2006){Teixeira}, {Lada}, {Young}, {Marengo},
  {Muench}, {Muzerolle}, {Siegler}, {Rieke}, {Hartmann}, {Megeath}, \&
  {Fazio}}]{2006ApJ...636L..45T}
{Teixeira}, P.~S., {Lada}, C.~J., {Young}, E.~T., {et~al.} 2006, \apjl, 636,
  L45

\bibitem[{{Tody}(1986)}]{1986SPIE..627..733T}
{Tody}, D. 1986, in Society of Photo-Optical Instrumentation Engineers (SPIE)
  Conference Series, Vol. 627, Instrumentation in astronomy VI, ed. D.~L.
  {Crawford}, 733

\bibitem[{{Trumpler}(1931)}]{1931PASP...43..255T}
{Trumpler}, R.~J. 1931, \pasp, 43, 255

\bibitem[{{Wenger} {et~al.}(2000){Wenger}, {Ochsenbein}, {Egret}, {Dubois},
  {Bonnarel}, {Borde}, {Genova}, {Jasniewicz}, {Lalo{\"e}}, {Lesteven}, \&
  {Monier}}]{2000A&AS..143....9W}
{Wenger}, M., {Ochsenbein}, F., {Egret}, D., {et~al.} 2000, \aaps, 143, 9

\bibitem[{{Werner} {et~al.}(2004){Werner}, {Roellig}, {Low}, {Rieke}, {Rieke},
  {Hoffmann}, {Young}, {Houck}, {Brandl}, {Fazio}, {Hora}, {Gehrz}, {Helou},
  {Soifer}, {Stauffer}, {Keene}, {Eisenhardt}, {Gallagher}, {Gautier}, {Irace},
  {Lawrence}, {Simmons}, {Van Cleve}, {Jura}, {Wright}, \&
  {Cruikshank}}]{2004ApJS..154....1W}
{Werner}, M.~W., {Roellig}, T.~L., {Low}, F.~J., {et~al.} 2004, \apjs, 154, 1

\bibitem[{{Wright} {et~al.}(2010){Wright}, {Eisenhardt}, {Mainzer}, {Ressler},
  {Cutri}, {Jarrett}, {Kirkpatrick}, {Padgett}, {McMillan}, {Skrutskie},
  {Stanford}, {Cohen}, {Walker}, {Mather}, {Leisawitz}, {Gautier}, {McLean},
  {Benford}, {Lonsdale}, {Blain}, {Mendez}, {Irace}, {Duval}, {Liu}, {Royer},
  {Heinrichsen}, {Howard}, {Shannon}, {Kendall}, {Walsh}, {Larsen}, {Cardon},
  {Schick}, {Schwalm}, {Abid}, {Fabinsky}, {Naes}, \&
  {Tsai}}]{2010AJ....140.1868W}
{Wright}, E.~L., {Eisenhardt}, P.~R.~M., {Mainzer}, A.~K., {et~al.} 2010, \aj,
  140, 1868

\bibitem[{{Zacharias} {et~al.}(2013){Zacharias}, {Finch}, {Girard}, {Henden},
  {Bartlett}, {Monet}, \& {Zacharias}}]{2013AJ....145...44Z}
{Zacharias}, N., {Finch}, C.~T., {Girard}, T.~M., {et~al.} 2013, \aj, 145, 44

\end{thebibliography}

\clearpage
\begin{appendix}

\section{Data characteristics}
\label{app:data_characteristics}

Here, we want to provide additional information on the VISTA Orion~A source catalog, in particular on the photometric and astrometric quality of the data reduction, as well as present some quality control parameters.

\begin{figure}[tp]
        \resizebox{\hsize}{!}{\includegraphics{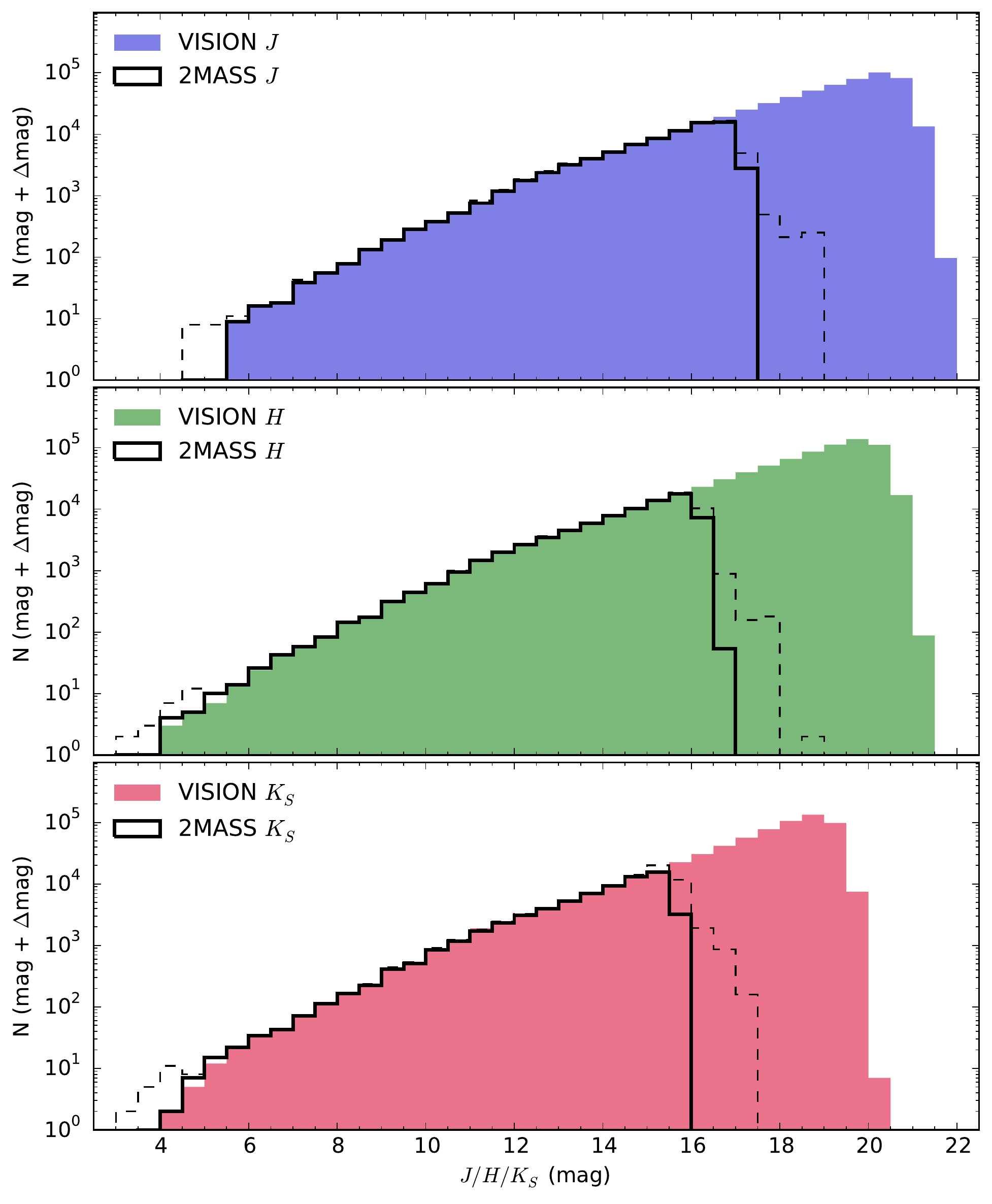}}
        \caption[]{Luminosity functions (histograms) for the three observed bands with bin widths of 0.5 mag. The solid black lines represent the 2MASS histograms for sources with a quality flag of at least C for the same coverage as the VISTA survey. The dashed lines show the complete 2MASS histograms.}
        \label{img:lumfunc}
\end{figure}

\begin{figure}[tp]
        \resizebox{\hsize}{!}{\includegraphics{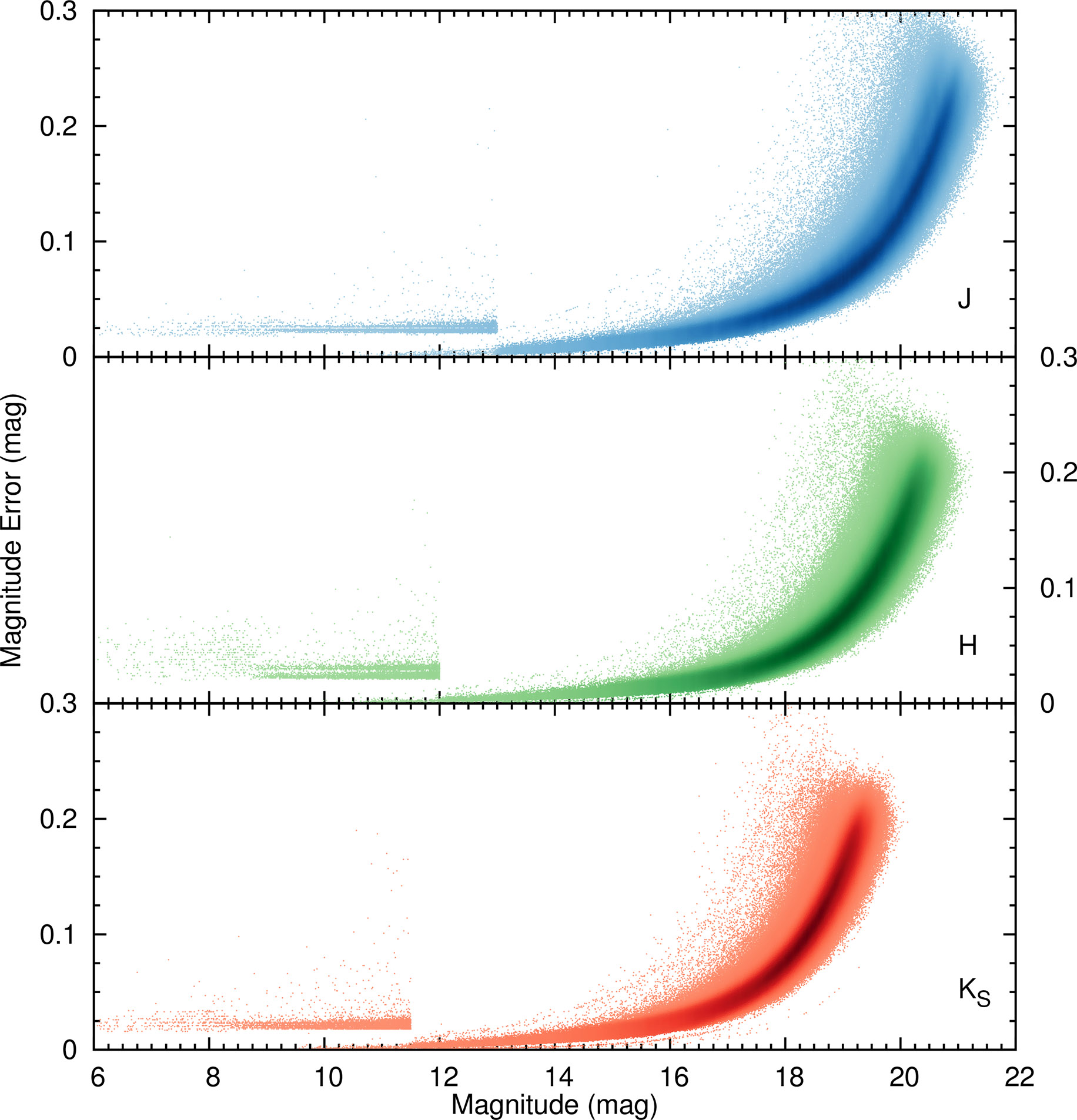}}
        \caption[]{Magnitudes and their associated errors in our survey. The discontinuity at the bright end is due to the replacement with 2MASS measurements. The shading indicates source density in a $0.1 \times 0.02$ mag box in this parameter space.}
        \label{img:magerr}
\end{figure}

\begin{figure*}[tp]
        \centering
        \resizebox{\hsize}{!}{\includegraphics[]{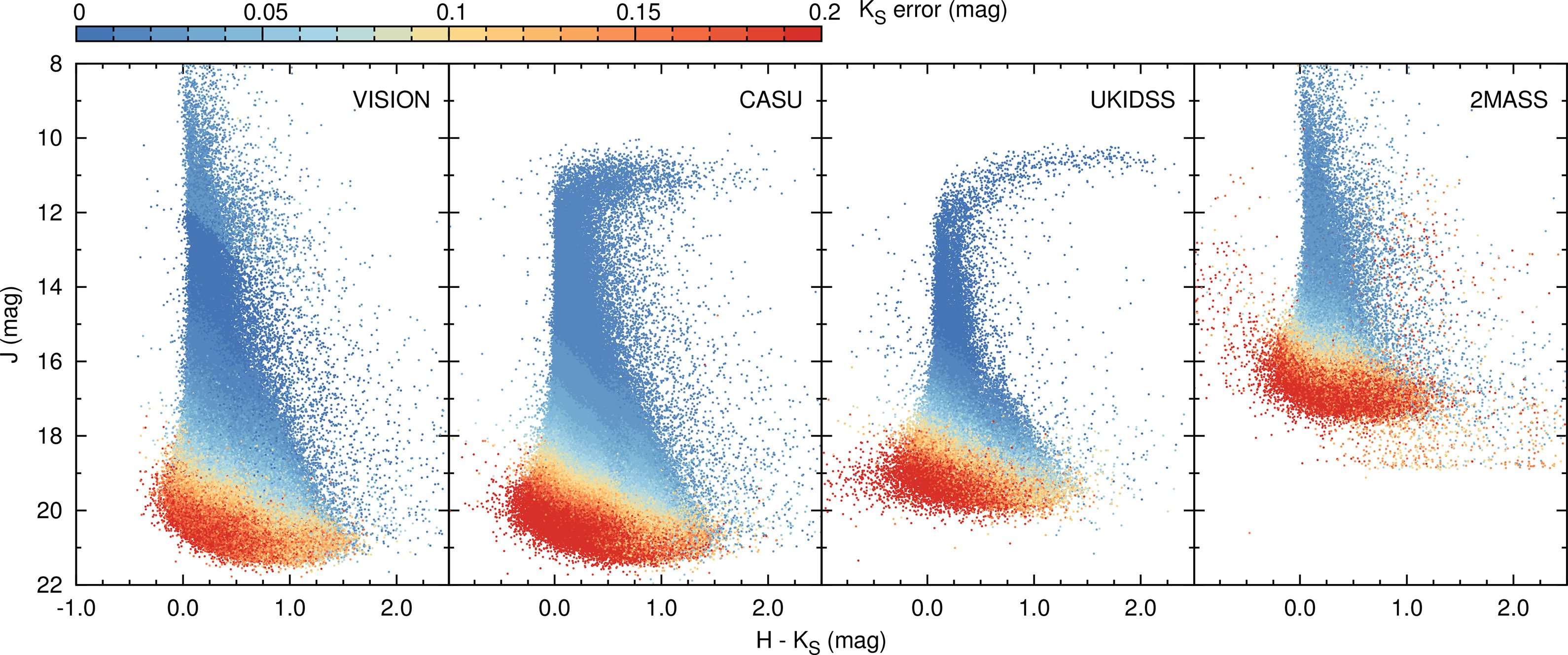}}
        \caption[]{$J$ vs. $H - K_S$ color-magnitude diagrams generated for four different data sets. Our data features the largest consistent dynamic range, as well as the lowest photometric error overall. Also, due to our optimized aperture photometry, the stellar sequence is slightly narrower at the faintest end. Note here, that UKIDSS and 2MASS data come from unoptimized surveys in this region and naturally do not reach equal depths.}
        \label{img:catalogs}
\end{figure*}

\begin{figure*}[tp]
        \centering
        \resizebox{\hsize}{!}{\includegraphics[]{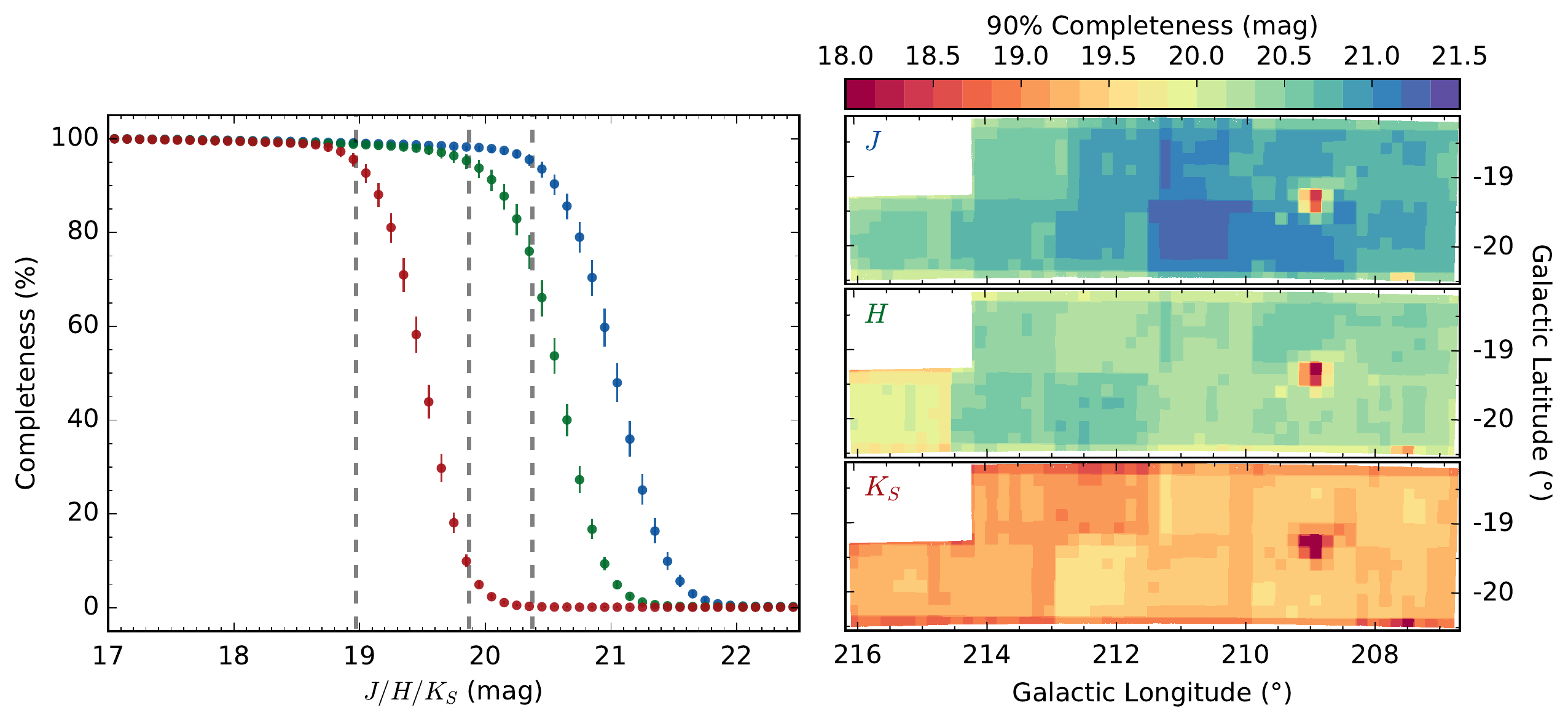}}
        \caption[]{Completeness estimates for the entire survey in all three observed bands. Left panel: The completeness estimate as a function of source magnitude. The blue, green, and red points represent the $J$, $H$, and $K_S$ bands respectively. The errorbars correspond to a 99\% bootstrap confidence interval ($\alpha = 0.01$) estimated with $10^5$ samples. The vertical gray lines correspond to 90\% completeness. Right panel: Spatial dependency of the completeness at the 90\% level. Clearly, the bright nebula near the ONC has a negative impact on source extraction. Well visible are the effects of observing conditions on the completeness. Clearly, better seeing (compare e.g. $J$ band with Fig. \ref{img:qc_quality}) results in more sensitive observations.}
        \label{img:completeness}
\end{figure*}

\begin{table}
        \caption{Tabulated completeness estimates from our artificial star tests for the entire Orion~A survey. The large gap between 99\% and 95\% is a consequence of the variable observing conditions.}
        \label{tab:completeness}
        \begin{tabular}{cccc}
        \hline\hline
        Completeness    & $J$   & $H$   & $K_S$ \\
        (\%)                    & (mag) & (mag) & (mag) \\
        \hline                                                                                                  
        99                              & 19.06 & 18.74 & 18.5  \\
        95                              & 20.38 & 19.87 & 18.97 \\
        90                              & 20.56 & 20.09 & 19.11 \\
        80                              & 20.74 & 20.29 & 19.26 \\
        50                              & 21.04 & 20.58 & 19.51 \\
        10                              & 21.45 & 20.94 & 19.85 \\
        \hline    
        \end{tabular}
\end{table}

\subsection{Photometric properties}
\label{app:phot}

The luminosity functions for all three bands are displayed in Fig. \ref{img:lumfunc}. Here we show both the complete 2MASS point source catalog histograms, as well as only sources with a quality flag of either A, B, or C since only these were added into the VISTA Orion~A catalog. At the bright end we closely match with 2MASS since essentially all bright sources originate from the reference catalog. The minor discrepancy between the VISTA histogram and the cleaned 2MASS histogram at the bright end mostly comes from the region around the ONC which we cleaned from bad detections by hand. Depending on the band, we gain between three and four magnitudes in dynamic range over the reference catalog. 

Magnitudes and their errors are shown in Fig. \ref{img:magerr}. The magnitude errors only start to increase significantly around 18 mag in all bands. One can clearly see the discontinuity at the bright end due to the catalog extension with 2MASS. There still are pure VISION sources below the cut-offs since some sources lie above the cleaning threshold in 2MASS (see Sect. \ref{sec:cat_clean_2mass} for details), but have a brighter magnitude in our survey. At a given magnitude, the error distribution does not follow a gaussian, but naturally has a longer tail towards larger errors. This effect is introduced by unequal coverage and if we select those sources with errors larger than the median error, we find that these indeed fall into regions with low effective exposure time. We again note here that these errors serve as lower limits only. In addition, we emphasize that the errors in the public catalog come from two different data sets (VISTA and 2MASS) which must be considered carefully for any application involving them.

Figure \ref{img:catalogs} shows the color-magnitude diagrams ($J$ vs. $H - K_S$) of our survey in comparison to several other data products including the CASU reduction, UKIDSS \citep[DR10,][]{2007MNRAS.379.1599L}, and the 2MASS point source catalog. The errors of the photometry in $K_S$ are shown in a discrete color-code. We note several things here: (a) our dedicated survey goes much deeper than UKIDSS and 2MASS; (b) VISION covers the largest dynamic range because we replaced the bright end of the luminosity function with 2MASS photometry; (c) The photometric errors also seem to improve over the standard CASU pipeline; (d) our optimized aperture photometry produces a narrower stellar sequence for faint magnitudes (compare e.g. the blue sources at $J \sim 20$ mag).

The completeness of our survey is expected to be spatially highly variable due to changing observing conditions. A reliable completeness estimate can only be determined from the same data on which source detection and extraction was performed. Thus, any test can not be performed on the  stacked pawprints, but must be applied to the Orion~A mosaics from which the final source catalogs were generated. For this reason we performed artificial star tests on 10 arcmin wide sub-fields of the full mosaics (the size of a VIRCAM detector amounts $\sim$11.6 $\times$ 11.6 arcmin). For each of these fields we applied multiple subsequent processing steps: We (1) constructed a PSF model with PSFEx, (2) performed source extraction and profile fitting with SExtractor to subtract all significant sources, (3) measured the source density, (4) constructed a set of artificial stars from the given PSF (with the calculated stellar density) with magnitudes ranging from 17 to 22.5 mag with Skymaker, (5) performed source extraction on the artificial sources placed on the PSF-subtracted fields, (6) calculated the completeness relative to the input source list of the artificial stars. Steps 4 -- 6 were repeated 50 times for statistical reasons. The final completeness estimate (as a function of magnitude) for each field was calculated as the mean of all iterations. The completeness-magnitude distribution for these sub-fields were very well fitted by a modified logistic function of the form:

\begin{equation}
f(x) = -\frac{1}{1 + e^{-k \cdot (x - x_0)}} + 1
\end{equation}

We combined the results of these individual sub fields to estimate the completeness for the entire survey. The results of this procedure are displayed in Fig. \ref{img:completeness}, where we show the completeness as a function of source magnitude as well as its spatial dependency (displayed at the 90\% level) and tabulate the results in Tab. \ref{tab:completeness}. We note here, that the completeness function for the entire cloud is not well described by the aforementioned function due to the spatial variations. While for individual fields we observe almost 100\% completeness up to the point where the function appears to drop, the distributions for the entire survey show a small continuous decline towards this point. As a consequence we see a large difference in completeness magnitude between e.g. 99\% and 95\%, even though large parts of the survey are essentially complete at the latter limit.

\begin{figure*}[p]
        \centering
        \includegraphics[width=16cm]{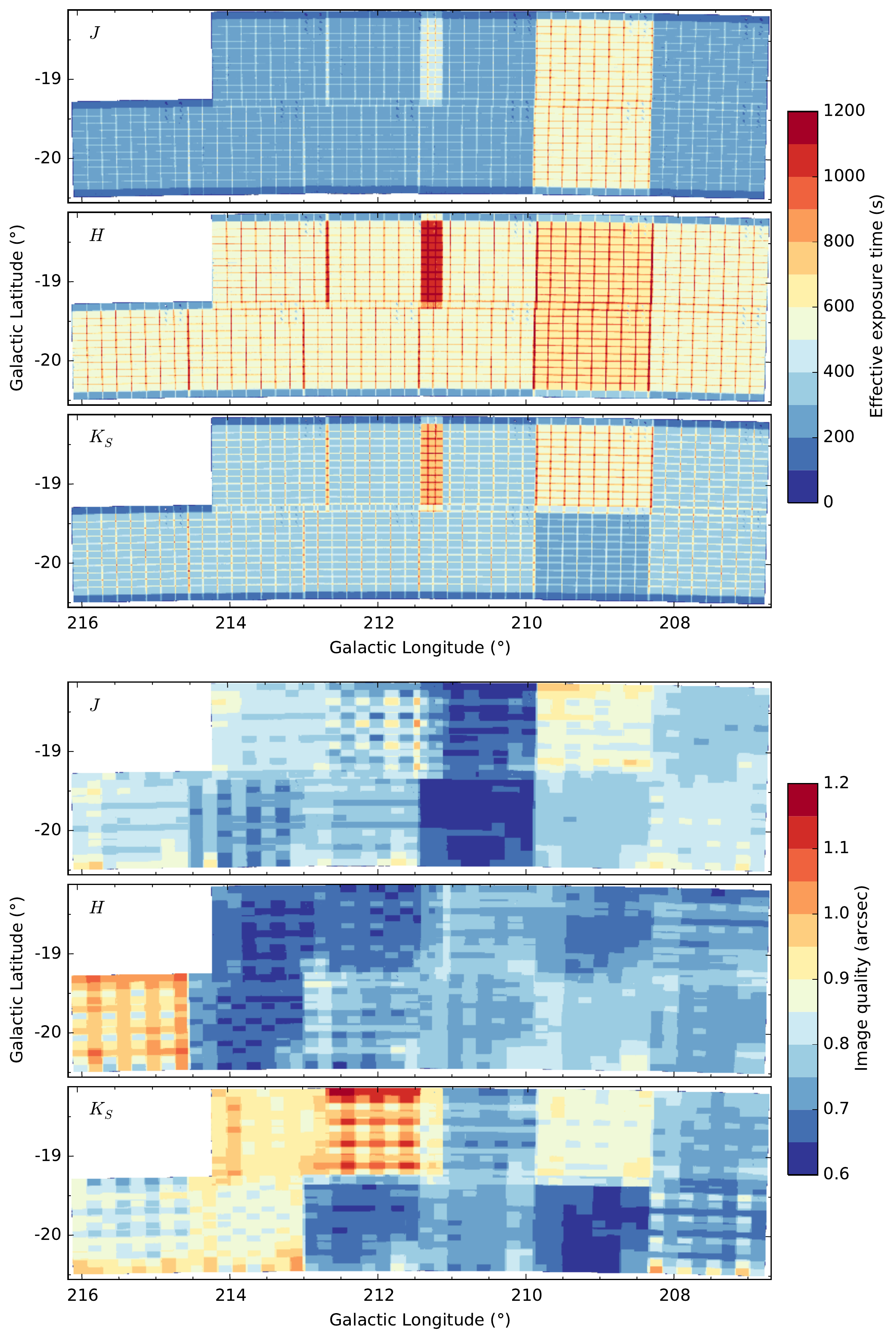} 
        \caption[]{Effective exposure time (top) and image quality (FWHM, bottom) across the Orion~A mosaic. Note the lower effective exposure time for tile S2 in $K_S$ and the very patchy FWHM structure.}
        \label{img:qc_quality}
\end{figure*}

The spatial variations of the completeness correlate well with the observing conditions (compare with Fig. \ref{img:qc_quality}) and the presence of bright nebulous emission near the ONC. For e.g. tile S6 (east-most tile) in $H$ band we observe PSF FWHMs exceeding 1 arcsec, and at the same time a noticeably shallower completeness. Tile S3 in $J$ featured the best observing conditions in our survey where we see an increase in sensitivity of $\sim$1 mag for this region.

\begin{figure}[tp]
        \resizebox{\hsize}{!}{\includegraphics{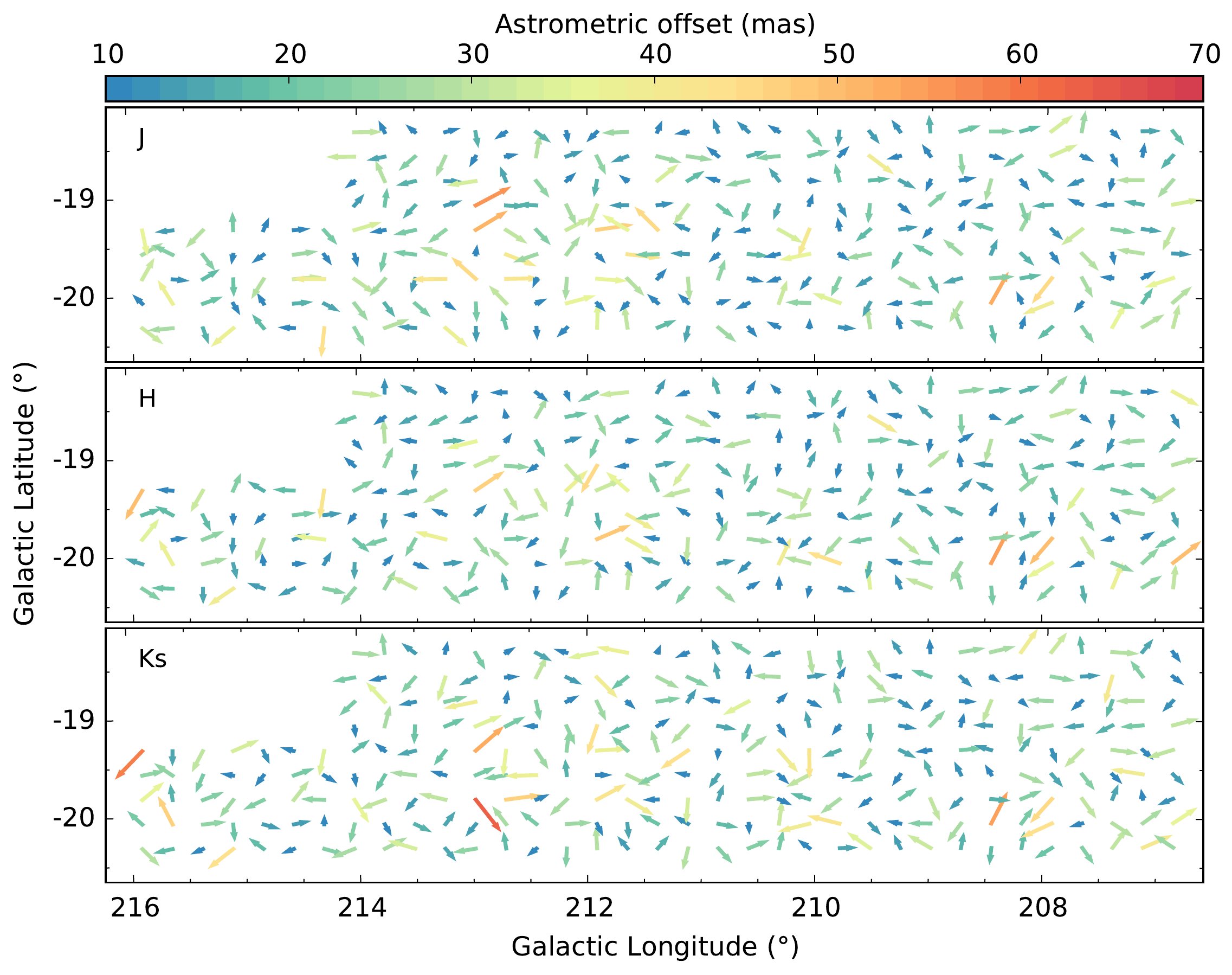}}
        \caption[]{Mean astrometric offsets relative to 2MASS for the final Orion~A source catalog in $15 \times 15$ arcmin boxes. No systematic trends across the entire field for all three bands are visible. The colors indicate the mean offset in the box which also is linearly proportional to the arrow lengths.}
        \label{img:astrometry}
\end{figure}

\subsection{Astrometric properties}
\label{app:astr}

The overall astrometric calibration of our data was done with Scamp using 2MASS as a reference catalog (see Sect. \ref{sec:astromcal} for details). The global error budget resulted in an RMS of about 70 mas with respect to reference sources and about 40 mas when considering internal source matches only. The most important factors contributing to this discrepancy are 2MASS S/N limits, general catalog errors, unresolved multiple sources, and unaccounted proper motions. To check for any remaining local systematic errors in VISION we used the final catalog, cross-matched again to 2MASS with a maximum allowed distance of 1 arcsec and subsequently calculated the mean astrometric offsets in boxes of $15 \times 15$ arcmin. The result is displayed in Fig. \ref{img:astrometry} which shows the sum of all offsets in a given box including magnitude and direction. No local systematic trend can be seen. The errors in this figure are generally smaller than the given global error due to the averaging in each given box. When decreasing the box size to below 10 arcmin we still see no systematics trends across the field, but the magnitude of the errors increase as expected.

\begin{figure}[tp]
        \centering
        \resizebox{\hsize}{!}{\includegraphics{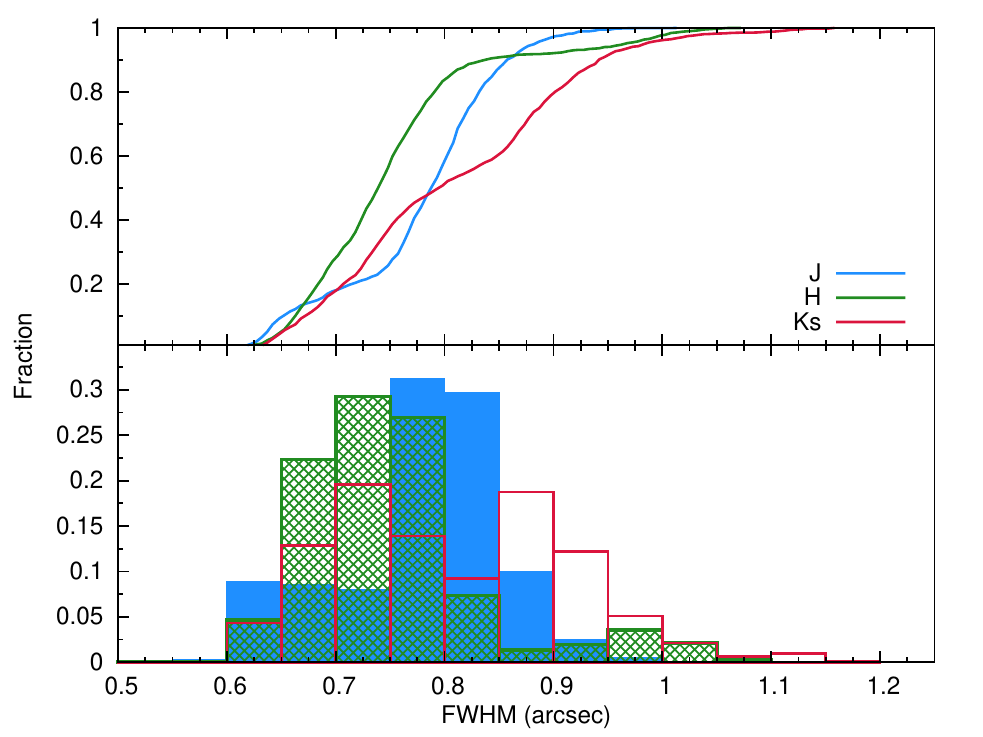}}
        \caption[]{Seeing statistics for the entire Orion~A survey. The bottom row shows the normalized histogram for all three bands in bin widths of 0.05 arcsec whereas the top graph displays the cumulative histograms with a refined resolution. Most data have been taken during very good seeing conditions resulting in FWHMs measured on point-sources in the range between 0.7 and 0.85 arcsec for $J$ and $H$; only $K_S$ features a significant amount of data with point-source FWHM values $\ga 0.85$ arcsec.
    
    }
        \label{img:seeing}
\end{figure}

\subsection{Quality control parameters}
\label{app:qc}

As part of the data calibration we also generated several quality control parameters for the entire survey. The most important among them are the image quality and survey coverage. Each source in the catalog is also supplemented with its associated effective exposure time, frame coverage, observing date (MJD), source FWHM and local seeing conditions. Figure \ref{img:qc_quality} visualizes effective exposure time and image quality for our survey. The image quality here refers to the measured seeing value for each photometrically stable subset (see Sect. \ref{sec:coadd} for details). In the exposure time maps the overlaps between the detectors and tiles are clearly visible. Note here tile S2 in $K_S$ for which we only included one of the observed sequences due to the large discrepancy in observing conditions. Fortunately the other sequence features one of the best image qualities of the entire survey and no obvious decrease in the completeness is observed with respect to the other tiles (compare with Fig. \ref{img:completeness}). We can also see that the image quality can vary by a few tens of percent even within tiles. This complex structure is simply a consequence of the observing strategy with VISTA where the same position is observed by multiple detectors spread out over the entire length of the OB. 

Most data were taken under excellent seeing conditions. The mean seeing, determined as the FWHM from a bright, high S/N subset, was 0.78, 0.75, 0.8 arcsec for Orion~A, and 0.69, 0.77, 0.76 for the CF in $J, H$ and $K_S$, respectively. Figure \ref{img:seeing} displays histograms of the image quality statistics. The $K_S$ data show a bi-modal distribution which can (coincidentally) also clearly be seen in Fig. \ref{img:qc_quality} in a roughly east-west oriented gradient.

\section{Tables}
\label{app:tables}

Here we provide supplementary tables for the main text. In addition to Tab. \ref{tab:vision_columns}, which shows the published columns of the source catalog, Tab. \ref{tab:vision_sample} contains a sample of 12 out of the 35 total columns from the final catalog. Table \ref{tab:newobj} contains coordinates and photometry for  all sources discussed in Sect. \ref{sec:results_newobjects}.

\begin{sidewaystable*}
        \caption{Sample data taken from the VISTA Orion~A catalog. Only 12 out of the 35 total columns are shown.}
        \label{tab:vision_sample}
        \centering
        \begin{tabular}{c c c c c c c c c c c c c} 
        \hline\hline             
        VISION                  &RAJ2000        &DEJ2000	    &Class\_cog     &Class\_sex     &$K_S$  &$K_S$\_err     &$K_S$\_mjd     &$K_S$\_exptime &$K_S$\_fwhm    &$K_S$\_seeing  &$K_S$\_coverage\\
                                &(hh:mm:ss)                     &(dd:mm:ss)             &\{0,1\}        &[0,1]          &(mag)  &(mag)          &                       &(s)                    &(arcsec)               &(arcsec)               &                               \\
        \hline
        05362896-0409496&05:36:28.96            &-04:09:49.6    &0              &1.0                         &19.289 &0.184          &56319.047      &400                    &1.55                   &0.76                   &10                             \\     
        05362896-0711209&05:36:28.96            &-07:11:20.9    &0              &1.0                         &18.396 &0.103          &56322.035      &440                    &1.85                   &0.7                         &11                             \\     
        05362896-0724448&05:36:28.96            &-07:24:44.8    &1              &0.973                  &17.443 &0.044          &56322.043      &560                    &0.79                   &0.73                   &14                             \\     
        05362896-0652099&05:36:28.96            &-06:52:09.9    &0              &0.324                  &17.627 &0.058          &56322.03         &400                    &1.34                   &0.72                   &10                             \\     
        05362896-0822224&05:36:28.96            &-08:22:22.4    &1              &0.959                  &18.868 &0.198          &56332.07         &200                    &0.88                   &0.74                   &5                                 \\     
        05362896-0450050&05:36:28.96            &-04:50:05.0    &0              &0.004                  &19.096 &0.174          &56320.555      &630                    &1.54                   &0.86                   &21                             \\     
        05362896-0555005&05:36:28.96            &-05:55:00.5    &0              &1.0                         &19.084 &0.168          &56281.82       &450                    &1.13                   &0.85                   &15                             \\     
        05362896-0521375&05:36:28.96            &-05:21:37.5    &0              &1.0                         &18.583 &0.105          &56320.555      &750                    &2.37                   &0.87                   &25                             \\     
        05362896-0808415&05:36:28.96            &-08:08:41.5    &1              &0.979                  &17.534 &0.059          &56332.07         &400                    &0.9                    &0.83                   &10                             \\     
        05362896-0411396&05:36:28.96            &-04:11:39.6    &1              &0.61                         &19.143 &0.155          &56319.043      &480                    &1.19                   &0.77                   &12                             \\     
        05362896-0406475&05:36:28.96            &-04:06:47.5    &1              &0.995                  &18.981 &0.136          &56319.04         &400                    &1.16                   &0.76                   &10                             \\     
        05362896-0454271&05:36:28.96            &-04:54:27.1    &1              &0.983                  &17.672 &0.059          &56320.555      &600                    &0.86                   &0.85                   &20                             \\     
        05362896-0334519&05:36:28.96            &-03:34:51.9    &0              &1.0                         &17.548 &0.076          &56319.05       &200                    &2.77                   &0.75                   &5                                 \\     
        05362897-0740077&05:36:28.97            &-07:40:07.7    &1              &0.999                  &18.011 &0.066          &56322.035      &520                    &0.85                   &0.74                   &13                             \\     
        05362897-0544574&05:36:28.97            &-05:44:57.4    &0              &0.0                         &18.761 &0.127          &56320.55       &600                    &1.57                   &0.87                   &20                             \\     
        05362897-0354050&05:36:28.97            &-03:54:05.0    &1              &0.983                  &17.353 &0.048          &56319.04         &360                    &0.8                    &0.75                   &9                                 \\     
        05362897-0827114&05:36:28.97            &-08:27:11.4    &1              &0.996                  &14.922 &0.02                 &56332.066      &200                    &0.78                   &0.75                   &5                                 \\     
        05362897-0452245&05:36:28.97            &-04:52:24.5    &1              &0.98                         &17.89  &0.07           &56320.555      &600                    &0.94                   &0.85                   &20                             \\     
        05362897-0528390&05:36:28.97            &-05:28:39.0    &1              &0.978                  &17.14         &0.047          &56320.555      &600                    &0.91                   &0.85                   &20                             \\     
        05362897-0531284&05:36:28.97            &-05:31:28.4    &1              &0.983                  &15.278 &0.017          &56320.555      &600                    &0.88                   &0.85                   &20                             \\     
        05362897-0328278&05:36:28.97            &-03:28:27.8    &1              &0.943                  &18.101 &0.11                 &56319.055      &200                    &1.12                   &0.76                   &5                                 \\     
        05362898-0438323&05:36:28.98            &-04:38:32.3    &1              &0.069                  &17.997 &0.08                 &56320.555      &600                    &1.12                   &0.88                   &20                             \\     
        05362898-0407127&05:36:28.98            &-04:07:12.7    &1              &0.911                  &15.942 &0.021          &56319.043      &480                    &0.8                         &0.76                   &12                             \\     
        05362898-0814538&05:36:28.98            &-08:14:53.8    &0              &0.819                  &18.699 &0.115          &56332.08         &400                    &1.02                   &0.72                   &10                             \\     
        05362898-0449499&05:36:28.98            &-04:49:49.9    &1              &0.989                  &16.752 &0.034          &56320.555      &660                    &0.87                   &0.86                   &22                             \\     
        05362898-0502207&05:36:28.98            &-05:02:20.7    &1              &0.975                  &18.191 &0.068          &56320.555      &630                    &1.13                   &0.86                   &21                             \\     
        05362898-0700397&05:36:28.98            &-07:00:39.7    &1              &0.983                  &17.851 &0.073          &56322.04         &400                    &0.82                   &0.74                   &10                             \\     
        05362898-0748052&05:36:28.98            &-07:48:05.2    &1              &0.96                         &17.806 &0.052          &56327.055      &800                    &0.8                         &0.75                   &20                             \\     
        05362898-0434172&05:36:28.98            &-04:34:17.2    &1              &0.947                  &16.247 &0.025          &56320.555      &600                    &0.93                   &0.87                   &20                             \\     
        05362898-0654349&05:36:28.98            &-06:54:34.9    &1              &0.999                  &18.537 &0.1                 &56322.03       &400                    &0.82                   &0.72                   &10                             \\     
        05362898-0532286&05:36:28.98            &-05:32:28.6    &1              &0.975                  &18.349 &0.105          &56320.555      &600                    &1.02                   &0.85                   &20                             \\     
        05362898-0327518&05:36:28.98            &-03:27:51.8    &1              &0.994                  &15.162 &0.02                 &56319.055      &200                    &0.83                   &0.76                   &5                                 \\     
        05362899-0702345&05:36:28.99            &-07:02:34.5    &1              &1.0                         &19.029 &0.144          &56322.043      &520                    &0.92                   &0.73                   &13                             \\     
        05362899-0343035&05:36:28.99            &-03:43:03.5    &1              &1.0                         &19.647 &0.199          &56319.047      &560                    &0.68                   &0.77                   &14                             \\     
        05362899-0810323&05:36:28.99            &-08:10:32.3    &1              &0.968                  &16.264 &0.028          &56332.07         &400                    &0.81                   &0.83                   &10                             \\     
        05362899-0754268&05:36:28.99            &-07:54:26.8    &0              &1.0                         &18.765 &0.133          &56332.086      &400                    &1.19                   &0.79                   &10                             \\      
    \hline
        \end{tabular}
  \tablefoot{The full table is available in electronic form at the CDS.}
\end{sidewaystable*}

\begin{table*}[p]
        \centering
    \caption{Cross identifications, coordinates, and magnitudes for all the objects identified in Sect. \ref{sec:results_newobjects}. These objects are also shown in Fig. \ref{img:matrix_ysos} and \ref{img:matrix_new_ysos}.}
    \label{tab:newobj}
    \begin{tabular*}{\textwidth}{c @{\extracolsep{\fill}} c c c c c c c c}
    \hline\hline
    VISION              &Label\tablefootmark{a} &       Right Ascension & Declination                    &$J$    &$H$    &$K_S$  &Size\tablefootmark{b}          & ID\tablefootmark{c}     \\
                                &               &       (hh:mm:ss)              &(dd:mm:ss)                             &(mag)  &(mag)  &(mag)  &(arcsec, pc)     &               \\
        \hline
    \multicolumn{9}{c}{\textit{Class I}}                                                                                                                                                        \\
        \hline  
    05381810-0702259&a          &05:38:18.10            &-07:02:25.9                    &9.253  &6.762  &5.154  &330, 0.66    &670    \\
    05402745-0727300&b          &05:40:27.45            &-07:27:30.0                    &-              &-              &9.694  &180, 0.36    &536    \\
    05424707-0817070&c          &05:42:47.07            &-08:17:07.0                    &15.998 &14.158 &11.063 &33, 0.07    &287    \\
    05413419-0835274&d          &05:41:34.19            &-08:35:27.4                    &16.232 &14.430 &12.103 &68, 0.14    &257    \\
    05352985-0626583&e          &05:35:29.85            &-06:26:58.3                    &14.957 &14.491 &12.302 &68, 0.14    &879    \\
    05363034-0432170&f          &05:36:30.34            &-04:32:17.0                    &15.496 &13.641 &12.648 &50, 0.10    &2748   \\
    05410201-0806019&g          &05:41:02.01            &-08:06:01.9                    &17.494 &14.606 &13.118 &68, 0.14    &362    \\
    05412474-0754081&h          &05:41:24.74            &-07:54:08.1                    &-              &17.745 &13.416 &60, 0.12    &439    \\
    05412398-0753421&i          &05:41:23.98            &-07:53:42.1                    &-              &17.437 &13.731 &60, 0.12    &445    \\
    05350554-0551541&j          &05:35:05.54            &-05:51:54.1                    &18.628 &16.003 &14.108 &66, 0.13    &1165   \\
    05344909-0541419&k          &05:34:49.09            &-05:41:41.9                    &-              &16.780 &15.240 &53, 0.11    &1294   \\
    05402095-0756240&l          &05:40:20.95            &-07:56:24.0                    &-              &-              &15.540 &57, 0.11    &423    \\
    05325056-0534424&m          &05:32:50.56            &-05:34:42.4                    &19.486 &15.958 &16.066 &45, 0.09    &1433   \\
    05361721-0638016&n          &05:36:17.21            &-06:38:01.6                    &-              &17.089 &-              &50, 0.10    &823    \\
    -                           &o              &-                                      &-                                              &-              &-              &-              &53, 0.11    &1504   \\
    \hline
    \multicolumn{9}{c}{\textit{Class II}}                                                                                                                                                       \\
        \hline  
    05362543-0642577&a          &05:36:25.43            &-06:42:57.7                    &8.107  &6.964  &5.947  &180, 0.36    &796    \\
    05384279-0712438&b          &05:38:42.79            &-07:12:43.8                    &10.823 &9.283  &8.124  &83, 0.17    &618    \\
    05384322-0658089&c          &05:38:43.22            &-06:58:08.9                    &12.046 &9.978  &8.617  &75, 0.15    &726    \\
    05404806-0805587&d          &05:40:48.06            &-08:05:58.7                    &9.914  &9.140  &8.751  &60, 0.12    &364    \\
    05404662-0807128&e          &05:40:46.62            &-08:07:12.8                    &11.928 &10.203 &9.333  &120, 0.24    &351    \\
    05353163-0500141&f          &05:35:31.63            &-05:00:14.1                    &14.517 &11.964 &10.129 &45, 0.09    &2450   \\
    05362378-0623113&g          &05:36:23.78            &-06:23:11.3                    &15.377 &13.225 &11.111 &60, 0.12    &925    \\
    05410413-0923194&h          &05:41:04.13            &-09:23:19.4                    &13.497 &13.033 &12.637 &27, 0.05    &129    \\
    05363700-0614579&i          &05:36:37.00            &-06:14:57.9                    &16.731 &14.379 &13.258 &45, 0.09    &996    \\
    05384652-0705375&j          &05:38:46.52            &-07:05:37.5                    &16.978 &15.933 &13.550 &45, 0.09    &649    \\
    \hline      
    \multicolumn{9}{c}{\textit{Class III}}                                                                                                                                                      \\      
        \hline          
    05350906-0614200&a          &05:35:09.06            &-06:14:20.0                    &9.612  &9.292  &9.188  &30, 0.06    &125    \\
    05375451-0656455&b          &05:37:54.51            &-06:56:45.5                    &10.704 &9.916  &9.697  &30, 0.06    &1040   \\
    05352974-0548450&c          &05:35:29.74            &-05:48:45.0                    &10.739 &10.067 &9.893  &30, 0.06    &765    \\
    05345803-0612238&d          &05:34:58.03            &-06:12:23.8                    &10.893 &10.163 &9.974  &30, 0.06    &137    \\
    05431072-0831500&e          &05:43:10.72            &-08:31:50.0                    &10.823 &10.297 &10.125 &30, 0.06    &289    \\
    \hline
    \multicolumn{9}{c}{\textit{New YSO candidates}}                                                                                                                                     \\
        \hline  
    05312709-0427593&a          &05:31:27.09            &-04:27:59.3                    &13.383 &11.086 &9.425  &135, 0.27    &-              \\
    05315171-0523082&b          &05:31:51.71            &-05:23:08.2                    &12.006 &10.492 &9.709  &90, 0.18    &-              \\
    05324165-0535461&c          &05:32:41.65            &-05:35:46.1                    &17.965 &13.797 &11.389 &45, 0.09    &-              \\
    05324165-0536115&d          &05:32:41.65            &-05:36:11.5                    &-              &18.889 &14.060 &30, 0.06    &-              \\
    05305155-0410348&e          &05:30:51.55            &-04:10:34.8                    &-              &-              &14.415 &45, 0.09    &-              \\
    05305129-0410322&e          &05:30:51.29            &-04:10:32.2                    &-              &-              &13.835 &45, 0.09    &-              \\
    \hline
    \multicolumn{9}{c}{\textit{New galaxy cluster candidates}}                                                                                                          \\
    \hline      
    -                           &a              &05:40:44                       &-09:57:56                              &-              &-              &-              &150, -               &-              \\
    -                           &b              &05:47:05                       &-08:55:17                              &-              &-              &-              &150, -               &-              \\
    -                           &c              &05:41:51                       &-09:06:33                              &-              &-              &-              &150, -               &-              \\
    -                           &d              &05:39:46                       &-08:47:38                              &-              &-              &-              &150, -               &-              \\
    -                           &e              &05:44:53                       &-08:03:58                              &-              &-              &-              &120, -               &-              \\
    -                           &f              &05:42:36                       &-06:59:34                              &-              &-              &-              &200, -               &-              \\
    -                           &g              &05:40:25                       &-05:59:36                              &-              &-              &-              &160, -               &-              \\
    -                           &h              &05:32:06                       &-06:05:21                              &-              &-              &-              &150, -               &-              \\
    -                           &i              &05:31:08                       &-05:30:48                              &-              &-              &-              &160, -               &-              \\
    -                           &j              &05:30:28                       &-04:14:42                              &-              &-              &-              &170, -               &-              \\
        \hline
        \end{tabular*}
\tablefoot{
        \tablefoottext{a}{Refers to the labels in the bottom left corners of the sub-figures in Fig. \ref{img:matrix_ysos} and \ref{img:matrix_new_ysos}.}
    \tablefoottext{b}{Size of the postage stamp in Fig. \ref{img:matrix_ysos} and \ref{img:matrix_new_ysos}. The physical sizes were calculated with the adopted distance of 414 pc.}
    \tablefoottext{c}{For Class I/II sources the ID refers to the internal numbering of \cite{2012AJ....144..192M}, for Class III to \cite{2013ApJ...768...99P}.}
}
\end{table*}

\end{appendix}

\end{document}